\documentclass[prd,12pt,nofootinbib,showpacs,tightenlines,longbibliography]{revtex4-1}

\usepackage{hyperref}
\usepackage{amsmath,amssymb,amsthm,graphicx}
\usepackage[mathscr]{eucal}
\newcommand{\no}{\nonumber}
\usepackage{xcolor}
\usepackage{soul}
\usepackage{abraces}
\usepackage{multirow}

\begin{document}
\title{Analogous Hawking radiation and quantum entanglement in two-component
	Bose-Einstein condensates: The gapped excitations}
\date{\today}
\author{Wei-Can Syu}
\email{syuweican@gmail.com}
\affiliation{Department of
	Physics, National Dong-Hwa University, Hualien 974301, Taiwan, R.O.C.}
\author{Da-Shin Lee}
\email{dslee@gms.ndhu.edu.tw}
\affiliation{Department of
	Physics, National Dong-Hwa University, Hualien 974301, Taiwan, R.O.C.}
\author{Chi-Yong Lin}
\email{lcyong@gms.ndhu.edu.tw}
\affiliation{Department of
Physics, National Dong-Hwa University, Hualien 974301, Taiwan, R.O.C.}

\begin{abstract}
	The condensates of cold atoms at zero temperature in the tunable binary Bose-Einstein condensate system  are studied with  the Rabi transition between  atomic hyperfine states where the system can be represented by a coupled two-field model of gapless excitations and gapped excitations. We set up the configuration of the supersonic and subsonic regimes with the acoustic horizon between them in the elongated two-component Bose-Einstein condensates, trying to mimic Hawking radiations, in particular due to the gapped excitations. The simplified steplike sound speed change is adopted for the subsonic-supersonic transition so that the model can be analytically treatable. The effective-energy gap term in the dispersion relation of the gapped excitations introduces the threshold frequency $\omega_\text{min}$ in the subsonic regime, below which the propagating modes do not exist.
	Thus, the particle spectrum of the Hawking modes significantly deviates from that of the gapless cases near the threshold frequency due to the modified gray-body factor, which vanishes as the mode frequency is below $\omega_\text{min}$. The influence from the gapped excitations to the quantum entanglement of the Hawking mode and its partner of the gapless excitations is also studied according to the Peres-Horodecki-Simon (PHS)  criterion. It is found that the presence of the gapped excitations will deteriorate the quantumness of the pair modes of the gapless excitations when the frequency of the pair modes in particular is around  $\omega \sim \omega_\text{min}$. On top of that, when the coupling constant between the gapless and gapped excitations becomes large enough, the huge particle density of the gapped excitations in the small $\omega$ regime will significantly disentangle the pair modes of the gapless excitations. The detailed time-dependent PHS criterion will be discussed.
\end{abstract}


\maketitle
\newpage
\section{Introduction}

The program of the analog models of gravity is an attempt to implement laboratory systems to mimic various phenomena that happen  in the interplay between general relativity and quantum field theory such as  in black holes and the early Universe.
The aim is to devise experiments of  real laboratory tests that provide insights in the phenomena and further probe the structure of curved-space quantum field theory.
The beginning of analog gravity dates back to the pioneering work of Unruh \cite{Unruh1981}, who used the sound waves in a moving fluid as an analogue to light waves in curved spacetime and further showed that supersonic fluid can generate  a ``dumb hole'', an acoustic analogue of a ``black hole''.
From there, the existence of analog photonic Hawking radiations due to the presence of the acoustic horizon can be theoretically demonstrated.
Since its development, the analog gravity program has received much attention to explore fundamental physics through interdisciplinary research among particle/astrophysicists and condensed matter physicists (for a review see \cite{Visser2005}).
Despite the early start of  theoretic investigations, great progress has also been made recently in experimental  analog gravity to advance its technologies for realizing the Hawking effects.
One of the most successful systems is the Bose-Einstein condensations (BECs).
The work of \cite{Jeff2019} is a first experimental observation of Hawking radiation extracted from the correlations of the collective excitations that agree with a thermal spectrum with the temperature estimated from  analog surface gravity.
Also, the time dependence of the Hawking radiation in an analogue black hole is observed in \cite{Kolobov:2021wd}.

Most of the works for BEC analogous black holes  consider the Hawking radiations due to the gapless excitations analogous to the massless scalar particles.
As the Hawking temperature rises to the order of the mass scale of some massive particles, the production of these massive particles also becomes significant \cite{Diatlyk2021}.
The aspects of the analogue models of the gapped excitations of the BECs have been explored in \cite{Jannes2011, Antonin2012, Dudley2018}, where the energy-gap term is either induced from the transverse wave number inversely proportional to the size in the perpendicular direction  in elongated quasi-one-dimensional system or added by hand as a toy model.
One of the main features for the gapped excitations  includes the existence of the minimum frequency $\omega_{\rm min}$, below which in the subsonic regime the propagating modes do not survive resulting in the total reflection of radiation coming from the supersonic regime at the horizon.
The other feature is that solving the dispersion relation can find the nonvanishing zero-frequency mode where their density-density correlation functions reveal the undulation phenomenon in the supersonic regime.
In this work, we will focus on the gapped excitations created from the binary BECs.
Two-component BECs   have been experimentally studied using the mixture of atoms with two hyperfine states of $^{87}\text{Rb}$~\cite{Tojo2010} or the mixture of two different species of atoms~\cite{Modugno2002,Thalhammer2008,Papp2008,McCarron2011,Cipriani2013}.
In the analogue gravity program,   the class of two-component BECs subject to laser- or radio-wave induced Rabi transition between different atomic hyperfine states  has been proposed to serve as an ``emergent'' spacetime model, which provides  very rich spacetime geometries, such as a specific class of pseudo-Finsler geometries, and both bimetric pseudo-Riemannian geometries and single-metric pseudo-Riemannian geometries of interest in cosmology and general relativity~\cite{Fischer2004,Visser2005,Liberati2006,Liberati2006_2,Weinfurtner2007}.
In fact, this class of the two-component BEC systems with the Rabi interaction exhibits two types of  excitations on condensates: the gapless excitations due to the ``in phase'' oscillations between  two respective density waves of the binary system and  the gapped excitation stemming from the ``out-of-phase'' oscillations of the density waves in the presence of the Rabi transition, which are respectively analogous of the Goldstone modes and the Higgs modes in particle physics.
In addition, in \cite{Syu2020}  the dynamics of collective atomic motion by choosing tunable scattering lengths through Feshbach resonances has been studied with the introduced effective parameter characterizing the miscible or immiscible regime of binary condensates and their stabilities.

In this system, we plan to set up the configuration of the supersonic and subsonic regimes with the acoustic horizon  between them.
We consider the simplified steplike sound speed profile to implement the subsonic-supersonic transition in the elongated two-component Bose-Einstein condensates using the tunable couplings between the atoms through the Feshbach resonances \cite{Hamner2011, Hamner2013} so that the configuration can be analytically treatable.
Although the problem of this sudden transition gives the infinity acoustic surface gravity,  the corresponding Hawking temperature still can be read off from the obtained spectrum of the Hawking modes.
%
More realistic transitions can be considered in the waterfall configuration that relies on full numerical studies to explore its physics. The experimentally spatial variation of the interaction strengths to fit into the configuration  is challenged but feasible \cite{Clark2015,Arunkumar2019,Carli2020}.
We first find the dispersion relation of the gapped excitations and identify the various modes in both supersonic and subsonic regimes.
The energy-gap term  is governed by the Rabi-coupling constant $\Omega$ between two hyperfine states of Bose-Einstein condensates \cite{Visser2005},
$m_\text{eff}\propto \sqrt{\Omega}$, which is tunable experimentally.
In addition, a spatial varying of the Rabi-coupling constant is also doable experimentally \cite{Arunkumar2019}, which allows to design two different effective-energy gaps on the subsonic and supersonic regimes.
The matching between two sets of the wave functions in sub/supersonic regimes allows to define the $S$-matrix, whose elements can be experimentally determined from the density-density correlation functions \cite{Jeff2019}.
One of the main results in this paper then comes from the study of the quantum entanglement between the Hawking mode and its partner of the gapped excitations by employing the Peres-Horodecki-Simon (PHS)  criterion \cite{Horodecki1996,Horodecki1997,Simon2000}.
On top of that, to this binary Bose-Einstein condensate system, it provides the opportunity to study the analogue gravity phenomena in the open quantum system, where  the gapless excitations are  treated as the system and the gapped excitations as an environment.
The same idea has been employed to examine the effect of quantum fluctuations due to the gapped excitations on phonon propagation in the binary BECs system to build up an analogous model of the light cone fluctuations induced by quantum gravitational effects in \cite{Syu2019}.
Here we study the entanglement  between the  Hawking mode and its partner of the gapless excitations under the influence of the  environment of the gapped excitations.

We organize this paper as follows.  In Sec.~\ref{ABH2}, we introduce the model of interest and construct the  gapped excitations in the supersonic/subsonic configuration where the acoustic horizon is present between them in a binary BECs system.
In Sec.~\ref{ABH3},  the matching of the mode functions of two sides of the acoustic horizon is carried out to obtain the scattering coefficients for three outgoing channels.
In Sec.~\ref{ABH4},  we study density-density correlation functions.
Section~\ref{ABH5} is devoted to the study of the nonseparability of the gapless excitations  influenced by the gapped excitations.
We conclude the work in Sec.~\ref{ABH6}.

\section{The model}\label{ABH2}
\subsection{The Bogoliubov-de Gennes equations in coupled Bose condensates}
We consider the binary BECs of the same atoms in two different internal hyperfine states.
We then assume that a strong cigar-shape trap potential is used where the size of the trap $L_x$ along the axial direction, say in the $x$-direction is much larger than the size of $L_r$ along the radial direction.
Therefore, the system (in units of $\hbar=k_B=1$) can be treated in the pseudo-one-dimension with the Lagrangian given by
\begin{align}
L_\text{1D}=&\int dx\Bigg\{\sum_{a=A,B} \left[\frac{i}{2}\left(\hat\Psi_a^\dagger {\partial_t \hat\Psi_a}-\hat\Psi_a{\partial_t \hat\Psi_a^\dagger }\right)-\left(\frac{1}{2 m}\partial_x\hat\Psi_a^\dagger \partial_x\hat\Psi_a+ V_{a} \hat\Psi_a^\dagger\hat\Psi_a+\frac{U_{aa}}{2}\hat\Psi_a^\dagger\hat\Psi_a^\dagger\hat\Psi_a\hat\Psi_a\right)\right]\nonumber\\
&\qquad\qquad\qquad
- \, U_{AB}\hat\Psi_A^\dagger\hat\Psi_B^\dagger\hat\Psi_A\hat\Psi_B+\frac{\Omega}{2} (\hat\Psi_A\hat \Psi_B^\dagger+\hat \Psi_A^\dagger \hat\Psi_B)\Bigg\}\; ,
\label{Lag}
\end{align}
where $m$ is atomic mass and $V_A,\,V_B$ are the external potential along the axial direction on the hyperfine states $A$ and $B$.
The field operators obey the equal-time commutation relations 
\begin{equation}
[\hat{\Psi}_a(x,t),\hat{\Psi}_b^\dagger(x',t)]=\delta_{ab}\delta(x-x')\,.
\end{equation}
The interaction strengths of atoms between the same hyperfine states and different hyperfine states  are given by $U_{AA},\,U_{BB}$ and $U_{AB}$, respectively.
The coupling strengths  are related with the scattering lengths $a$ as $U=2 {a}/m L_r^2$.
Experimentally, the values of  scattering lengths can be tuned using Feshbach resonances such as two hyperfine states of $^{87}\text{Rb}$~\cite{Thalhammer2008,Tojo2010}.
In addition, we introduce a Rabi-coupling term by shining the laser field or applying the radio wave with the strength given by the Rabi frequency $\Omega$.

The corresponding  time-dependent equations of motion are obtained as
\begin{subequations} \label{FE}
\begin{align}
&i\partial_t\hat{\Psi}_A=\left[-\frac{1}{2m}\partial_x^2+V_A(x)+U_{AA}\hat{\Psi}_A^\dagger\hat{\Psi}_A+U_{AB}\hat{\Psi}_B^\dagger\hat{\Psi}_B\right]\hat{\Psi}_A-\frac{\Omega}{2}\hat{\Psi}_B\, ,\label{FE1}\\
&i\partial_t\hat{\Psi}_B=\left[-\frac{1}{2m}\partial_x^2+V_B(x)+U_{BB}\hat{\Psi}_B^\dagger\hat{\Psi}_B+U_{AB}\hat{\Psi}_A^\dagger\hat{\Psi}_A\right]\hat{\Psi}_B-\frac{\Omega}{2}\hat{\Psi}_A \, .
\label{FE2}
\end{align}
\end{subequations}
The condensate wave functions are given by the expectation value of the field operator $\langle\hat{\Psi}_a \rangle=\psi_ae^{-i\mu t}$, where they are governed by the stationary  Gross-Pitaevskii (GP)  equations \cite{pethick2008}
\begin{subequations} \label{GPEs}
	\begin{align}
		&\mu{\psi}_A=\left[-\frac{1}{2m}\partial_x^2+V_A(x)+U_{AA}\vert{\psi}_A\vert^2+U_{AB}\vert{\psi}_B\vert^2\right]{\psi}_A-\frac{\Omega}{2}{\psi}_B\, ,\label{GPEs1}\\
		&\mu{\psi}_B=\left[-\frac{1}{2m}\partial_x^2+V_B(x)+U_{BB}\vert{\psi}_B\vert^2+U_{AB}\vert{\psi}_A\vert^2\right]{\psi}_B-\frac{\Omega}{2}{\psi}_A \, ,
		\label{GPEs2}
	\end{align}
\end{subequations}
with the chemical potential $\mu$.
The condensate wave functions define the density $\rho_a$ and the phase $\theta_a$
\begin{align}
	\psi_a(x)=\sqrt{\rho_a(x)}e^{i\theta_a(x)}.
\end{align}
Meanwhile, the continuity equation gives the relations
\begin{align}\label{v_rho_c}
 v_a (x)\rho_a (x)=\text{constant}
\end{align}
with the condensate-flow velocities $\partial_x\theta_a (x)/m=v_a (x)$.

Here we choose the scattering parameters in this binary systems so as to have a stable and miscible state of the background condensates where $\rho_A=\rho_B=\rho $ and $\theta_A=\theta_B=\theta$, and also $U_{AA}=U_{BB}=U$ \cite{Hamner2011,Hamner2013}.
The detailed analysis of the choice of the parameters can be found in our previous work \cite{Syu2019}.
We further assume that $\rho=\rho_0$ is a constant across the whole condensate.
We also consider the constant condensate-flow velocity $v_A=v_B=-v\,(v>0)$ from the positive $x$ to the negative $x$.
For the interaction strengths, we consider the steplike change for $U(x)$, namely
\begin{align}
U(x)=\left\{
    \begin{array}{ll}
        U_{l}\,,\qquad x<0, \\
        U_{r}\,,\qquad x\ge 0, \\
    \end{array}
    \right.
    \label{u}
\end{align}
giving the steplike change of the sound speed across $x=0$ while keeping $U_{AB}$ and $\Omega$ uniform across the condensate.
The experimentally spatial variation of the interaction strengths is challenging but feasible \cite{Clark2015,Arunkumar2019,Carli2020}.
Additionally, the external potential is chosen to satisfy \cite{Carusotto2008,Recati2009}
\begin{align}
V_l+(U_l+U_{AB})\rho_{0}-\Omega/2 =V_r+(U_r+U_{AB})\rho_{0}-\Omega/2 \,.
\label{constrain}
\end{align}

The perturbations around the stationary wave function are defined through
\begin{align}
\hat{\Psi}_a=\langle\hat{\Psi}_a \rangle(1+\hat{\phi}_a) \, ,
\label{psi}
\end{align}
where the perturbed fields  $\hat{\phi}_a$ obey the equal-time commutation relations
\begin{equation}\label{cr}
\left[\hat{\phi}_{a}(x,t),\hat{\phi}^\dagger_{b}(x',t)\right]=\frac{1}{\rho_0}\delta_{ab}\delta(x-x')
\end{equation}
with $a=A,B$.
Substituting (\ref{psi}) into \eqref{FE} and using the GP equations we found the Bogoliubov-de Gennes equations
\begin{subequations} \label{Bogo_DeGenne}
\begin{align}
i\partial_t{\hat{\phi}}_A=&-\frac{1}{2m}\partial_x^2 \hat{\phi}_A-\frac{i}{m}\partial_x\theta\partial_x\hat{\phi}_A+U\rho_0(\hat{\phi}_A+
\hat{\phi}_A^\dagger)
+U_{AB}\rho_0(\hat{\phi}_B+
\hat{\phi}_B^\dagger)+\frac{\Omega}{2}(\hat{\phi}_A-
\hat{\phi}_B),\label{Bogo_DeGenne1}\\
i\partial_t{\hat{\phi}}_B=&-\frac{1}{2m}\partial_x^2\hat{\phi}_B-\frac{i}{m}\partial_x\theta\partial_x\hat{\phi}_B+U\rho_0(\hat{\phi}_B+
\hat{\phi}_B^\dagger)+U_{AB}\rho_0(\hat{\phi}_A+
\hat{\phi}_A^\dagger)+\frac{\Omega}{2}(\hat{\phi}_B-
\hat{\phi}_A)\, ,
\label{Bogo_DeGenne2}
\end{align}
\end{subequations}
where $U$ has a steplike form across $x=0$.

The system of equations (\ref{Bogo_DeGenne}) can be further decoupled by means of the field transformation
\begin{subequations} \label{decouple}
 \begin{align}
 &\hat{\phi}_{d}=\frac{1}{\sqrt{2}}(\hat{\phi}_A+\hat{\phi}_B)\, ,\label{d}\\
&\hat{\phi}_p=\frac{1}{\sqrt{2}}(\hat{\phi}_A-\hat{\phi}_B)\,,\label{p}
\end{align}
\end{subequations}
where the fields $\hat{\phi}_d$ and $\hat{\phi}_p$ are due to the density and polarization fluctuations, respectively.
It will be seen that the field $\hat{\phi}_{d}$ of the gapless excitations and the field $\hat{\phi}_p$ of the gapped excitations are analogous to the Goldstone and Higgs modes in particle physics.

In this time-translational invariant system, the field operators $\hat{\phi}_d$ and  $\hat{\phi}_p$ can be decomposed in the frequency  domain
\begin{subequations} \label{mode_expand}
\begin{align}
	&\hat{\phi}_d(x,t)=\sum_j \int d\omega\left[\hat{a}_{\omega j} \phi_{d\,\omega j}(x) e^{-i\omega t}+\hat{a}^{\dagger}_{\omega j} \varphi^\ast_{d\,\omega j}(x) e^{i\omega t}\right],\label{gapless}\\
	&\hat{\phi}_p(x,t)=\sum_j \int d\omega\left[\hat{b}_{\omega j}\phi_{p\,\omega j}(x) e^{-i\omega t}+\hat{b}^{\dagger}_{\omega j} \varphi^\ast_{p\,\omega j}(x) e^{i\omega t}\right]\,.\label{gapped}
\end{align}
\end{subequations}
Apart from the integration over the various frequencies $\omega$, for each $\omega$, there  exist either the outgoing channels or incoming channels $j$ to be summed over  that will be discussed later. 
 Notice that in the above decomposition, the propagating modes are involved where the range of the frequency for each channel $j$  of the gapless and gapped excitations for having the propagating modes can be different and relies on the dispersion of relation to be seen next.
The detailed expansions will be shown in a precise manner  below when all the modes in each side of $x$ are determined.
 In \eqref{mode_expand}, the creation and annihilation operators satisfy the canonical commutation relations, namely
\begin{equation}\label{ab_cr}
\left[\hat{a}_{\omega j},\hat{a}^\dagger_{\omega'j'}\right]=\left[\hat{b}_{\omega j},\hat{b}^\dagger_{\omega' j'}\right]=\delta_{jj'}\delta(\omega-\omega').
\end{equation}
Substituting \eqref{decouple} into \eqref{Bogo_DeGenne}, one obtains the wave equations for  the mode functions $\phi_{\omega}(x)$ and $\varphi_{\omega}(x)$
\begin{subequations} 	\label{phi_gapl_eq}
\begin{align}
	&\left[\left(\omega-iv\partial_x\right)+\frac{1}{2m\rho_0}\rho_0\partial^2_x\right]\,{\phi}_{d\,\omega}
	-\left[\left({U}+{U}_{AB}\right)\rho_0\right]\left({\phi}_{d\,\omega}+{\varphi}_{d\,\omega}\right)=0,\label{phi_gapl_eq1}\\
	&\left[\left(\omega-iv\partial_x\right)-\frac{1}{2m\rho_0}\rho_0\partial^2_x\right]\,{\varphi}_{d\,\omega}+\left[\left({U}+{U}_{AB}\right)\rho_0\right]\left({\phi}_{d\,\omega}+{\varphi}_{d\,\omega}\right)=0,
	\label{phi_gapl_eq2}
\end{align}
\end{subequations}
and
\begin{subequations} 	\label{phi_gap_eq}
\begin{align}
	&\left[\left(\omega-iv\partial_x\right)+\frac{1}{2m\rho_0}\rho_0\partial_x^2\right]\,{\phi}_{p\,\omega}-\left[\left({U}-{U}_{AB}\right)\rho_0+\frac{{\Omega}}{2}\right]\left(\phi_{p\,\omega}+\varphi_{p\,\omega}\right)-\frac{\Omega}{2}\left(\phi_{p,\omega}-\varphi_{p\,\omega}\right)=0,\label{phi_gap_eq1}\\
	&\left[\left(\omega-iv\partial_x\right)-\frac{1}{2m\rho_0}\rho_0\partial_x^2\right]\,\varphi_{p\,\omega}+\left[\left({U}-{U}_{AB}\right)\rho_0+\frac{{\Omega}}{2}\right]\left(\phi_{p,\omega}+\varphi_{p\,\omega}\right)-\frac{\Omega}{2}\left(\phi_{p,\omega}-\varphi_{p\,\omega}\right)=0\, .
	\label{phi_gap_eq2}
\end{align}
\end{subequations}
Together with (\ref{cr}), the normalization of the mode functions is given by
\begin{equation}\label{nc}
\int dx \, [\phi_{s\,\omega j}(x)\phi^*_{s\,\omega' j'}(x)-\varphi^\ast_{s\,\omega j}(x) \varphi_{s\,\omega' j'}(x) ]=\frac{\delta_{jj'}\delta(\omega-\omega')}{\rho_0}\,
\end{equation}
with $s=d,p$.

\subsection{Plane wave solutions and dispersion relations}
Assuming the plane wave solutions for each side of $x=0$ in the background of the homogenous condensate given by
\begin{subequations} \label{gap_sol2}
\begin{align}
&\phi_{s\,\omega}(x,t)=A_{s\,k} e^{-i\omega t+ikx},\\
&\varphi_{s\,\omega}(x,t)=B_{s\,k} e^{-i\omega t+ikx},
\end{align}
\end{subequations}
we obtain the dispersion relation for the gapless excitations
\begin{align}
(\omega+vk)^2=c_d^2k^2+\frac{k^4}{4m^2}\,
\label{dsp_gapl}
\end{align}
with the sound velocity
\begin{align}\label{cd}
c_d=\sqrt{\frac{\left({U}+{U}_{AB}\right)\rho_0}{m}}\, .
\end{align}
Also the dispersion relation for the gapped excitations are found to be
\begin{align}
(\omega+vk)^2=c_p^2k^2+\frac{k^4}{4m^2}+m_\text{eff}^2
\label{dsp_gap}
\end{align}
with the speed of the excitations
\begin{align}
c_p=\sqrt{\frac{\left({U}-{U}_{AB}\right)\rho_0+\Omega}{m}}
\label{cs}
\end{align}
and the effective-energy gap term
\begin{align}
m_\text{eff}=\sqrt{2\left({U}-{U}_{AB}\right)\rho_0\Omega+\Omega^2}\,,
\label{meff}
\end{align}
due to the Rabi-coupling effects \cite{Cominotti2022}.
The  healing length of two types of the excitations is given respectively by
\begin{align}\label{cl}
\xi_s=\frac{1}{mc_s}\;, \;\;\; s=d,p \; .
\end{align}
The steplike change of the parameters in~(\ref{u}) reflects in
the sudden change of
sound speeds ~(\ref{cd}), (\ref{cs}) and effective-energy gap in~(\ref{meff})
 at $x=0$.
\begin{figure}[t]
	\centering
	\includegraphics[width=0.6\textwidth]{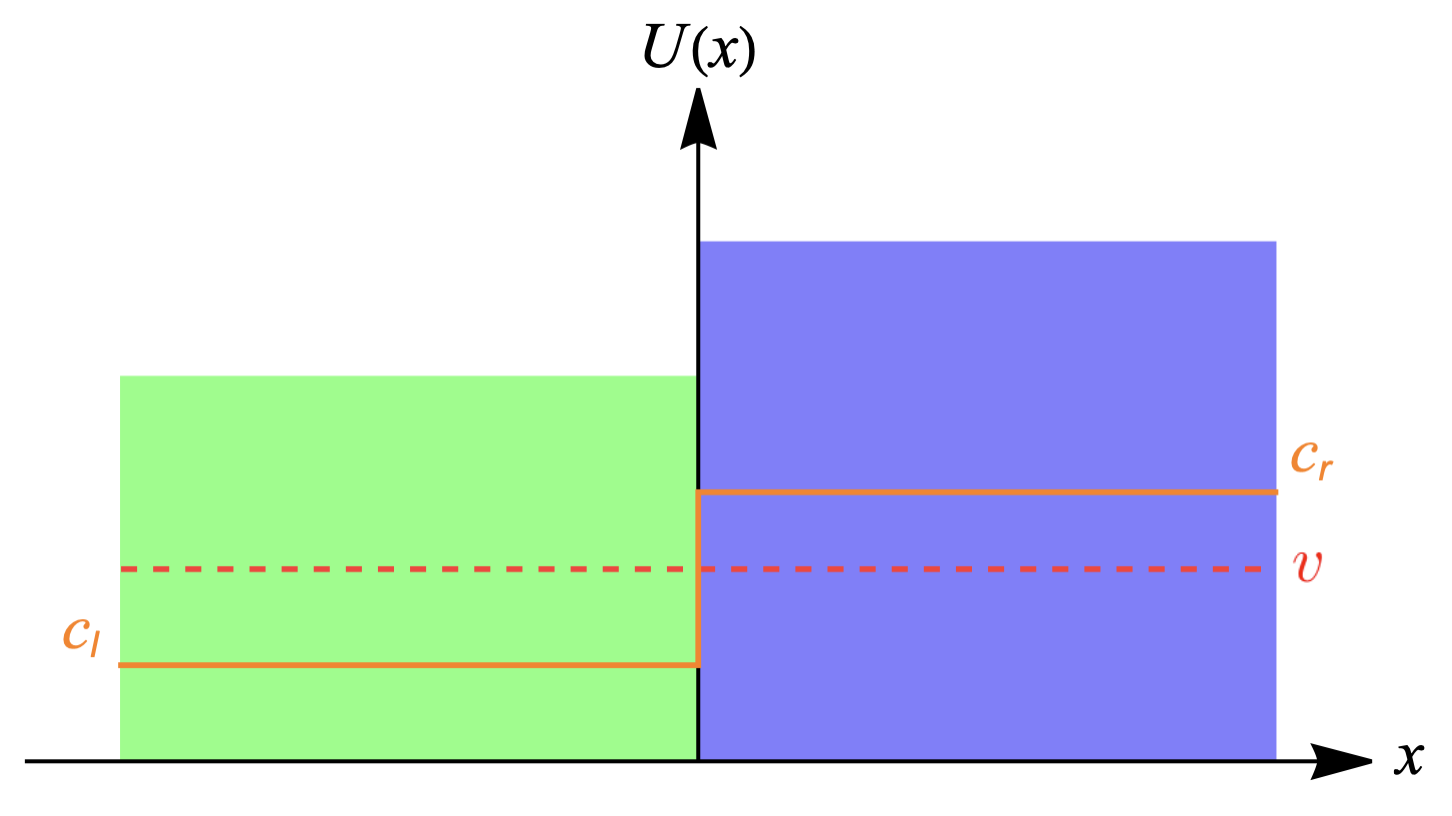}
	\caption{The schematic plot depicts two sides of $x=0$. The supersonic region $x<0$ has the sound speed $c_l<v$, while the subsonic region $x>0$ has the sound speed  $c_r>v$. This setting is achieved by tuning the intraspecies interaction as $U_r>U_l$ in \eqref{u}. This works for both gapless and gapped excitations.}
	\label{fig_dsp}
\end{figure}

The mode amplitudes are obtained from the wave equations \eqref{phi_gapl_eq}-\eqref{phi_gap_eq}. According to the normalization conditions \eqref{nc}, the coefficients of the plane wave solutions in \eqref{gap_sol2} follow the relation
\begin{align}
	\vert A_{s\,k}\vert^2-\vert B_{s\,k}\vert^2=\pm\frac{1}{2\pi\rho_0}\left\vert\frac{dk}{d\omega}\right\vert.
	\label{norm}
\end{align}
After some algebra we find
\begin{align}
&A_{d\,k}=\frac{\omega+vk+k^2/2m}{\sqrt{8\pi \rho_0 (k^2/2m)v_g(\omega+vk)}},\qquad B_{d\,k}=\frac{-(\omega+vk)+k^2/2m}{\sqrt{8\pi \rho_0 (k^2/2m)v_g(\omega+vk)}},\label{mode_amplitude}\\[7pt]
&A_{p\,k}=\frac{\omega+vk+k^2/2m+\Omega}{\sqrt{8\pi \rho_0 (k^2/2m+\Omega)v_g(\omega+vk)}},\qquad B_{p\,k}=\frac{-(\omega+vk)+k^2/2m+\Omega}{\sqrt{8\pi \rho_0 (k^2/2m+\Omega)v_g(\omega+vk)}},
\label{mode_amplitude2}
\end{align}
where $v_g$ is the group velocity $v_g=d\omega/dk$.
In \eqref{norm}, the positive(negative) sign corresponds to positive(negative) norm branch.

In order to create Hawking radiations, one needs the supersonic and subsonic configuration such that
\begin{subequations}
\begin{align}\label{setting}
&c_{d,l}<v<c_{d,r},\\
&c_{p,l}<v<c_{p,r},
\end{align}
\end{subequations}
through the choice of $U_l<U_r $ in each side of $x=0$ while keeping the flow velocity of the condensates a constant.
The schematic plot of the setup is in Fig.~\ref{fig_dsp}.
The Mach numbers are defined as  \cite{Larre2012,Michel2016}
\begin{align}
m_{s,l}=\frac{v}{c_{s,l}},\qquad\text{and} \qquad m_{s,r}=\frac{v}{c_{s,r}}\;,\qquad s=d,p\,.
\label{mach}
\end{align}
The requirement $m_{s,r}<1<m_{s,l}$ leads to the subsonic region ($x>0$) and the supersonic region ($x<0$) separated by a sudden change of the speed at $x=0$ so that the acoustic horizon emerges.
Analogous Hawking radiations given by the gapless excitations have been studied extensively \cite{Carlos2011}.
Here we would like to focus on the gapped excitations instead, which themselves can create analogous Hawking radiations.
Additionally, turning on the interaction between the gapless and gapped excitations opens the possibility to study how the gapped excitations influence the analogous Hawking radiations from the gapless excitations.

\subsection{Gapped excitation modes}
\begin{figure}[t]
	\centering
	\includegraphics[width=1\textwidth]{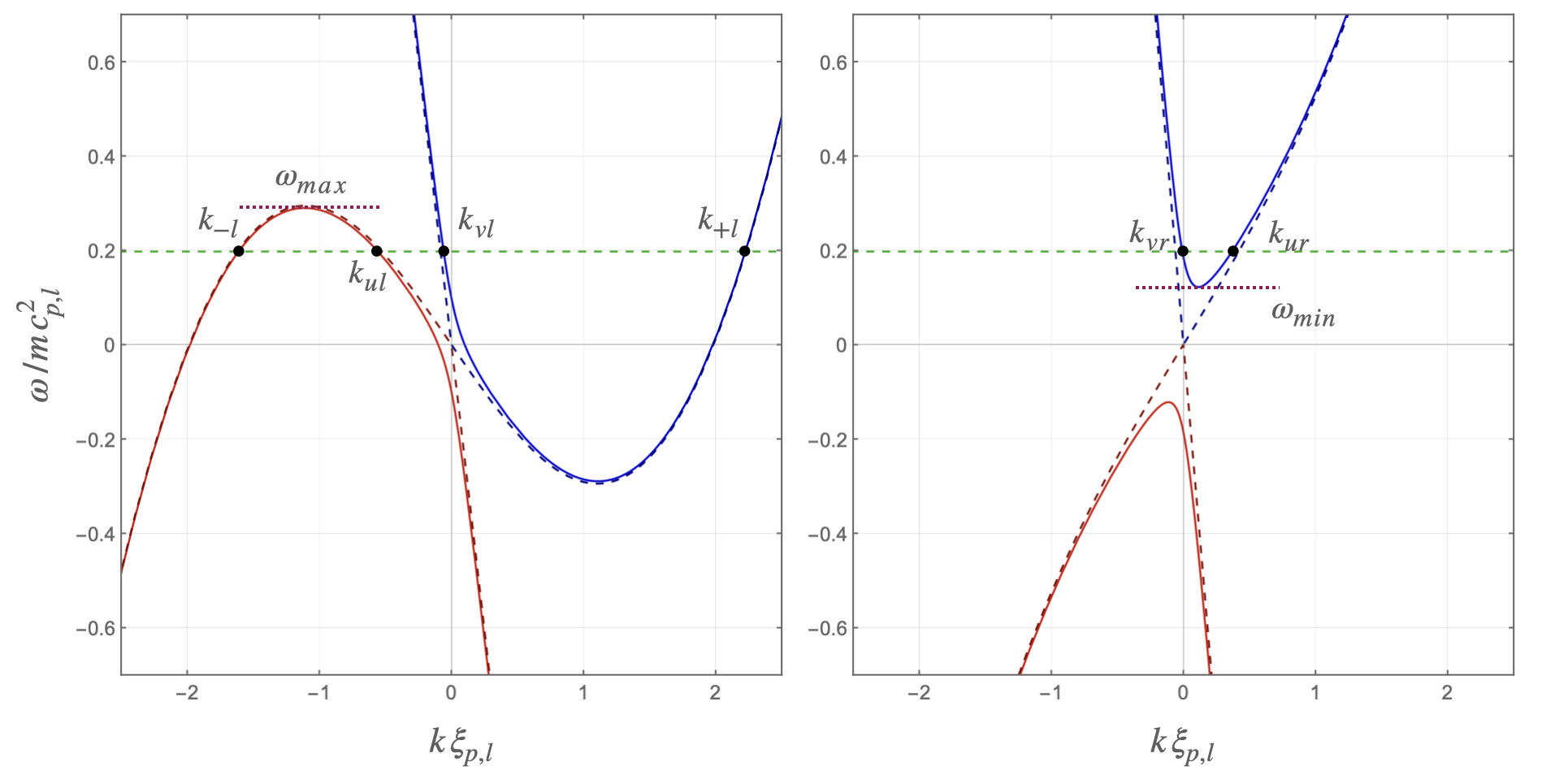}
	\caption{The frequency $\omega$ varies with wave number $k$ according to the dispersion relation of gapped excitation \eqref{dsp_gap}  (solid line with lighter color) and gapless excitation \eqref{dsp_gapl} (dashed line with darker color). In the left panel, a given $\omega$ gives four real-valued roots when $\omega<\omega_\text{max}$. However, in the right panel two real-valued roots exist only for $\omega\ge\omega_\text{min}$. We consider the single-metric geometry $c_p=c_d$.}
	\label{fig_disp}
\end{figure}

We now investigate the wave numbers of gapped excitations with a fixed frequency $\omega$ from solving the dispersion relation \eqref{dsp_gap}.
Dispersive effects of the system lead to four solutions for the wave numbers.
In the supersonic region or downstream region  in $x<0$ with $c_{p,l} <v$, the solutions are qualitatively illustrated in Fig.~\ref{fig_disp}.
Two of the solutions $k_{+l}, k_{vl}$ are from the dispersion of the relation of  the positive comoving frequencies branches and  the other two  $k_{-l},k_{ul}$ are from the negative comoving frequencies branches when $\omega < \omega_\text{max}$.
When $\omega > \omega_\text{max}$, the solutions of the wave numbers $k_{-l},k_{ul}$ will change to complex values, leaving two real-number solutions $k_{+l}, k_{vl}$ only.
The  wave number $k_\text{max}$ and the frequency  $\omega_\text{max}$ are determined by requiring $\frac{d\omega}{dk}\vert_{k=k_\text{max}}=0$, from the dispersion of relation in the negative comoving frequencies branch, namely
\begin{equation}
\omega=-v k-\sqrt{c_{p,l}^2 k^2+\frac{k^4}{4m^2}+m_{\text{eff},l}^2} \;\; .
\label{nom}
\end{equation}
%
To have analytical expression of $\omega_\text{max}$ and $k_\text{max}$, we consider small $m_{\text{eff},l}$ limit, where
$m c^2_{p,l} \approx m v^2 \gg m_{\text{eff},l}$, leading to $U \rho_0 \gg \Omega$.
Then, to order $m^2_{\text{eff},l}$, the solution of $k_\text{max}$ can be expressed as $k_\text{max}=k_\text{max}^{\scriptscriptstyle(0)}+k_\text{max}^{\scriptscriptstyle(1)}$, where
\begin{align}\label{k_0}
k_\text{max}^{\scriptscriptstyle(0)}=-\frac{m\sqrt{v \sqrt{8 c_{p,l}^2+v^2}-4 c_{p,l}^2+ v^2}}{\sqrt{2}},
\end{align}
corresponds to the gapless case \cite{Carlos2011}, and the correction due to $m_{\rm eff}$ is obtained as
\begin{align}
k^{\scriptscriptstyle(1)}_{\max}=\frac{2 m^2 \left[2 c_{p,l}^2 m^2+(k_\text{max}^{\scriptscriptstyle(0)})^2\right]}{(k_\text{max}^{\scriptscriptstyle(0)})^3 \left[6 c_{p,l}^2 m^2+(k_\text{max}^{\scriptscriptstyle(0)})^2\right]}m_{\text{eff},l}^2+\mathcal{O}(m_{\text{eff},l}^3)\, .
\end{align}
The resulting  $\omega_\text{max}$  can be then approximated by $\omega_\text{max}=\omega_\text{max}^{\scriptscriptstyle(0)}+\omega_\text{max}^{\scriptscriptstyle(1)}$ with
\begin{align}\label{omega_0}
&\omega_\text{max}^{\scriptscriptstyle(0)}=-v k_\text{max}^{\scriptscriptstyle(0)}-\sqrt{c_{p,l}^2 (k_\text{max}^{\scriptscriptstyle(0)})^2+\frac{(k_\text{max}^{\scriptscriptstyle(0)})^4}{4m^2}}\,,\\[7pt]
&\omega_\text{max}^{\scriptscriptstyle(1)}=\frac{2m^2[2 c_{p,l}^2 m^2 \omega_\text{max}^{\scriptscriptstyle(0)}+2 (k_\text{max}^{\scriptscriptstyle(0)})^3 v+3 (k_\text{max}^{\scriptscriptstyle(0)})^2 \omega_\text{max}^{\scriptscriptstyle(0)}]}{(k_\text{max}^{\scriptscriptstyle(0)})^4[6c_{p,l}^2m^2+(k_\text{max}^{\scriptscriptstyle(0)})^2]}m_{\text{eff},l}^2+\mathcal{O}(m_{\text{eff},l}^3)\,.
\end{align}
As long as  $c_{p,l}<v$, $\omega_\text{max}^{\scriptscriptstyle(1)}$ is negative, leading to a smaller value of $\omega_\text{max}$ due to the effective-energy gap term $m_{\text{eff},l}$ as compared with $\omega_\text{max}^{\scriptscriptstyle(0)}$ corresponding to the gapless cases.
The existence of the two real-number solutions of the wave numbers in the negative comoving frequencies branch allows the quantum states with negative norm that  open a window to trigger the subsequent analogous Hawking radiation.
Roughly speaking, the above analytical results can provide an estimate of  the value of $\omega_\text{max}$.

\begin{figure}
\centering
\includegraphics[width=\textwidth]{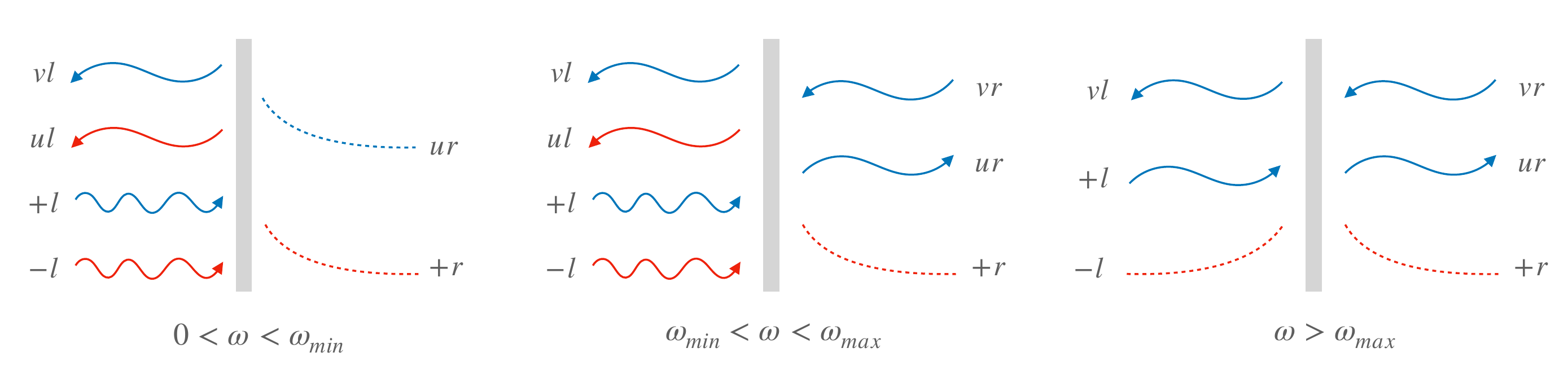}
\caption{Schematic representation of the scattering modes and the decaying modes. In the downstream (supersonic) region, there are four plane wave modes, while in the  upstream (subsonic) region, there are only two plane wave modes and one decaying mode. The amplitudes for each mode can {be solved by the  matching conditions in (\ref{M}).}
}
\label{fig_mover}
\end{figure}

In the subsonic region ($c_{p,r} > v$) and also in the upstream for $x>0$, when $\omega>\omega_\text{min}$, there are two solutions  $k_{ur},k_{vr}$ of the real numbers obtained from the positive comoving frequencies branch and the other two wave numbers $k_{\pm r}$ of the complex numbers from the negative comoving frequencies branch. See Figs.~\ref{fig_disp} and \ref{fig_mover} for more details.

In fact, from the supersonic to subsonic region,  the solution of $k_{u}$  shifts from the negative comoving frequencies branch to the positive comoving frequencies branch  whereas the solution $k_+$ shifts in stead from the positive comoving frequencies branch    to the negative comoving frequencies branch, then becoming complex valued in the subsonic region.
On the contrary, the solutions of $k_{v}$  and $k_{-}$ remain in the positive and negative comoving frequencies branches on both regions, but $k_-$ becomes complex valued in the subsonic region.
The minimum frequency $\omega_\text{min}$ can be obtained, when two real wave number solutions merge, with  the approximate value in the small $m_{\text{eff}}$ limit, namely $m c^2_{p} \approx m v^2 \gg m_{\text{eff}}$
 as \cite{Antonin2012,Roberto2019}
\begin{align}
\omega_{\text{min}}\simeq\sqrt{\frac{c_{p,r}^2-v^2}{c_{p,r}^2}}m_{\text{eff},r}.
\label{omin}
\end{align}
For $\omega < \omega_\text{min}$, all four wave numbers are complex values, where two of them are the decaying modes and the other two are growing modes.
The propagating modes together with the decaying modes will be taken into account on the matching at $x=0$
between two sides of the modes. Here we assume that the growing modes will not be excited.
In Fig.~\ref{fig_mover}, we draw the corresponding moving direction of each mode  according to the sign of the group velocity $v_g$.


%
The analytical solutions of the wave numbers can only be explored in the limits of small frequency $\omega$ and the effective-energy gap $m_\text{eff}$  when
 $\mu \equiv m_\text{eff}/mc_p^2 \ll1 $ and $\nu\equiv \omega/mc_p^2 \ll 1$.
In the case of the supersonic wave, for  $\omega < \omega_\text{max}$, two of the solutions with small wave numbers of order much smaller than $1/\xi_p$ in (\ref{cl}) are obtained by treating the term $k^4/4m^2$ in the dispersion relation (\ref{dsp_gap}) perturbatively to yield
\begin{align}
&k_v=\frac{ v\, \omega+\sqrt{(v^2-c_p^2)m_\text{eff}^2+\omega^2c_p^2}}{c_p^2-v^2}\Bigg\{1-\frac{c_p^4\left[v\, \nu+\sqrt{(v^2-c_p^2)\mu^2+\nu^2 c_p^2}\right]^3}{8(c_p^2-v^2)^3\sqrt{(v^2-c_p^2)\mu^2+\nu^2c_p^2}}+\mathcal{O}{(\nu^4,\mu^4)}\Bigg\},\label{kv}\\[10pt]
&k_u=\frac{v\, \omega-\sqrt{(v^2-c_p^2)m_\text{eff}^2+\omega^2c_p^2}}{c_p^2-v^2}\Bigg\{1+\frac{c_p^4\left[v\, \nu-\sqrt{(v^2-c_p^2)\mu^2+\nu^2 c_p^2}\right]^3}{8(c_p^2-v^2)^3\sqrt{(v^2-c_p^2)\mu^2+\nu^2c_p^2}}+\mathcal{O}{(\nu^4,\mu^4)}\Bigg\}.
\label{ku}
\end{align}

The other two solutions with large wave numbers of the order of $1/\xi_p$, where the dispersive term in the dispersion relation is dominant, are approximated by
\begin{align}
k_\pm = &\frac{v\,\omega}{v^2-c_p^2}\Bigg\{1+\frac{c_p^4\mu^2}{2(v^2-c_p^2)^2}+\frac{(v^2+c_p^2)c_p^4\nu^2}{2(v^2-c_p^2)^3}+\mathcal{O}{(\nu^4,\mu^4)}\Bigg\}\no\\[10pt]
&\pm 2m\sqrt{v^2-c_p^2}\Bigg\{1-\frac{c_p^4\mu^2}{8(v^2-c_p^2)^2}-\frac{(2v^2+c_p^2)c_p^4\nu^2}{8(v^2-c_p^2)^3}+\mathcal{O}{(\nu^4,\mu^4)}\Bigg\}\, .
\label{kpm}
\end{align}
In the subsonic region, both $k_{\pm}$ become complex values. Notice that the nonzero zero-frequency modes will play an important role in forming the undulations in the density-density correlation functions to be discussed later.

If we switch off the Rabi frequency $\Omega=0$ ($m_\text{eff}=0$), and replace $c_p$ by $c_d$, the resulting ~(\ref{kv})-(\ref{kpm}) are identical to the wave numbers obtained in \cite{Carlos2011} for the gapless excitations.
For $c_p<v$ of the supersonic region, and in the case of $c_p \approx v$ under consideration,  $k_\text{max} \sim m c_p$ in (\ref{k_0}) gives the value $\omega_\text{max} \sim m c_p^2$ obtained from (\ref{omega_0}).
Thus the above approximate solutions in \eqref{kv}-\eqref{kpm} valid in particular for $\omega \ll m c_p^2$ fail to give the value of $\omega_\text{max}$.
Nevertheless, for $c_p > v$ of the subsonic region, the value of $\omega_\text{min}$ in (\ref{omin}), below which all four solutions become complex valued, can  be obtained by setting $k_u=k_v$ using the above approximate expressions.

\section{Mode functions and scattering matrices}\label{ABH3}
%
\subsection{Matching of mode functions}
We proceed by considering the matching of the wave functions at $x=0$, in which  both wave functions at $x>0$ and $x<0$ must change smoothly
\begin{align}
&\phi^j (0^-,t)=\phi^j (0^+,t), \quad\varphi^j (0^-,t)=\varphi^j (0^+,t),\no\\
& \partial_x{\phi^j}(0^-,t)=\partial_x\phi^j (0^+,t),\quad\partial_x\varphi^j (0^-,t)=\partial_x\varphi^j (0^+,t)\, ,
\label{matching_con}
\end{align}
where again $j$ denotes outgoing or incoming channels.
According to the dispersion relation of the gapped excitation, for a given frequency there exist four wave numbers for each side of $x=0$ (\ref{dsp_gap}).

For each channel $j$, the general solution is  a linear superposition of the four solutions
\begin{align}
\phi(t,x)=\Bigg\{
\begin{array}{ll}
C_{ul}\phi^{{ul}}+C_{vl}\phi^{{vl}}+C_{+l}\phi^{{+l}}+C_{-l}\phi^{{-l}}\qquad \text{for}\quad x<0\\
C_{ur}\phi^{{ur}}+C_{vr}\phi^{{vr}}+C_{+r}\phi^{{+r}}+C_{-r}\phi^{{-r}}\qquad \text{for}\quad x>0\\
   \end{array}.
   \label{phi}
\end{align}
Similarly we have
\begin{align}
\varphi(t,x)=\Bigg\{
\begin{array}{ll}
C_{ul}\varphi^{{ul}}+C_{vl}\varphi^{{vl}}+C_{+l}\varphi^{{+l}}+C_{-l}\varphi^{{-l}}\qquad \text{for}\quad x<0\\
C_{ur}\varphi^{{ur}}+C_{vr}\varphi^{{vr}}+C_{+r}\varphi^{{+r}}+C_{-r}\varphi^{{-r}}\qquad \text{for}\quad x>0\\
    \end{array}.
    \label{variphi}
\end{align}
Using (\ref{phi}) and (\ref{variphi}) together with (\ref{gap_sol2}) in the matching condition ~(\ref{matching_con}), one can write the result in the following matrix form
\begin{align}
\begin{pmatrix}
C_{ul}\\
C_{vl}\\
C_{+l}\\
C_{-l}
\end{pmatrix}
=M_\text{scat}
\begin{pmatrix}
C_{ur}\\
C_{vr}\\
C_{+r}\\
C_{-r}
\end{pmatrix}
\label{matching_eq},
\end{align}
where
\begin{equation}
M_\text{scat}=W_l^{-1} W_r
\label{M}
\end{equation}
 with $W_l$ and $W_r$ given by
\begin{align}
W_{h}=
\begin{pmatrix}
A^{uh} & A^{vh} & A^{+h} & A^{-h}\\
B^{uh} & B^{vh} & B^{+h} & B^{-h}\\
ik_{uh}A^{uh} & ik_{vh}A^{vh} & ik_{+h}A^{+h} & ik_{-h}A^{-h}\\
ik_{uh}B^{uh} & ik_{vh}B^{vh} & ik_{+h}B^{+h} & ik_{-h}B^{-h}
\end{pmatrix},\qquad h=l, \,r.
\label{M_explicit}
\end{align}
To figure out $M_\text{scat}$, we adopt the results of the mode functions as well as their wave numbers in the small $\omega$ expansion
and consider the leading terms only.
The coefficients $C's$ will be connected to the elements of the $S$-matrix to be determined later.

\subsection{Construction of the $S$-matrix}
The density fluctuation field operator for the gapped excitation are then decomposed in terms of the in or the out-basis.
In this paper, we mainly focus on the effects of the modes  from the negative comoving frequencies branch  with $\omega < \omega_\text{max}$.
In the region $\omega_\text{min}\le\omega\le\omega_\text{max}$, as shown in Fig.~\ref{fig_mover},
three modes with the wave functions $\phi_p^{vr},\phi_p^{\pm l}$  move toward $x=0$ and the other three modes with the wave functions $\phi_p^{ur},\phi_p^{vl},\phi_p^{ul}$ move away from $x=0$, which form the respective incoming and outgoing channels.
For the three outgoing (incoming) channels, the corresponding out (in) states involving the linear superposition of all relevant plane wave solutions in their respective channels form a out (in) basis \cite{Carlos2011}.
The details of particular modes involved in each of outgoing channels will be specified below.
Although the decaying modes are considered on the matching of the wave functions, they will decay and thus will not contribute to the asymptotic states in the scattering processes.
Additionally, the existence of the solutions $k_{ul},k_{-l}$ from the negative comoving frequencies branch in the supersonic region leads to the negative norm states that destabilize the vacuum state of the system with the corresponding creation/annihilation operators,  denoted by $(\hat{b}_{-\omega}^{-l,in})^\dagger, \,\hat{b}_{-\omega}^{-l,in}$ and $(\hat{b}_{-\omega}^{ul,out})^\dagger,\,\hat{b}_{-\omega}^{ul,out}$.
The mode expansion can then be expressed as
\begin{subequations}
\begin{align}
\hat{\phi}_p(x,t)=&\int_{\omega_\text{min}}^{\omega_\text{max}}d\omega \,\bigg\{\left[\hat{b}_{\omega}^{vr,in}\phi_p^{vr,in}+\hat{b}_{\omega}^{+l,in}\phi_p^{+l,in}+(\hat{b}_{-\omega}^{-l,in})^\dagger\phi_p^{-l,in}\right]\no\\
&\qquad\qquad\quad +\left[(\hat{b}_{\omega}^{vr,in})^\dagger\varphi_p^{vr,in}+(\hat{b}_{\omega}^{+l,in})^\dagger\varphi_p^{+l,in}+\hat{b}_{-\omega}^{-l,in}\varphi_p^{-l,in}\right]\bigg\}\\[10pt]
=&\int_{\omega_\text{min}}^{\omega_\text{max}}d\omega \,\bigg\{\left[\hat{b}_{\omega}^{ur,out}\phi_p^{ur,out}+\hat{b}_{\omega}^{vl,out}\phi_p^{vl,out}+(\hat{b}_{-\omega}^{ul,out})^\dagger\phi_p^{ul,out}\right]\no\\
&\qquad\qquad\quad +\left[(\hat{b}_{\omega}^{ur,out})^\dagger\varphi_p^{ur,out}+(\hat{b}_{\omega}^{vl,out})^\dagger\varphi_p^{vl,out}+\hat{b}_{-\omega}^{ul,out}\varphi_p^{ul,out}\right]\bigg\}.
\label{gp_expand}
\end{align}
\end{subequations}

The relation of the in basis to the out basis is through the $S$-matrix, which can be defined  for the wave function $\phi$ as \cite{Macher2009,Recati2009}
\begin{align}
\begin{pmatrix}
\phi_p^{vr,in}\\
\phi_p^{+l,in}\\
\phi_p^{-l,in}
\end{pmatrix}
=S\cdot\begin{pmatrix}
\phi_p^{ur,out}\\
\phi_p^{vl,out}\\
\phi_p^{ul,out}
\end{pmatrix}=
\begin{pmatrix}
S_{ur,vr} &S_{vl, vr} &S_{ul,vr}\\
S_{ur,+l} &S_{vl, +l} &S_{ul,+l}\\
S_{ur,-l} &S_{vl,-l} &S_{ul,-l}
\end{pmatrix}
\begin{pmatrix}
\phi_p^{ur,out}\\
\phi_p^{vl,out}\\
\phi_p^{ul,out}
\end{pmatrix}\, .
\label{bogo_n}
\end{align}
The same transformation also applies to the wave function $\varphi$.
This in turn gives the Bogoliubov transformation,
\begin{align}
\begin{pmatrix}
\hat{b}_{\omega}^{ur,out}\\
\hat{b}_{\omega}^{vl,out}\\
(\hat{b}_{-\omega}^{ul,out})^\dagger
\end{pmatrix}
=
\begin{pmatrix}
S_{ur,vr} &S_{ur,+l} &S_{ur,-l}\\
S_{vl,vr} &S_{vl,+l} &S_{vl,-l}\\
S_{ul,vr} &S_{ul,+l} &S_{ul,-l}
\end{pmatrix}
\begin{pmatrix}
\hat{b}_{\omega}^{vr,in}\\
\hat{b}_{\omega}^{+l,in}\\
(\hat{b}_{-\omega}^{-l,in})^\dagger.
\end{pmatrix}.
\label{bogo_b}
\end{align}
Those scattering amplitudes are obtained by solving the matching equations in (\ref{matching_eq}) for all mode functions from each incoming/outgoing mode channels.
The current conservation in an asymptotic region requires that
\begin{align}\label{S_matrix_tran}
S^\dagger\eta S=\eta,\quad \text{with}\quad \eta=\text{diag}(1,1,-1).
\end{align}
Correspondingly, the scattering coefficients have the relations
\begin{subequations} \label{uni}
\begin{eqnarray}
&& \vert S_{ur,vr}\vert^2+\vert S_{ur,+l}\vert^2-\vert S_{ur,-l}\vert^2=1, \label{ur_c}\\
&& \vert S_{ul,vr}\vert^2+\vert S_{ul,+l}\vert^2-\vert S_{ul,-l}\vert^2=-1, \label{ul_c}\\
&& \vert S_{vl,vr}\vert^2+\vert S_{vl,+l}\vert^2-\vert S_{vl,-l}\vert^2=1.\label{vl_c}
\end{eqnarray}
\end{subequations}
The minus sign in the left-hand side of the relations is because of the incoming modes  $k_{-l}$  of the negative norm states, and  the minus sign in the right-hand side is due to the  outgoing modes $k_{ul}$ of also the negative norm state.
Both modes are in the supersonic regime.

For $0<\omega<\omega_\text{min}$, since in the subsonic region the propagating modes do not exist,  there are then two incoming modes and outgoing modes in the supersonic region \cite{Jannes2011} with the mode expansion given by
\begin{align}
\hat{\phi}_p(x,t)=&\int_{0}^{\omega_\text{min}}d\omega \,\bigg\{\left[\hat{b}_{\omega}^{+l,in}\phi_p^{+l,in}+(\hat{b}_{-\omega}^{-l,in})^\dagger\phi_p^{-l,in}\right]+\left[(\hat{b}_{\omega}^{+l,in})^\dagger\varphi_p^{+l,in}+\hat{b}_{-\omega}^{-l,in}\varphi_p^{-l,in}\right]\bigg\}\no\\[10pt]
=&\int_{0}^{\omega_\text{min}}d\omega \,\bigg\{\left[\hat{b}_{\omega}^{vl,out}\phi_p^{vl,out}+(\hat{b}_{-\omega}^{ul,out})^\dagger\phi_p^{ul,out}\right]+\left[(\hat{b}_{\omega}^{vl,out})^\dagger\varphi_p^{vl,out}+\hat{b}_{-\omega}^{ul,out}\varphi_p^{ul,out}\right]\bigg\}.
\label{gp_expand2}
\end{align}
The $S$-matrix can be constructed from the wave function $\phi$ below or the wave function $\varphi$ by
\begin{align}
\begin{pmatrix}
\phi_p^{+l,in}\\
\phi_p^{-l,in}
\end{pmatrix}
=S\cdot\begin{pmatrix}
\phi_p^{vl,out}\\
\phi_p^{ul,out}
\end{pmatrix}=
\begin{pmatrix}
S_{vl, +l} &S_{ul,+l}\\
S_{vl,-l} &S_{ul,-l}
\end{pmatrix}
\begin{pmatrix}
\phi_p^{vl,out}\\
\phi_p^{ul,out}
\end{pmatrix}
\label{bogo_n2}
\end{align}
giving the Bogoliubov transformation,
\begin{align}
\begin{pmatrix}
\hat{b}_{\omega}^{vl,out}\\
(\hat{b}_{-\omega}^{ul,out})^\dagger
\end{pmatrix}
=
\begin{pmatrix}
S_{vl,+l} &S_{vl,-l}\\
S_{ul,+l} &S_{ul,-l}
\end{pmatrix}
\begin{pmatrix}
\hat{b}_{\omega}^{+l,in}\\
(\hat{b}_{-\omega}^{-l,in})^\dagger.
\end{pmatrix}.
\label{bogo_b2}
\end{align}
The current conservation in this case becomes
\begin{align} \label{S_matrix_cc}
S^\dagger\eta S=\eta,\quad \text{with}\quad \eta=\text{diag}(1,-1).
\end{align}
Explicitly, the scattering matrix elements thus have the relations
\begin{subequations}
\begin{align}
&\vert S_{vl,+l}\vert^2-\vert S_{vl,-l}\vert^2=1,\\
&\vert S_{ul,+l}\vert^2-\vert S_{ul,-l}\vert^2=-1.
\end{align}
\end{subequations}
Apparently the existence of the negative norm states results in the Bogoliubov transformations
involving the mixture of the creation and annihilation operators giving particle production.

Finally, when $\omega>\omega_\text{max}$ all the negative norm states disappear, the mode expansion is given in terms of the positive norm states as
\begin{align}
\hat{\phi}_p(x,t)=&\int_{\omega_\text{max}}^{\infty}d\omega \,\bigg\{\left[\hat{b}_{\omega}^{vr,in}\phi_p^{vr,in}+\hat{b}_{\omega}^{+l,in}\phi_p^{+l,in}\right]+\left[(\hat{b}_{\omega}^{vr,in})^\dagger\varphi_p^{vr,in}+(\hat{b}_{\omega}^{+l,in})^\dagger\varphi_p^{+l,in}\right]\bigg\}\no\\[10pt]
=&\int_{\omega_\text{max}}^{\infty}d\omega \,\bigg\{\left[\hat{b}_{\omega}^{ur,out}\phi_p^{ur,out}+\hat{b}_{\omega}^{vl,out}\phi_p^{vl,out}\right]+\left[(\hat{b}_{\omega}^{ur,out})^\dagger\varphi_p^{ur,out}+(\hat{b}_{\omega}^{vl,out})^\dagger\varphi_p^{vl,out}\right]\bigg\}.
\label{gp_expand3}
\end{align}
The scattering matrix $S$ is thus defined from the wave function $\phi$ below or the wave function $\varphi$ by
\begin{align}
\begin{pmatrix}
\phi_p^{vr,in}\\
\phi_p^{+l,in}
\end{pmatrix}
=S\cdot\begin{pmatrix}
\phi_p^{ur,out}\\
\phi_p^{vl,out}
\end{pmatrix}=
\begin{pmatrix}
S_{ur, vr} &S_{vl,vr}\\
S_{ur,+l} &S_{vl,+l}
\end{pmatrix}
\begin{pmatrix}
\phi_p^{ur,out}\\
\phi_p^{vl,out}
\end{pmatrix},
\label{bogo_n3}
\end{align}
where the corresponding  Bogoliubov transformation becomes
\begin{align}
\begin{pmatrix}
\hat{b}_{\omega}^{ur,out}\\
\hat{b}_{\omega}^{vl,out}
\end{pmatrix}
=
\begin{pmatrix}
S_{ur,vr} &S_{ur,+l}\\
S_{vl,vr} &S_{vl,+l}
\end{pmatrix}
\begin{pmatrix}
\hat{b}_{\omega}^{vr,in}\\
\hat{b}_{\omega}^{+l,in}
\end{pmatrix}
\label{bogo_b3}
\end{align}
with no mixture of the creation and annihilation operators, showing no particle production. Thus, the scattering matrix elements satisfy the relations
\begin{subequations}
\begin{align}
&\vert S_{ur,vr}\vert^2+\vert S_{ur,+l}\vert^2=1,\\
&\vert S_{vl,vr}\vert^2+\vert S_{vl,+l}\vert^2=1.
\end{align}
\end{subequations}
We shall consider the process outgoing channels in, $\omega_\text{min} < \omega < \omega_\text{max}, \,0< \omega < \omega_\text{min}$ regimes, respectively, using the approximate formulas of the wave functions and their wave numbers obtained in the previous discussions.

\subsection{$ur$ outgoing channel}

\begin{figure}[t]
	\centering
	\includegraphics[scale=0.3]{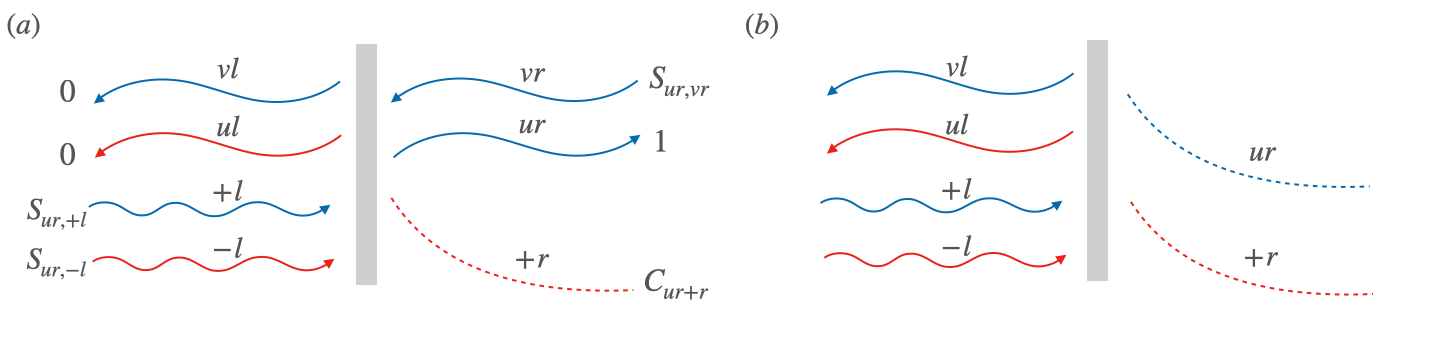}
	\caption{Schematic representation of $ur$ outgoing channel scattering processes when $\omega_{r}(=\omega_\text{min})<\omega<\omega_\text{max}$ in (a) and when $\omega <\omega_{r}\,(=\omega_\text{min})$ in (b).
		\label{fig_ur_ch}
	}
\end{figure}
%
To construct the $ur$ outgoing ($ur$,out) channel in the region $\omega_\text{min} < \omega < \omega_\text{max}$,  one might consider that an outgoing $\phi^{ur}_p$ wave  with unit amplitude leaves away $x=0$ due to the scattering of the incoming $\phi_p^{vr}$ mode moving toward $x=0$  with amplitude $S_{ur,vr}$ in subsonic region, the $\phi^{+l}_p$ modes in the supersonic region   with $S_{ur,+l}$, and $\phi^{-l}_p$ in a particular negative norm state with $S_{ur,-l}$, as shown in Fig.~\ref{fig_ur_ch}.
In addition, the decaying mode $\phi_p^{+r}$ will be also taken into account on the matching calculations.
However, in the region $\omega < \omega_\text{min}$, the $\phi^{ur}_p$  mode becomes a decaying mode with the vanishing amplitude in the asymptotic region so all $S_{ur,+l},S_{ur,-l},S_{ur,vr}$ vanish.
According to the above description, the matching equations can be written as
\begin{align}
\begin{pmatrix}
0\\
0\\
S_{ur,+l}\\
S_{ur,-l}
\end{pmatrix}
=
M_\text{scat}\begin{pmatrix}
1\\
S_{ur,vr}\\
C_{ur,+r}\\
0
\end{pmatrix},
\label{ur-eq}
\end{align}
where we have replaced  $C_j$ by the specific expression $S_{in,out}$.

Using the result of $M_\text{scat}$ given in (\ref{M}) and \eqref{M_explicit} one solves easily the system of equations \eqref{ur-eq}.
Up to two leading-order terms in the small $\omega$ expansion for $\omega <\omega_\text{max}$, to guarantee the unitary relation (\ref{ur_c}), we find that
%
\begin{align}
S_{ur,-l}=&\Bigg\{-
\frac{\sqrt{2mc_{p,r} \eta_r} \left(v^2-c_{p,l}^2\right){}^{3/4}  \sqrt{c_{p,r}^2-v^2}\left(\sqrt{c_{p,r}^2-v^2}+i \sqrt{v^2-c_{p,l}^2}\right)}{ \left(c_{p,r}^2-c_{p,l}^2\right) \left(c_{p,r}+v \eta_r\right)} \frac{1}{\sqrt{\omega}}\,\no\\
&+\left[2\sqrt{2m} v \left(c_{p,r}^2-c_{p,l}^2\right) \left(v^2-c_{p,l}^2\right){}^{3/4} \left(c_{p,r}^2-v^2\right) \left(c_{p,r}+v \eta_r\right)\right]^{-1}\no\\
 &\times\sqrt{{c_{p,r} \eta_r}\omega}\,\bigg[2 v^6+v^4 \left(c_{p,r}^2-4 c_{p,l}^2\right)+2 v^2 \left(2 c_{p,l}^4+c_{p,r}^4-3 c_{p,l}^2 c_{p,r}^2\right)+c_{p,l}^4 c_{p,r}^2\no\\
&-i \Sigma  \left(v^2-c_{p,l}^2\right) \left(c_{p,r}^2+2 v^2\right)
\bigg]+\mathcal{O}(\omega^{3/2})\Bigg\} \theta(\omega-\omega_{r}) ,\label{surml}
\end{align}
\begin{align}
&S_{ur,vr}=
\Bigg[\frac{v\eta_r-c_{p,r}}{c_{p,r}+v\eta_r}\,+\mathcal{O}(\omega^{3/2})\Bigg] \theta(\omega-\omega_{r}) ,\label{survr}
\end{align}
\begin{align}
S_{ur,+l}=&
\Bigg\{ \frac{ \sqrt{2mc_{p,r} \eta_r} \left(v^2-c_{p,l}^2\right){}^{3/4}  \sqrt{c_{p,r}^2-v^2}\left(\sqrt{c_{p,r}^2-v^2}-i \sqrt{v^2-c_{p,l}^2}\right)}{ \left(c_{p,r}^2-c_{p,l}^2\right) \left(c_{p,r}+v \eta_r\right)} \frac{1}{\sqrt{\omega}}\,\no\\
&+\left[2\sqrt{2m} v \left(c_{p,r}^2-c_{p,l}^2\right) \left(v^2-c_{p,l}^2\right){}^{3/4} \left(c_{p,r}^2-v^2\right) \left(c_{p,r}+v \eta_r\right)\right]^{-1}\no\\
&\times\sqrt{{c_{p,r} \eta_r}\omega}\,\bigg[2 v^6+v^4 \left(c_{p,r}^2-4 c_{p,l}^2\right)+2 v^2 \left(2 c_{p,l}^4+c_{p,r}^4-3 c_{p,l}^2 c_{p,r}^2\right)+c_{p,l}^4 c_{p,r}^2\no\\
&+i \Sigma  \left(v^2-c_{p,l}^2\right) \left(c_{p,r}^2+2 v^2\right)
\bigg]+\mathcal{O}(\omega^{3/2})\Bigg\} \theta(\omega-\omega_{r})
,\label{surpl}
\end{align}
where $\eta_r=\sqrt{1-\omega_r^2/\omega^2}$,  $\Sigma= \sqrt{(v^2-c_{p,l}^2)(c_{p,r}^2-v^2)}$, and $\omega_r$ is the minimum frequency $\omega_\text{min}$ \eqref{omin}.

\begin{figure}[t]
\centering
	\includegraphics[width=0.7\textwidth]{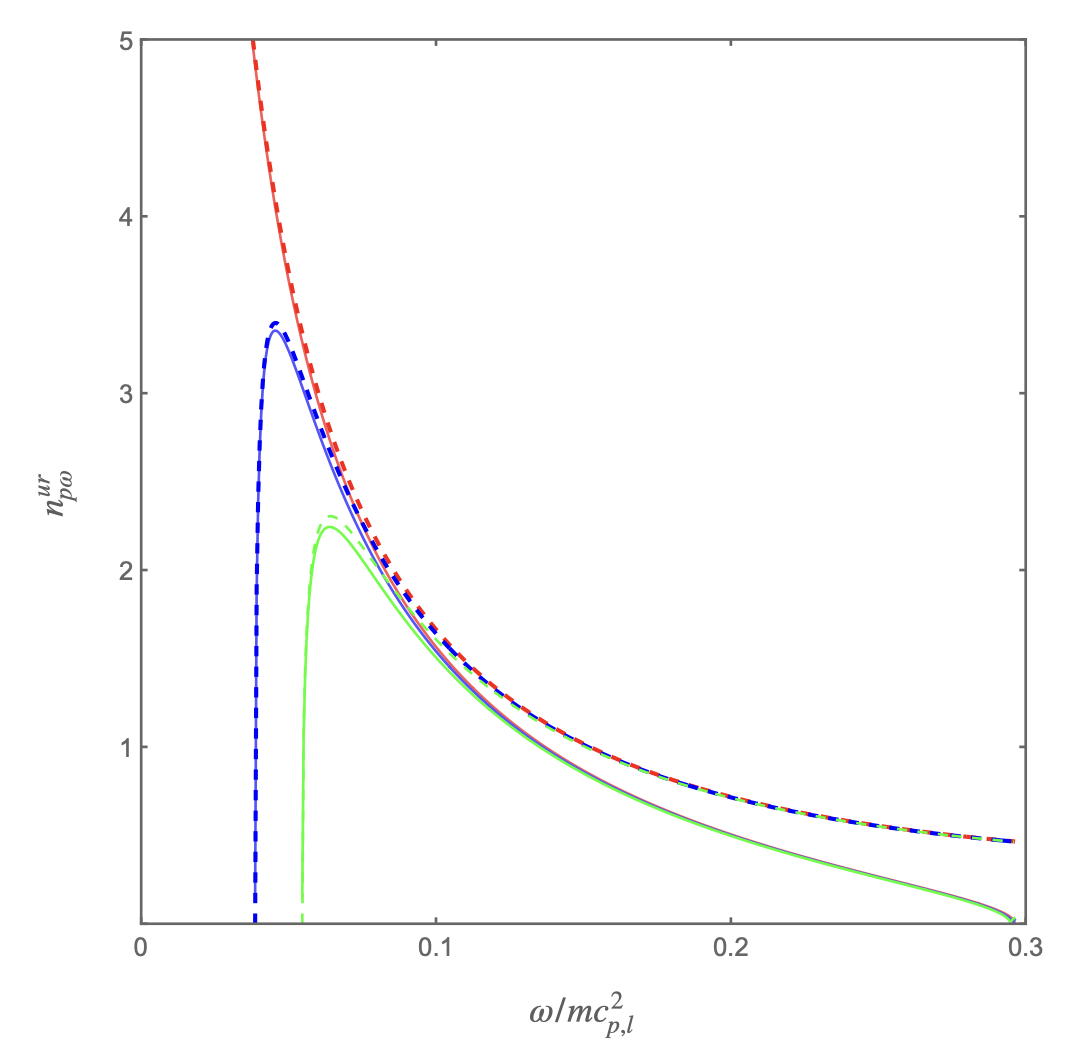}
	\caption{The Hawking spectrum $n_{p\,\omega}^{ur}$ varies as a function of $\omega$ (in the unit of $mc_{p,l}^2$) with different Rabi frequencies: $\Omega/\rho_0U_{l}=0.0$ (red), $3.3\times 10^{-4}$ (blue) and $6.6\times 10^{-4}$ (green). The solid lines are based on the numerical computation with {details stated in the text}, while dashed lines are due to the analytical prediction in~(\ref{np_ur}). We  consider $m_{p,l}=7/5,\,m_{p,r}=3/4$, with $U_r/U_{l}=8/3,\,U_{AB}/U_{l}=1/3$.
		\label{fig_surml}
	}
\end{figure}
%

%
These scattering coefficients can also be obtained numerically from the  matching equations \eqref{ur-eq}
together with the normalization condition \eqref{norm}.
Those scattering coefficients will be used as the inputs to numerically produce Figs.~\ref{fig_surml} and \ref{fig_nul_nvl}.
Moreover, in these analytical expressions, the step function $\theta(\omega-\omega_r)$ is added in each equation, when $\omega <\omega_r(=\omega_\text{min})$ all the modes of the subsonic (upstream) region will become decaying or growing waves.
The growing modes are ignored and the decaying modes can not reach the asymptotic region $x \rightarrow \infty$.
One expects that particle production occurs in the region ($x<0$) as a result of encountering the  total reflection at $x=0$.

To realize the radiation due to particle production from the negative norm states, we first consider the particle distribution function $n_{p\,\omega}^{ur}=\langle 0_{in} \vert\hat{b}_\omega^{ur \dagger}\hat{b}_\omega^{ur}\vert 0_{in}\rangle$.
We then apply the relations in \eqref{bogo_b}, where the mode mixing occurs in the $ur$ outgoing channel from the negative norm states $\phi_p^{-l}$, to give $n_{p\,\omega}^{ur}=\vert S_{ur,-l} \vert^2$, yielding
\begin{align}
n_{p \,\omega}^{ur}=\vert S_{ur,-l} \vert^2=\left[\frac{2 m c_{p,r} \eta_r  \left(v^2-c_{p,l}^2\right)^{3/2} \left(c_{p,r}^2-v^2\right)}{\left(c_{p,r}^2-c_{p,l}^2\right) \left(c_{p,r}+v \eta_r\right){}^2}\frac{1}{\omega}-\frac{2c_{p,r}v\eta_r}{(c_{p,r}+v\eta_r)^2}+\mathcal{O}(\omega)\right] \theta(\omega-\omega_{r})\, .
\label{np_ur}
\end{align}
The $\phi_p^{ur}$ modes are called the Hawking modes.
The small $\omega$ expansion of the particle distribution function of the Hawking radiation satisfies the Planck distribution
\begin{align}
\frac{\Gamma}{\exp{\left(\frac{ \omega}{T_H}\right)}-1},
\end{align}
accompanying by the gray-body factor $\Gamma$ \cite{Anderson2014,Fabbri2016,Antonin2018,Belgiorno2020} approximated as
\begin{align}
\Gamma\left(\frac{T_H}{ \omega}-\frac{1}{2}+\cdots\right).
\label{urplanck}
\end{align}
Then the effective Hawking temperature and its gray-body factor can be read off from the small $\omega$ expansion of (\ref{np_ur})  in terms of Mach numbers (\ref{mach}) as follows \cite{Larr2013}
\begin{align}
T_H=\frac{mc_{p,r}^2 \left(1-m_{p,r}^2\right) m_{p,r}^2 \left(m_{p,l}^2-1\right)^{3/2}}{2 m_{p,l} \left(m_{p,l}^2-m_{p,r}^2\right)},
\label{Th}
\end{align}
\begin{align}
\Gamma=\frac{4m_{p,r}\eta_r}{\left(1+m_{p,r}\eta_r\right)^2}.
\label{gray-body}
\end{align}
%
%
The effective temperature is monotonically decreased  as $m_{p,r}$ ($m_{p,l}$) increases (decreases) toward unity.
Especially $T_H\rightarrow 0$ as $m_{p,r}\rightarrow 1$ or/and $m_{p,l}\rightarrow 1$ where the analogous horizon disappears.
As for the gray-body factor, the relation $\Gamma=1-\vert S_{ur,vr}\vert^2$ is justified
showing that some flux of the particles moves through $x=0$ from the subsonic regime.
As compared with the gray-body factor of the gapless cases by sending $\omega_r \rightarrow 0$ giving $\eta_r \rightarrow 1$ \cite{Anderson2014,Antonin2018}
\begin{align}
\Gamma_{\omega_r=0}=\frac{4m_{p,r}}{(1+m_{p,r})^2}.
\end{align}
Here in (\ref{gray-body}) of the gapped cases the Mach numbers seem to be dressed by the parameters $\eta_{l/r}$ becoming frequency dependent.
But these effective Mach numbers have no effects to the Hawking temperature. In particular, when $\omega\rightarrow \omega_r$, $\eta_r\rightarrow 0$ leads  to the gray-body factor  $\Gamma\rightarrow 0$ consistent with \cite{Roberto2019}.
This is just the critical value of frequency, below which the  excitations in the supersonic regime move toward $x=0$ and then are totally reflected  away from $x=0$, leading to no Hawking radiation to be observed in the subsonic regime.
Thus, for $\omega< \omega_{{r}}$, all $S_{ur,+l},\,S_{ur,-l},\,S_{ur,vr}$ vanish.


\subsection{$ul$ outgoing channel}
\begin{figure}[t]
\centering

\includegraphics[scale=0.3]{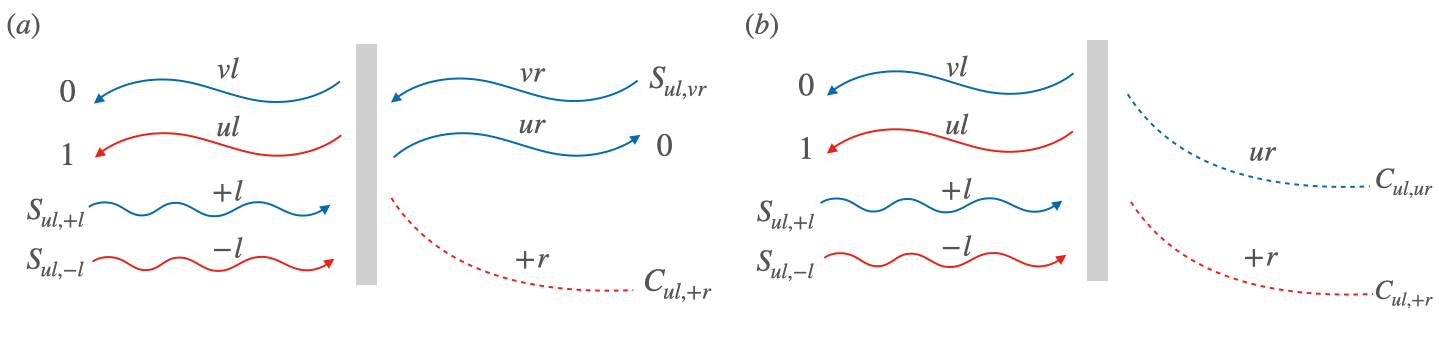}
\caption{Schematic representation of $ul$, out-channel scattering processes when $\omega_\text{r}<\omega<\omega_\text{max}$ in (a) and $\omega <\omega_r$ in (b).
\label{fig_ul_ch}
}
\end{figure}
%
We now discuss the $ul$ outgoing channel in the downstream region of $x<0$, which reveals information about the $\phi_p^{ul}$ modes, the partner modes of the Hawking radiations.
The $ul$ outgoing channel  involves the mode $\phi_p^{ul}$  of the negative norm state with the unit amplitude plus three incoming modes, namely  $\phi_p^{-l}$ with the amplitude $S_{ul,-l}$, $\phi_p^{+l}$ with the amplitude $S_{ul,+l}$, and $\phi_p^{vr}$ with the amplitude $S_{ul,vr}$ as shown in Fig.~\ref{fig_ul_ch}.
The decay mode $\phi_p^{+r}$ with the amplitude $C_{ul,+r}$ is included also in this calculation.
Now the matching equations become
\begin{align}
\begin{pmatrix}
1\\
0\\
S_{ul,+l}\\
S_{ul,-l}
\end{pmatrix}
=
M_\text{scat}\begin{pmatrix}
0\\
S_{ul,vr}\\
C_{ul,+r}\\
0
\end{pmatrix}.
\label{ul-eq}
\end{align}
As before we solve
the scattering coefficients for $\omega < \omega_\text{max}$ in the small $\omega$ expansion
\begin{align}
S_{ul,-l}=&-\frac{ \sqrt{m \left(c_{p,r}^2-v^2\right)}\left(v^2-c_{p,l}^2\right){}^{3/4} \left(c_{p,r} \eta_l+c_{p,l} \eta_r\right) \left(\sqrt{c_{p,r}^2-v^2}+i \sqrt{v^2-c_{p,l}^2}\right)}{\sqrt{2\,c_{p,l}\eta_l\,\omega} \left(c_{p,r}^2-c_{p,l}^2\right) \left(c_{p,r}+v \eta_r\right)}\no\\
&+\left[4 \sqrt{2} \sqrt{m} v \left(c_{p,r}^2-c_{p,l}^2\right) \left(v^2-c_{p,l}^2\right){}^{3/4} \left(c_{p,r}^2-v^2\right)\sqrt{c_{p,l}\,\eta_l} \left(c_{p,r}+v \eta_r\right)\right]^{-1}\no\\
&\times\sqrt{\omega} \Big\{c_{p,r} \left[-2 v c_{p,l} \left(c_{p,l}^2-c_{p,r}^2\right){}^2+\eta_l \left(v^2-c_{p,l}^2\right) \left(c_{p,r}^2+2 v^2\right) \left(-c_{p,l}^2-i \Sigma +v^2\right)\right]\no\\
&+\eta_r \left[-2 v^3 \eta_l \left(c_{p,l}^2-c_{p,r}^2\right){}^2+c_{p,l} \left(c_{p,l}^2-v^2\right) \left(c_{p,r}^2+2 v^2\right) \left(c_{p,l}^2+i \Sigma -v^2\right)\right]
\Big\}\no\\
&+\mathcal{O}(\omega^{3/2})\,,
\label{sulml}
\end{align}
\begin{align}
S_{ul,vr}=&\frac{ v \eta_l-c_{p,l}}{c_{p,r}+v \eta_r}\sqrt{\frac{c_{p,r} \eta_r}{c_{p,l} \eta_l}}\,\theta(\omega-\omega_r)+\mathcal{O}(\omega^{3/2}),\label{sulrv}
\end{align}
\begin{align}
S_{ul,+l}=&\frac{ \sqrt{m \left(c_{p,r}^2-v^2\right)}\left(v^2-c_{p,l}^2\right){}^{3/4} \left(c_{p,r} \eta_l+c_{p,l} \eta_r\right) \left(\sqrt{c_{p,r}^2-v^2}-i \sqrt{v^2-c_{p,l}^2}\right)}{\sqrt{2\,c_{p,l}\eta_l\,\omega} \left(c_{p,r}^2-c_{p,l}^2\right) \left(c_{p,r}+v \eta_r\right)}\no\\
&+\left[4 \sqrt{2} \sqrt{m} v \left(c_{p,r}^2-c_{p,l}^2\right) \left(v^2-c_{p,l}^2\right){}^{3/4} \left(c_{p,r}^2-v^2\right)\sqrt{c_{p,l}\,\eta_l} \left(c_{p,r}+v \eta_r\right)\right]^{-1}\no\\
&\times\sqrt{\omega} \Big\{c_{p,r} \left[-2 v c_{p,l} \left(c_{p,l}^2-c_{p,r}^2\right){}^2+\eta_l \left(v^2-c_{p,l}^2\right) \left(c_{p,r}^2+2 v^2\right) \left(-c_{p,l}^2+i \Sigma +v^2\right)\right]\no\\
&+\eta_r \left[-2 v^3 \eta_l \left(c_{p,l}^2-c_{p,r}^2\right){}^2+c_{p,l} \left(v^2-c_{p,l}^2\right) \left(c_{p,r}^2+2 v^2\right) \left(-c_{p,l}^2+i \Sigma +v^2\right)\right]
\Big\}\no\\
&+\mathcal{O}(\omega^{3/2})\,,\label{sulpl}
\end{align}
where $\eta_l=\sqrt{1+\omega_l^2/\omega^2}$, and $\omega_l=m_{\text{eff},l}\sqrt{(v^2-c_{p,l}^2)/c_{p,l}^2}$.
In particular, when $\omega < \omega_r$ the $\phi_p^{vr}$ mode turns out to be growing modes to be ignored giving the vanishing $S_{ul,vr}$.
These coefficients satisfy the unitary relation \eqref{ul_c}.

The number density $n_{{p\,-\omega}}^{ul}=\langle 0_{in}\vert \hat{b}_{-\omega}^{ul\dagger}\hat{b}_{-\omega}^{ul}\vert 0_{in}\rangle$ for the partner modes of the Hawking radiations is
\begin{align}
n_{p \,-\omega}^{ul}=&\vert S_{ul,+l}\vert^2+\vert S_{ul,vr}\vert^2\no\\
=&\frac{\left(m_{p,l} \eta_l +m_{p,r} \eta_r \right){}^2}{ m_{p,l} \eta_l \left(1+m_{p,r} \eta_r\right){}^2}\left(\frac{T_{H}}{\omega}-\frac{1+  m_{p,l} \eta_l\, m_{p,r} \eta_r}{2 \left( m_{p,l} \eta_l+ m_{p,r} \eta_r\right)}\right)+\frac{m_{p,r} \eta_r \left(1- m_{p,l} \eta_l\right)^2}{ m_{p,l}\eta_l \left(1-m_{p,r}\eta_r \right)^2}\,\theta(\omega-\omega_r)\no\\
&+\mathcal{O}(\omega)\hspace{7.5cm}\qquad \text{for}\qquad\omega_{r}<\omega\le\omega_\text{max}\label{np_ul}\\
&=\left(\frac{m_{p,l}^2\eta _l^2 +m_{p,r}^2 \vert\eta _r\vert^2}{ m_{p,l}\eta _l+m_{p,l}\eta _l  m_{p,r}^2 \vert\eta _r\vert^2}\right)\frac{T_{H}}{\omega}+\frac{\sqrt{m_{p,l}^2-1}\, m_{p,r}^2 \vert\eta _r\vert \left(m_{p,l}^2\eta _l^2 -1\right)}{2 \sqrt{1-m_{p,r}^2}\, m_{p,l}^2 \eta _l \left(m_{p,r}^2 \vert\eta _r\vert^2+1\right)}-\frac{1}{2}+\mathcal{O}(\omega)\no\\
&\hspace{10cm} \text{for}\qquad 0<\omega\le\omega_\text{r}\,.\label{np_ul2}
\end{align}
Since $\phi_p^{ul}$ mode itself is the negative norm states with the mode mixing of the incoming modes of the positive norm states, namely, the $\phi_p^{+l},\,\phi_p^{vr}$ modes,
we then write the particle density $n_{{p\,-\omega}}^{ul}$ in terms of the Hawking temperature $T_H$ obtained above.
Note that both  $\vert S_{ul,+l}\vert^2$ and $\vert S_{ul,vr}\vert^2$ can not be cast into the Planck distribution in the small $\omega$ expansion leading to the nonthermal nature of $n_{{p\,-\omega}}^{ul}$, the result also found in the work  \cite{Roberto2019}.
In general, $\vert S_{ul,vr}\vert^2\ll\vert S_{ul,+l}\vert^2$.
The above features are also true in the gapless cases by setting $\eta_l=\eta_r=1$.

\subsection{$vl$ outgoing channel}
\begin{figure}[h]
\centering
\includegraphics[scale=0.3]{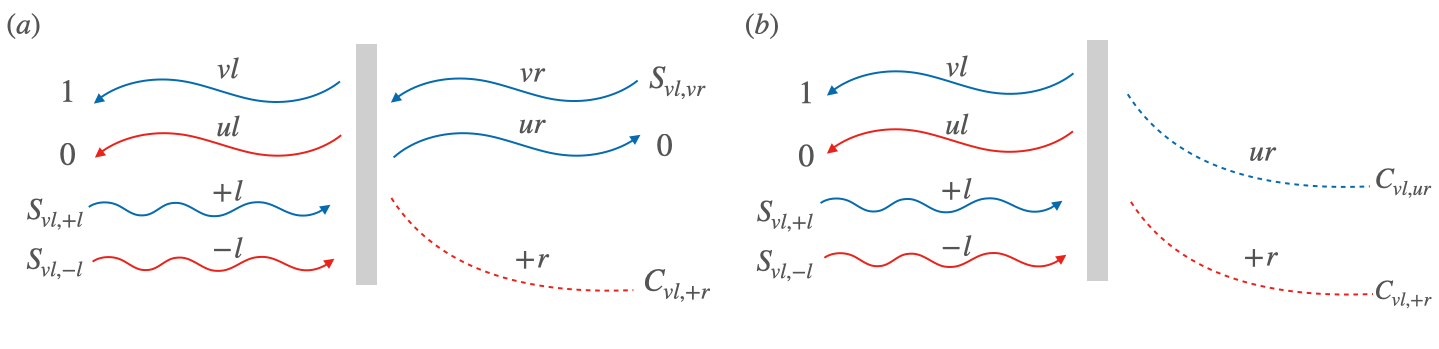}
\caption{Schematic representation of $vl$, out-channel scattering processes when $\omega_{{r}}<\omega<\omega_\text{max}$ in (a), and $\omega<\omega_{{r}}$ in (b).
\label{fig_vl_ch}
}
\end{figure}

%
Finally we compute the $vl$ outgoing channel of the positive-norm $\phi_p^{vl}$ mode with unit amplitude plus  incoming modes, which are  $\phi_p^{-l}$ of the negative norm state with the amplitude $S_{vl,-l}$, $\phi_p^{+l}$ with the amplitude $S_{vl,+l}$, and
   $\phi_p^{vr}$ with the amplitude $S_{vl,vr}$ (see Fig.~\ref{fig_vl_ch}).
The corresponding matching equations are
\begin{align}
\begin{pmatrix}
0\\
1\\
S_{vl,+l}\\
S_{vl,-l}
\end{pmatrix}
=
M_\text{scat}\begin{pmatrix}
0\\
S_{vl,vr}\\
C_{vl,+r}\\
0
\end{pmatrix},
\label{vl-eq}
\end{align}
giving the following scattering coefficients
\begin{align}
S_{vl,-l}=&-\frac{\left(v^2-c_{p,l}^2\right){}^{3/4} \sqrt{m(c_{p,r}^2-v^2)}  \left(c_{p,r} \eta_l-c_{p,l} \eta_r\right) \left(\sqrt{c_{p,r}^2-v^2}+i \sqrt{v^2-c_{p,l}^2}\right)}{\sqrt{2{c_{p,l} \eta_l}\,\omega} \left(c_{p,r}^2-c_{p,l}^2\right) \left(c_{p,r}+v \eta_r\right)}\no\\
&+\left[4 \sqrt{2} \sqrt{m} v \left(c_{p,r}^2-c_{p,l}^2\right) \left(v^2-c_{p,l}^2\right){}^{3/4} \left(c_{p,r}^2-v^2\right)\sqrt{c_{p,l}\,\eta_l} \left(c_{p,r}+v \eta_r\right)\right]^{-1}\no\\
&\times\sqrt{\omega}\, \Big\{c_{p,r} \left(-2 v c_{p,l} \left(c_{p,l}^2-c_{p,r}^2\right){}^2-\eta_l \left[v^2-c_{p,l}^2\right) \left(c_{p,r}^2+2 v^2\right) \left(-c_{p,l}^2+i \Sigma +v^2\right)\right]\no\\
&+\eta_r \left[2 v^3 \eta_l \left(c_{p,l}^2-c_{p,r}^2\right){}^2+c_{p,l} \left(v^2-c_{p,l}^2\right) \left(c_{p,r}^2+2 v^2\right) \left(-c_{p,l}^2+i \Sigma +v^2\right)\right]
\Big\}+\mathcal{O}(\omega^{3/2}),
\label{svlml}
\end{align}
\begin{align}
S_{vl,vr}=&\frac{ c_{p,l}+v \eta_l}{c_{p,r}+v \eta_r}\sqrt{\frac{c_{p,r} \eta_r}{c_{p,l} \eta_l}}\,\theta(\omega-\omega_r)+\mathcal{O}(\omega^{3/2}),\label{svlvr}
\end{align}
\begin{align}
S_{vl,+l}=&\frac{\left(v^2-c_{p,l}^2\right){}^{3/4} \sqrt{m(c_{p,r}^2-v^2)}  \left(c_{p,r} \eta_l-c_{p,l} \eta_r\right) \left(\sqrt{c_{p,r}^2-v^2}-i \sqrt{v^2-c_{p,l}^2}\right)}{\sqrt{2{c_{p,l} \eta_l}\,\omega} \left(c_{p,r}^2-c_{p,l}^2\right) \left(c_{p,r}+v \eta_r\right)}\no\\
&+\left[4 \sqrt{2} \sqrt{m} v \left(c_{p,r}^2-c_{p,l}^2\right) \left(v^2-c_{p,l}^2\right){}^{3/4} \left(c_{p,r}^2-v^2\right)\sqrt{c_{p,l}\,\eta_l} \left(c_{p,r}+v \eta_r\right)\right]^{-1}\no\\
&\times\sqrt{\omega}\, \Big\{c_{p,r} \left[-2 v c_{p,l} \left(c_{p,l}^2-c_{p,r}^2\right){}^2-\eta_l \left(v^2-c_{p,l}^2\right) \left(c_{p,r}^2+2 v^2\right) \left(-c_{p,l}^2-i \Sigma +v^2\right)\right]\no\\
&+\eta_r \left[2 v^3 \eta_l \left(c_{p,l}^2-c_{p,r}^2\right){}^2+c_{p,l} \left(c_{p,l}^2-v^2\right) \left(c_{p,r}^2+2 v^2\right) \left(c_{p,l}^2+i \Sigma -v^2\right)\right]
\Big\}\no\\
&+\mathcal{O}(\omega^{3/2})\,,\label{svlpl}
\end{align}
\begin{figure}[t]
	\centering
	\includegraphics[width=\textwidth]{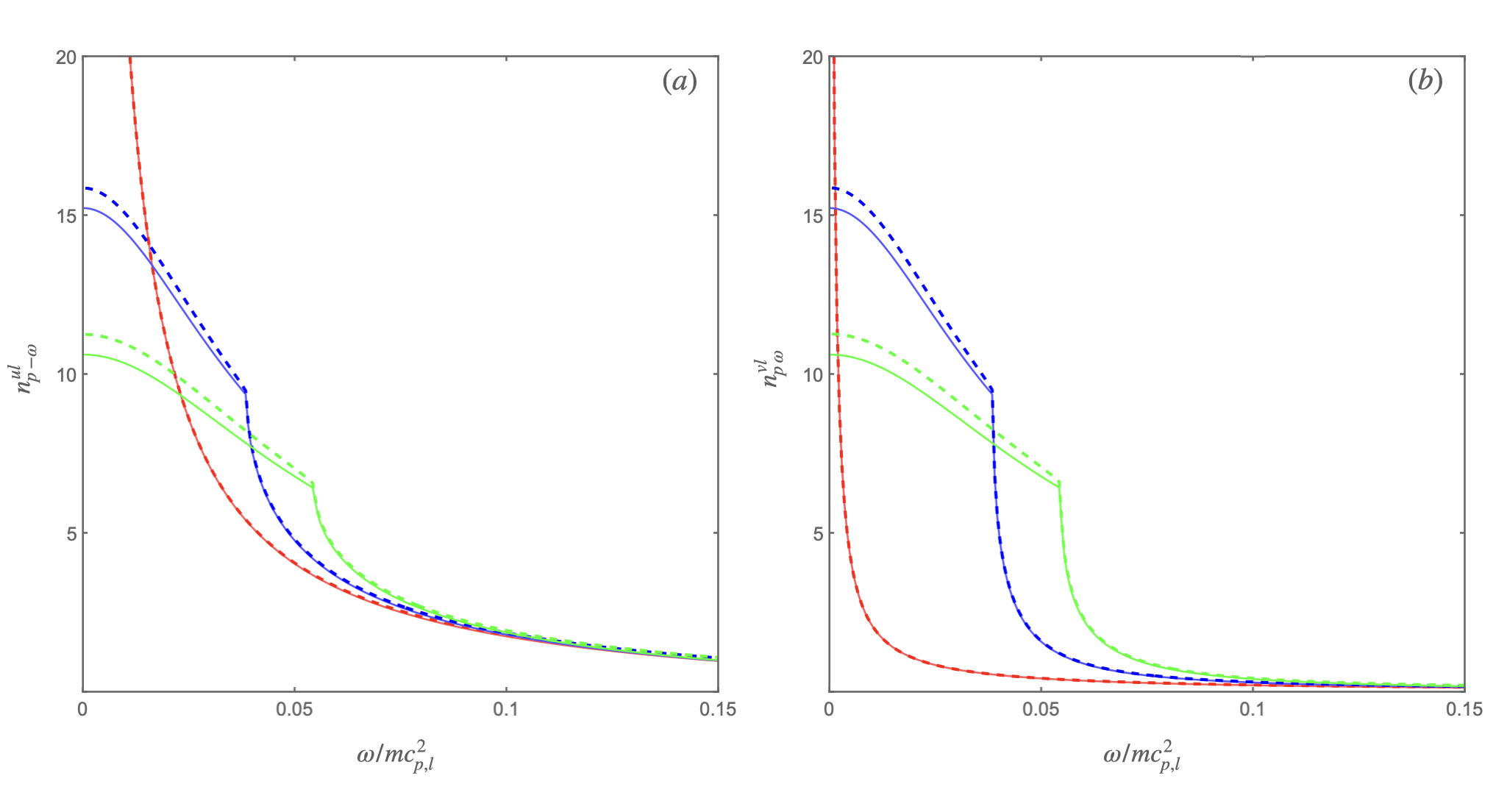}
	\caption{The spectrums of $n_{p\,-\omega}^{ul}$ in (a) and $n_{p\,\omega}^{vl}$ in (b) vary as a function of $\omega$ (in units of $mc_{p,l}^2$)  with different Rabi frequencies: $\Omega/\rho_0U_{l}=0.0$ (red), $3.3\times 10^{-4}$ (blue) and $6.6\times 10^{-4}$ (green). The analytical predictions drawn as  dashed lines are obtained from \eqref{np_ul}-\eqref{np_ul2} and \eqref{np_vl} while the {numerical results, whose details are stated in the text,} are presented with solid lines. The parameters are the same as that used in Fig.~\ref{fig_surml}.
	}
\label{fig_nul_nvl}
\end{figure}
%
which obey the unitary relation \eqref{vl_c}.
The particle density $n_{p\,\omega}^{vl}$ defined as $\langle 0_{in}\vert \hat{b}_{\omega}^{vl\dagger}\hat{b}_{\omega}^{vl}\vert 0_{in}\rangle$ turns out to be
\begin{equation}\label{np_vl}
n_{p \, \omega}^{vl}=\vert S_{vl,-l}\vert^2
\end{equation}
due to the mode mixing from the negative norm state $\phi_p^{-l}$.
Another consistency check comes from the fact that $n_{p \, -\omega}^{ul}-n_{p \,\omega}^{vl}=n_{p \, \omega}^{ur}$, given by the current conservation requirement in (\ref{S_matrix_cc}).
When $\omega<\omega_r=\omega_\text{min}$, since there is no Hawking radiation emission $n_{p \, \omega}^{ur}=0$, then $n_{p\, -\omega}^{ul}=n_{p\, \omega}^{vl} $.
Once the frequency is larger than $\omega_r$, the emergence of Hawking radiation will drastically decrease production of $n_{p\,\omega}^{vl}$. See Fig.~\ref{fig_nul_nvl}-($b$) for detail.

\section{Density-density correlations}\label{ABH4}

Density-density correlation functions of the gapped excitations can  assess the analog Hawking radiation by measuring the correlation between modes in the supersonic and subsonic regimes.
The approximate expressions of scattering coefficients for each outgoing channels obtained in the last section can help realize the density-density correlation function.
To proceed, the density and the associated phase operators can be defined in terms of the fields $\hat{\phi}_s$ and $\hat{\phi}_{s}^\dagger$ as
\begin{align}
&\delta \hat{n}_s=\frac{\delta\hat{\rho}_s}{2\rho_s}=\frac{\hat{\phi_s}+\hat{\phi}^\dagger_s}{2},\quad \delta \hat{\theta}_s=\frac{\hat{\phi}_s-\hat{\phi}_s^\dagger}{2i}, \qquad s=d,p.
\label{theta}
\end{align}
Then the density fluctuation operator  for the gapped excitations can be expanded as
\begin{align}
\delta \hat{n}_p(x,t)=&\int_{\omega_\text{max}}^\infty d\omega\left[\hat{b}_\omega^{ur,out}\chi_{p }^{ur,out}(x,t)+\hat{b}_\omega^{vl,out}\chi_{p}^{vl,out}(x,t)\right]\no\\
&+\int_{\omega_\text{min}}^{\omega_\text{max}}d\omega \left[\hat{b}_\omega^{ur,out}\chi_{p}^{ur,out}(x,t)+\hat{b}_\omega^{vl,out}\chi_{p}^{vl,out}(x,t)+(\hat{b}_{-\omega}^{ul,out})^\dagger\chi_{p}^{ul,out}(x,t)\right],\no\\
&+\int_0^{\omega_\text{min}} d\omega \left[\hat{b}_\omega^{vl,out}\chi_{p}^{vl,out}(x,t)+(\hat{b}_{-\omega}^{ul,out})^\dagger\chi_{p}^{ul,out}(x,t)\right]\no\\&+\text{h.c.},
\label{np}
\end{align}
where
\begin{equation}\label{chi_phi}
\chi_{p}^{j}=(\phi_{p}^j+\varphi_{p}^j)/2
\end{equation}
have the formulas given in (\ref{phi}) and (\ref{variphi}) according to different out basis.
The equal-time correlation function is defined as
\begin{align}
G_p(x,x')=&\langle 0_{in}\vert \{\delta\hat{n}_p(t,x),\delta\hat{n}_p(t,x')\}\vert 0_{in}\rangle,
\label{co_fun_p}
\end{align}
where we consider $in$-vacuum initial state, and $\{\,,\,\}$ is the anticommutation bracket.
Employing the Bogoliubov transforms ~(\ref{bogo_b}), \eqref{bogo_b2}, we have  \eqref{co_fun_p} written as
\begin{align}\label{nn_c}
&\langle\{\delta \hat{n}_{p}(x,t),\,\delta \hat{n}_{p}(x',t)\}\rangle_\omega\no\\
&={\left(\vert S_{ur,vr}\vert^2+\vert S_{ur,+l}\vert^2+\vert S_{ur,-l}\vert^2\right)}\,\chi_{p}^{{ur},out}(x)\chi_{p}^{{ur},out\ast}(x')\no\\
&\quad+{\left(\vert S_{vl,vr}\vert^2+\vert S_{vl,+l}\vert^2+\vert S_{vl,-l}\vert^2\right)}\,\chi_{p}^{{vl},out}(x)\chi_{p}^{{vl},out\ast}(x')\no\\
&\quad+{\left(\vert S_{ul,vr}\vert^2+\vert S_{ul,+l}\vert^2+\vert S_{ul,-l}\vert^2\right)}\,\chi_{p}^{ul,out}(x)\chi_{p}^{ul,out\ast}(x')\no\\
&\quad+{\left(S_{ur,vr}S_{ul,vr}^\ast+S_{ur,+l}S_{ul,+l}^\ast+S_{ur,-l}S_{ul,-l}^\ast\right)} \left(
\chi_{p}^{ur,out}(x) \chi_{p}^{ul,out\ast}(x')+\chi_{p}^{ur,out}(x') \chi_{p}^{ul,out\ast}(x)\right)\no\\
&\quad+{\left(S_{ur,vr}S_{vl,vr}^\ast+S_{ur,+l}S_{vl,+l}^\ast+S_{ur,-l}S_{ul,-l}^\ast\right)}\left(
\chi_{p}^{ur,out}(x) \chi_{p}^{vl,out\ast}(x')+\chi_{p}^{ur,out}(x') \chi_{p}^{vl,out\ast}(x)
\right)\no\\
&\quad+{\left(S_{ul,vr}S_{vl,vr}^\ast+S_{ul,+l}S_{vl,+l}^\ast+S_{ul,-l}S_{vl,-l}^\ast\right)}\left(\chi_{p}^{ul,out}(x) \chi_{p}^{vl,out\ast}(x')+\chi_{p}^{ul,out}(x') \chi_{p}^{vl,out\ast}(x)\right)\no\\
&\quad+\text{c.c.}.
\end{align}
In the end, we will sum over all possible Fourier frequencies $\omega$.
\begin{figure}[t]
\centering
\includegraphics[scale=1]{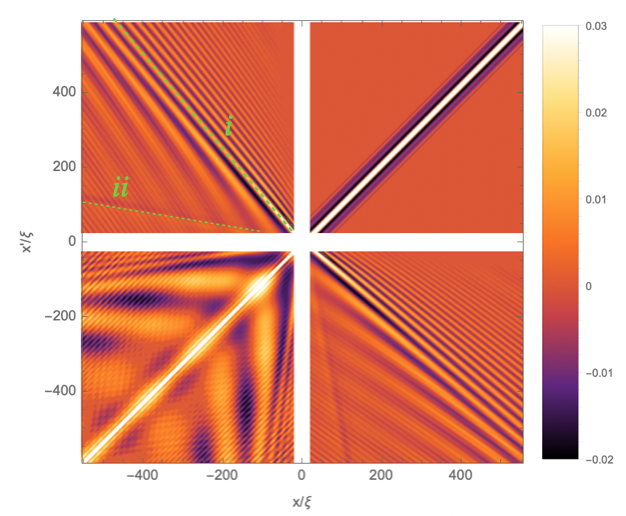}
\caption{The density-density correlation function pattern is obtained from numerical computations of the integral \eqref{co_fun_p} with the scattering coefficients, wave numbers and mode amplitudes also numerically obtained with the details stated in the text. We  have used the parameters $m_{p,l}=7/5,\,m_{p,r}=3/4$, and $\Omega/\rho_0U_{l}=6.6\times 10^{-4}$.}
\label{fig_co_fun_sum}
\end{figure}
%
In Fig.~\ref{fig_co_fun_sum}, we present the full numerical result of the correlation function at equal time, which has the same pattern as in paper \cite{Dudley2018}.
Our result  can be compared with the pattern of the density-density correlation function  of the gapless excitations in \cite{Recati2009,Larr2013}.
The upper-right quadrant $x>0,x'>0$ reveals the correlation of the right-moving positive norm modes $\phi_p^{ur}$ with themselves,  which shows
a clear peak along $x=x'$ line, and the correlations vanish when $x\neq x'$ due to the atom-atom repulsive interactions.
This peak pattern of the gapped excitation is almost the same as for the gapless excitation. The major difference in the density correction function  is manifested in the other three quadrants discussed separately below.

\subsection{The quadrant of correlation function within $x<0,x'>0$ or $x>0,x'<0$}
For  $x>0,\, x'<0$,  the density-density correlation function $G_p(x,x')$ involves the correlations of  the right-moving positive norm modes $\phi_p^{ur}$ outside the horizon at $x'>0$  and the left-moving negative norm modes $\phi_p^{ul}$   inside the horizon at $x<0$ as well as  the left-moving positive norm modes $\phi_p^{vl}$ also inside the horizon at $x<0$.
Apart from the peaks along the two lines denoted by $i$ and $ii$ respectively in Fig.~\ref{fig_co_fun_sum}, which appear in both gapped and gapless cases \cite{Recati2009,Dudley2018,Larr2013}, the ripples are found in the region below the line $i$  \cite{Dudley2018,Antonin2012}. The pattern of {{$G_p(x,x')$ in }}the quadrant $x<0,x'>0$ is the same as in the quadrant $x>0,x'<0$ according to the symmetry along the diagonal line $x=x'$.

The lines of the peaks and the ripples  can be  analytically studied as follows.
Applying (\ref{surml}) and (\ref{sulml}) to (\ref{nn_c}) and collecting the relevant terms in the region of $x<0, x'>0$, the equal-time correlation function  of the upper-left quadrant   turns out to be
\begin{align}
&G_p(x<0,x'>0)=\no\\
&\qquad\int_{\omega_\text{min}}^{\omega_\text{max}}d\omega\,\left[2
S_{ur,-l}S_{ul,-l}^\ast\chi_{p}^{ur,out}(x')\chi_{p}^{ul,out\ast}(x)+2S_{ur,-l}S_{vl,-l}^\ast\chi_{p}^{ur,out}(x')\chi_{p}^{vl,out\ast}(x)
\right]\no\\
&\qquad\qquad\qquad+\text{c.c.}\no\\
&\qquad\sim\int_{\omega_\text{min}}^{\omega_\text{max}}d\omega \bigg[ \frac{m_{p,r}^2 \left(m_{p,l}\eta _l+1\right) \left(m_{p,l}\eta _l +m_{p,r} \eta _r\right)\sqrt{m_{p,l}^2-1} }{4 \pi  \rho_0  v \eta _l \left(m_{p,l}^2-m_{p,r}^2\right)\left(m_{p,r} \eta _r+1\right)}\,\cos{\left(k_{ul}x-k_{ur}x'\right)}\no\\[7pt]
&\qquad\qquad\qquad\qquad\qquad -\frac{m_{p,r}^2 \left(m_{p,l}\eta _l -1\right) \left(m_{p,l}\eta _l -m_{p,r} \eta _r\right)\sqrt{m_{p,l}^2-1} }{4 \pi  \rho_0 v \eta _l \left(m_{p,l}^2-m_{p,r}^2\right) \left(m_{p,r} \eta _r+1\right)}\,\cos{\left(k_{vl}x-k_{ur}x'\right)}\bigg]\, .
\label{co_fun_upleft}
\end{align}
Here the lines $i,\,ii$ can be determined by the above integral evaluated at $\omega=\omega_\text{max}$.
One can approximate the integrand with
$\eta_r=\eta_l=1$ in ~(\ref{co_fun_upleft}) as $\omega\rightarrow \omega_\text{max}$ and then consider the integral at $\omega_\text{max}$ to find
\begin{align}\label{x-_x+}
G_p(x<0,x'>0)\sim &\frac{\left(m_{p,l}+1\right) m_{p,r}^2 \sqrt{m_{p,l}^2-1}}{4\pi \rho_0  v \left(m_{p,r}+1\right) \left(m_{p,l}-m_{p,r}\right)}\frac{\sin\left[\omega_\text{max}\left(\frac{x}{v-c_{p,l}}+\frac{x'}{c_{p,r}-v}\right)\right]}{\left(\frac{x}{v-c_{p,l}}+\frac{x'}{c_{p,r}-v}\right)}\no\\[7pt]
&-\frac{\left(m_{p,l}-1\right) m_{p,r}^2 \sqrt{ m_{p,l}^2-1}}{4 \pi  \rho_0m_{p,l}  \left(m_{p,r}+1\right) \left(m_{p,l}+m_{p,r}\right)}\frac{\sin\left[\omega_\text{max}\left(\frac{x}{v+c_{p,l}}+\frac{x'}{c_{p,r}-v}\right)\right]}{\left(\frac{x}{v+c_{p,l}}+\frac{x'}{c_{p,r}-v}\right)}.
\end{align}
The peaks occur in the lines,
\begin{align*}
&i\qquad\Rightarrow\qquad\frac{x}{v-c_{p,l}}=-\frac{x'}{c_{p,r}-v},\no\\
&ii\qquad\Rightarrow \qquad\frac{x}{v+c_{p,l}}=-\frac{x'}{c_{p,r}-v},
\end{align*}
which are attributed to the correlations of $\phi_p^{ul}-\phi_p^{ur}$ modes ( the line $i$) and $\phi_p^{vl}-\phi_p^{ur}$ modes (the line $ii$), respectively.
The same formula for determining the line $i$ is found for the gapless cases in \cite{{Carlos2011}} as long as $\omega_\text{max} \gg \omega_{l,r}$ leading to the gapless results \cite{Jeff2019}.
The line $ii$ is also obtained. In particular, the magnitude of the density correlations along the line $i$ is much larger than that in the line of $ii$, which is shown in Fig.~\ref{fig_co_fun_sum} and also in \cite{Dudley2018} for the gapped cases, and in \cite{Recati2009,Dudley2018,Larr2013} for the gapless cases.

The pattern reveals the ripples in the region of $\frac{x}{v-c_{p,l}}<-\frac{x'}{c_{p,r}-v}$ to be realized
by studying the phase  $\theta=k_{ul}x-k_{ur}x'$  ($\theta=k_{vl}x-k_{ur}x'$) of the integrand of the first (second) term in (\ref{co_fun_upleft}) as a function of $\omega$.
At the points of this region, the phase $\theta$ decreases with $\omega$ as $\omega$ starts from the lower limit of the integral, namely $\omega_r$, and then increases instead with $\omega$ as $\omega$ reaches $\omega_\text{max}$, developing the local minimum of $\theta$, around which values of $\omega$ give significant contributions to the respective integral (\ref{co_fun_upleft}).
The wavelength of the ripples can be estimated from the minimum value of the phase $\theta$ along the lines with the slops $\frac{v-c_{p,l}}{c_{p,r}-v}$ and $\frac{v+c_{p,l}}{c_{p,r}-v}$, which are perpendicular to $i$, $ii$
giving respectively
\begin{subequations}
\begin{align}
\lambda_i=
&\big\{2 \pi  v m_{p,l} \left(1-m_{p,l}^2\right) \left(m_{p,r}-1\right)^2 \left(m_{p,r}+1\right)\}/\{m_{p,l}^2 \left(m_{p,r}-1\right)^2 \left(m_{p,r}+1\right) \sqrt{\omega_l^2+\omega_r^2}\nonumber
\\&+\omega_r\left[2 m_{p,l}^3 m_{p,r}^3-m_{p,l}^2 m_{p,r}^2 \left(m_{p,l}+m_{p,r}\right)-m_{p,l} m_{p,r} \left(m_{p,l}^2+m_{p,r}^2\right)+m_{p,l}^3+m_{p,r}^3\right]\big\},\\
\lambda_{ii}=
&\big\{2 \pi  v m_{p,l} \left(1-m_{p,l}^2\right) \left(m_{p,r}-1\right)^2 \left(m_{p,r}+1\right)\}\{m_{p,l}^2 \left(m_{p,r}-1\right)^2 \left(m_{p,r}+1\right) \sqrt{\omega_l^2+\omega_r^2}\no\\
&+\omega_r\left[2 m_{p,l}^3 m_{p,r}^3-m_{p,l}^2 m_{p,r}^2 \left(m_{p,l}-m_{p,r}\right)-m_{p,l} m_{p,r} \left(m_{p,l}^2+m_{p,r}^2\right)+m_{p,l}^3-m_{p,r}^3\right]\big\}.
\end{align}
\end{subequations}
In general $\lambda_{ii}<\lambda_{i}$  can be seen in Fig.~\ref{fig_co_fun_sum}, and both wavelengths are inversely proportional to $\omega_l$ and $\omega_r$.
However, in the region of $\frac{x}{v-c_{p,l}}>-\frac{x'}{c_{p,r}-v}$ ripples disappeared because both phases $\theta=k_{ul}x-k_{ur}x'$ and  $\theta=k_{vl}x-k_{ur}x'$ monotonically decrease with $\omega$ that lead to the phase cancellation when integrated over $\omega$.

\subsection{The quadrant of correlation function within  $x<0,x'<0$}
In the lower-left quadrant of $x<0,x'<0$, the density correlations account for the negative norm mode $\phi_p^{ul}$ and the positive norm mode $\phi_p^{vl}$ in the supersonic regime.
The correlation pattern shows a cone shape of the group of the peaks in Fig.~\ref{fig_co_fun_sum}, where a similar behavior also appears in the gapless cases  \cite{Jeff2019}.
However, for the gapped  cases, again the peculiar pattern of the undulations due to the existence of the zero-frequency modes are also shown as in the paper \cite{Dudley2018}.
To understand the origin, notice that  the density correlation function obtained from the relevant terms of (\ref{nn_c}) in the region of $x<0,x'<0$ is
\begin{align}
&G_p(x<0,x'<0)=\no\\
&\int_0^{\omega_\text{max}} d\omega\Big\{\left(1+2\vert S_{vl,-l}\vert^2\right)\chi_{p}^{{vl},out}(x)\chi_{p}^{{vl},out\ast}(x')+{\left(1+2\vert S_{ul,vr}\vert^2+2\vert S_{ul,+l}\vert^2\right)}\chi_{p}^{ul,out}(x)\chi_{p}^{ul,out\ast}(x')\no\\
&\qquad\qquad\quad+ 2\,S_{ul,-l} S_{vl,-l}^\ast\left[ \chi_{p}^{ul,out}(x)\chi_{p}^{vl,out\ast}(x')+\chi_{p}^{ul,out}(x')\chi_{p}^{vl,out\ast}(x)\right]\no\\
&\qquad\qquad\quad +\text{c.c.}\Big\}\,,\label{co_fun_lowerleft}
\end{align}
where the first line contributes the peaks mainly along the line of $x=x'$, and again due to the atom-atom repulsive interaction the correlation vanishes when $x\neq x'$.
The second line manifests the correlations of the $\phi^{ul}$ and $\phi^{vl}$ modes expressed explicitly in terms of the mode functions $\phi$ and $\varphi$ through (\ref{chi_phi}) by
\begin{align}
\sim \int_0^{\omega_\text{max}} d\omega\,
&\frac{m_{p,l} \left(1-m_{p,r}^2\right) \left(m_{p,l}^2\eta _l^2 +1\right)\left( m_{p,l}^2\eta _l^2-m_{p,r}^2 \vert\eta _r\vert^2\right)}{8 \pi  \rho_0  v \eta _l^2 \sqrt{m_{p,l}^2-1} \left(m_{p,l}^2-m_{p,r}^2\right) \left(m_{p,r} \eta _r+1\right){}^2}\no\\
&\hspace{2cm}\times\left[\cos{\left(k_{ul}x-k_{vl}x'\right)}+\cos{\left(k_{ul}x'-k_{vl}x\right)}\right],\label{co_fun_lowerleft2}
\end{align}
where the magnitude is given by inserting the scattering coefficients in (\ref{sulml}) and (\ref{svlml}).
Thus  the cone shape boundary of the two lines  can be determined by evaluating the above integral at its upper limit $\omega_\text{max}$.
In this case, because $\omega_\text{max} \gg \omega_{l,r}$ where $\eta_{l,r} \rightarrow 1$ is considered, the integral evaluated at the upper limit leads to the similar formula as in (\ref{x-_x+}), giving the two lines to be
\begin{align}
\frac{x}{m_{p,l}-1}=\frac{x'}{m_{p,l}+1},\quad\text{and}\quad\frac{x}{m_{p,l}+1}=\frac{x'}{m_{p,l}-1},
\end{align}
which are the same as in the gapless cases in \cite{Recati2009}. The magnitude of the density correlations on the two lines are the same.
The slopes of the lines ($i$ and $ii$) only depend on $m_{p,l}$ since both modes are at $x<0,x'<0$.
The increasing $m_{p,l}$ can increase the effective temperature $T_H$ in (\ref{Th})
but decrease the angle of cone, namely, reducing the area of the undulations confined in the cone.

The undulations, which show up only for the gapped cases, can be understood analytically as follows.
Their typical wavelength $\lambda$  near the direction of $x=x'$ can be extracted directly form the phase $\theta=k_{ul}x-k_{vl}x'$ in the region of small $\omega$ given by
 \begin{align}
&\lambda \simeq \frac{ 2\pi}{(k_{ul}-k_{vl})} \Big \vert_{\omega=0} \simeq \frac{\pi(v^2-c_{p,l}^2)}{c_{p,l}\omega_l},\label{period1}
\end{align}
which is consistent with Fig.~\ref{fig_co_fun_sum}.
The zero-frequency modes in the solutions of the momentum $k_v$ and $k_u$ in (\ref{kv}) and (\ref{ku}) in the supersonic region for the gapped cases play a key role in determining the wavelength of the undulations.
The phase $\theta$ outside the cone decreases monotonically with $\omega$ that leads to the phase cancellation when integrating over $\omega$ so that the undulations disappear.

\section{Quantum entanglement between  Hawking modes and its partners}\label{ABH5}

\subsection{The PHS  criterion}
This section devotes to the discussion of quantum entanglement  in this system.
We first briefly introduce the criterion of the quantum entanglement between the Hawking mode and its partner.
The quantum entanglement and/or the nonseparability  of  the bipartite  systems can be explored based upon  the PHS  criterion \cite{Peres1996,Horodecki1996,Horodecki1997,Simon2000}, which then  is adapted   to assess the entanglement through the simple measure  \cite{Busch2014}
\begin{align}\label{delta_n_c}
&\Delta^{ur,ul}_{p\,\omega}=n_{p\,\omega}^{ur}n_{p\,-\omega}^{ul}-\vert c^{ur,ul}_{p\,\omega}\vert^2.
\end{align}
The formula uses the particle number density $n_\omega$ and their cross correlations  defined as
\begin{align}
&c^{ur,ul}_{p\,\omega}=\langle 0_{in}\vert \hat{b}_{\omega}^{ur,out}\hat{b}_{-\omega}^{ul,out}\vert  0_{in} \rangle =S_{ur,vr}S_{ul,vr}^\ast+S_{ur,+l}S_{ul,+l}^\ast=S_{ur,-l}S_{ul,-l}^\ast\,.\label{curul}
\end{align}
Here we consider the pair modes of the positive norm state $\phi^{ur}$, the analogous Hawking mode in the subsonic regime, and the negative norm state $\phi^{ul}$, the partner mode in the supersonic regime.
Using the unitary properties of the $S$-matrix, the measure then becomes
\begin{equation}
\Delta^{ur,ul}_{p\,\omega}=-\vert S_{ur,-l}\vert^2 \, .
\end{equation}
Apparently,  the values of $\Delta_\omega^{ur,ul}$ are always negative implying
that  two modes are quantum mechanical entanglement. The above discussions hold true for both gapless and gapped excitations.

Here is the side issue about how the nonzero particle  number distribution for the incoming modes in the subsonic regime affects the entanglement of the Hawking mode and its partner \cite{Antonin2018}.
It is quite straightforward to generalize the above formulas by considering the incoming $\phi_p^{vr}$ mode with the nonzero particle distribution function $n_{p\,\omega}^{vr,in}$.
In the gapped cases, it reads
\begin{align}
&\Delta^{ur,ul}_{{{p}}\, \omega}=\left(\vert S_{ul,vr} \vert^2-\vert S_{ur,-l}\vert^2\right)n_{p\,\omega}^{vr,in}-\vert S_{ur,-l}\vert^2,\qquad \omega>\omega_r\,,
\end{align}
which can be rewritten as
\begin{align}
&\frac{\Delta^{ur,ul}_{{{p}} \, \omega}}{\vert S_{ur,-l}\vert^2}=\delta_{ur,ul}n_{p\,\omega}^{vr,in}-1
\end{align}
with
\begin{align}
\delta_{ur,ul} =& \frac{\vert S_{ul,vr}\vert^2-\vert S_{ur,-l}\vert^2}{\vert S_{ur,-l}\vert^2}=\frac{\left(m_{p,l}\eta_l-m_{p,r}\eta_r\right)^2}{4m_{p,l}\eta_lm_{p,r}\eta_r}.
\label{delta}
\end{align}
In \eqref{delta}, $\delta_{ur,ul}$ is always positive, and thus can not possibly be set to vanishing since $m_{p,l}, \eta_l >1$ in the supersonic region and $m_{p,r}, \eta_r <1$ in the subsonic region. $\delta_{ur,ul}$ can be tuned to zero  so as to avoid the reduction of entanglement in the gapless cases ($\eta_{l,r} \rightarrow 1$) under the condition of $v_l c_{l}=v_r c_{r}$ \cite{Antonin2018}.

\subsection{Inhomogeneous equations}

As for the quantum entanglement in the BEC systems, which are affected by the omnipresent environment, there have been extensive studies in various  fluctuating environmental degrees of freedom \cite{Adamek2013,Busch2013,Robertson2015,Lang2020}.
Here the binary systems have the parameter window  to turn on  the interactions between gapless and gapped excitations.
More specifically, we would like to study how the gapped excitations treated as an environment affect the quantum entanglement between the Hawking modes and their partners of the gapless excitations.
This  is an extension of our previous work in \cite{Syu2019}, where the gapped excitations serve as the environmental degrees of freedom to induce the sound cone fluctuations, the analogous light cone fluctuations given by the quantum gravitational effects.
To turn on the interactions, we relax  the restriction of the parameters to have small difference between two intra-species interaction $U_{AA}$, $U_{BB}$ while  other restrictions such as $\rho_l=\rho_r=\rho_0$, $v_l=v_r={-v\,(v>0)}$, $\Omega_l=\Omega_r=\Omega$ still hold.
Then,  the interaction between gapless and gapped excitations starts at $t=0$ with the interaction term
\begin{align}
\mathcal{L}_{\text{int}}=-\alpha\rho_0^2\delta \hat{n}_p\delta \hat{n}_d \, .
\end{align}
The interaction strength $\alpha$ is given by
\begin{align}
\alpha=({U}_{AA}-{U}_{BB})\theta(t).
\end{align}

We assume that two degrees of freedom are completely decoupled when $t<0$, and  establish their own excitations under the super-subsonic configuration, namely $c_{{s},l} < v <c_{s,r}$ for $s=d,p$.
After turning on the interaction, the gapped excitations start to influence the quantum entanglement of the pair modes of the gapless excitations, resulting in  the time dependent  $\Delta_\omega^{jj'}$  developed above.
The coupled equations of motion for $\delta\hat{n}_d$ of the gapless excitations and  $\delta\hat{n}_p$ of the gapped excitations defined in (\ref{theta}), which  can be adapted from \eqref{phi_gapl_eq}-\eqref{phi_gap_eq} by including the above  interaction term, are obtained as
\begin{align}
\left(\partial_t-v\partial_x\right)^2\delta \hat{n}_d+\frac{1}{2m}\partial_x^2\left(\frac{1}{2m}\partial_x^2-2mc_d^2\right)\delta \hat{n}_d=\frac{\alpha \rho_0}{2m}\partial_x^2\delta \hat{n}_p,
\label{ddnd3}
\end{align}
\begin{align} \label{np_e}
\left(\partial_t-v\partial_x\right)^2\delta \hat{n}_p+\frac{1}{2m}\partial_x^2\left(\frac{1}{2m}\partial_x^2-2mc_p^2\right)\delta \hat{n}_p+m_\text{eff}^2\delta \hat{n}_p=\frac{\alpha \rho_0}{2m}(\partial_x^2-2m\Omega)\delta \hat{n}_d\,.
\end{align}
Thus, the general solution of \eqref{ddnd3} is the sum of two parts, the homogeneous solution $\delta n_{d,0}$ and the inhomogeneous solution $\delta n_{d 1}$, namely
\begin{align}
\delta \hat{n}_d(x,t)=\delta \hat{n}_{d,0}(x,t)+\delta \hat{n}_{d,1}(x,t).
\label{gdnd}
\end{align}

The homogeneous solution $\delta n_{d,0}$ obeying the source-free equation (\ref{ddnd3})  is very much the same as in the gapped excitations with the mode expansion (\ref{np}), where the creation/annihilation operators and also the mode functions  are replaced by the counterparts denoted by $ {\hat{a}_\omega^\dagger}/\hat a_\omega$ and $\chi_{d}(x,t)$  defined as in (\ref{chi_phi})  and   obtained from (\ref{gap_sol2}) and \eqref{mode_amplitude}.
The wave number $ k_d $ of the gapless cases for each incoming/outgoing modes are those in (\ref{ku}) and \eqref{kpm} by setting $m_\text{eff}=0$ and replacing $c_{p (l,r)}$ by $c_{d (l,r)}$.
The maximum wave number of the gapless case for having the negative norm states in the downstream at $x<0$ is found to be $ k_\text{max}^{\scriptscriptstyle(0)}$ in (\ref{k_0}), again by replacing $c_{p (l,r)}$ to $c_{d (l,r)}$.
The threshold momentum turns out to be zero for the gapless cases.
In the end, the density-density correlations used later to identify the quantities $n_\omega^j$ and $c_\omega^{jj'}$, with which to compute the criterion $\Delta_\omega^{jj'}$ in (\ref{delta_n_c}) in the gapless cases,  will have the same form as in (\ref{nn_c}). All the $S$-matrix elements in the gapless case are those of the gapped cases in (\ref{surml})-(\ref{surpl}) for the $ur$ outgoing channel, (\ref{sulml})-(\ref{sulpl}) for the $ul$ outgoing channel, and (\ref{svlml})-(\ref{svlpl}) for the $vl$ outgoing channel by taking the limit of $m_\text{eff}=0$ and replacing $c_{p (l,r)}$ by $c_{d (l,r)}$.

On the other hand, the inhomogeneous solution $\delta n_{d\, 1}$ obeys the Eq.~(\ref{ddnd3}) due to the source term from $\delta n_p$,  which is the solution of (\ref{np_e}).
The solution of $\delta n_p$ can also be written as the homogeneous solution $\delta n_{p 0}$   and the inhomogeneous solution that depends linearly on $\delta n_{d\,1}$.
According to \cite{Boyanovsky2017, Syu2021}, substituting the solution $\delta n_p=\delta n_{p\,0}+\delta n_{p\,1}$ back to the Eq.~(\ref{ddnd3}) of $\delta n_{d\,1}$ gives the damping term from the contribution of  $\delta n_{p\,1}$ and leaves $\delta n_{p \,0}$ in the right-hand side of the equation as a source term.
Since the interaction term of the gapless and gapped excitations involves second order spatial derivatives,  the damping effect of the momentum dependence can be parametrized as
 \begin{equation}
 \gamma(k)=(\alpha\rho_0)^2[k^2 (k^2+2m \Omega)/4m^2]/2m_{\text{eff}}^3
 \end{equation}
  \cite{Busch2013}.
In the hydrodynamical approximation  with small $k$ and $\omega$, the relevant parameter from  the gapped excitations to the damping term is $m_{\text{eff}}$, giving the correct dimension of $\gamma$.
Here we will focus on the saturated value of $\Delta_\omega^{jj'}$ whereas the effect of the damping term damps out the oscillatory time-dependent terms, leading to saturation.
The relaxation time scales can be estimated  from $1/\gamma$.
Strictly speaking, the effects from the gapped excitations to the gapless degrees of freedom can be obtained by integrating out the gapped excitations as in \cite{Syu2019}.
This then leads to the Langevin equation that takes into account not only the fluctuations of the gapped excitations manifested in the noise term  but also the damping effect. This deserves future studies.

So, the fluctuations of the gapped degrees of freedom will give the corrections to the density-density correlation function of the gapless excitations through the solution of $\delta \hat{n}_{d,1}$ due to the source term  $\delta \hat{n}_{p, 0}$
\begin{align}
\delta \hat{n}_{d,1}(x,t)=\int_{-\infty}^{\infty} dx'\int_{-\infty}^{\infty}dt' \,g_\text{ret}(x,t,x',t')\,\frac{\alpha \rho_0}{2m}\partial_{x'}^2\delta \hat{n}_{p, 0} (x',t')\,.
\end{align}
The Fourier transform of the retarded Green's function is defined as
\begin{align}
g_\text{ret}(x,t;x',t')=\int \frac{d\omega}{2\pi}\frac{dk}{2\pi}\,\, \tilde{g}_\text{ret}(\omega,k)e^{-i\omega(t-t')}e^{ik(
x-x')}
\end{align}
with the solution
\begin{align}
\tilde{g}_{\text{ret}}({\omega},{k})=\frac{-1}{\left({\omega+v_0{k}}\right)^2+ 2 i \gamma(k)  \omega-c_d^2{k}^2-{k}^4/4m^2} \,,
\end{align}
which includes the damping effect.
The inhomogeneous solution  for the small wave number consistent with the hydrodynamical approximation lying within   {$\gamma < vk$}  becomes
\begin{align}
\delta \hat{n}_{d,1}(x,t)=& \frac{\alpha\rho_0i}{2m}\int dx'dt'\theta(t')\theta(t-t')\no\\
&\times\int\frac{ d{k}}{2\pi} \left[\frac{e^{-i({\omega}_{d\,-}(k)-i\gamma(k))(t-t')}-e^{-i({\omega}_{d\,+} (k)-i\gamma(k))(t-t')}}{{\omega}_{d\,-}(k)-{\omega}_{d\,+} (k)}\right]e^{i{k}(x-x')}
\,\partial_{x'}^2\delta \hat{n}_{p,0}(x',t'),
\label{dndp}
\end{align}
where, ${\omega}_{d\, -,+}$ are  given by
\begin{subequations}
\begin{align}
&\omega_{d\, -} (k)\equiv -v\,k-\sqrt{c_d^2\,k^2+k^4/4m^2},\\
&\omega_{d\,+} (k)\equiv-v\,k+\sqrt{c_d^2\,k^2+k^4/4m^2}\, .
\end{align}
\end{subequations}
The frequencies $\omega_{d\, -}$ and $\omega_{d\,+}$ are obtained from the dispersion relation \eqref{dsp_gapl} of the gapless excitations of the negative and positive frequencies branches, respectively.
Substituting the mode expansion of $\delta \hat{n}_{p, 0}$  expressed in (\ref{np}) and integrating over the space, time and also the wave number, we end up with the formula
\begin{align}
\delta \hat{n}_{d,1}(x,t)=
&\int_{\omega_\text{max}}^\infty d{\omega}\{\,\hat{b}_\omega^{ur,out} w_+^{ur}+\hat{b}_\omega^{vl,out} w_+^{vl}\}\no\\
&+\int_{\omega_\text{min}}^{\omega_\text{max}}d{\omega}\{\,\hat{b}_\omega^{ur,out} w_+^{ur}+(\hat{b}_{-\omega}^{ul,out})^\dagger w_+^{ul}+\hat{b}_\omega^{vl,out} w_+^{vl}\}\no\\
&+\int_0^{\omega_\text{min}} d{\omega}\{\,(\hat{b}_{-\omega}^{ul,out})^\dagger w_+^{ul}+\hat{b}_\omega^{vl,out} w_+^{vl}\}\no\\
&+\text{H.c.},
\label{dndp6}
\end{align}
where we have defined the effective mode functions
\begin{subequations} \label{dndp4}
\begin{align}
w_+^{j}(x,t)=&-\frac{\alpha\rho_0}{2m}\,\left(
\frac{1-e^{-\gamma[k_j]}e^{-i({\omega}_{d\,-}[{k}_j]-{\omega})t}}{({\omega}_{d\,-}[{k}_j]-i\gamma[k_j]-{\omega})}-
\frac{1-e^{-\gamma[k_j]}e^{-i(\omega_{d\, +}[{k}_j]-{\omega})t}}{(\omega_{d\, +}[{k}_j]-i\gamma[k_j]-{\omega})}\right)\\
&\times\frac{{k}_j^2}{\Delta\omega_{d\, \pm}[{k}_j]}\,\chi_{p}^{j,out}(x,t),\no\\[9pt]
w_-^{j}(x,t)=&-\frac{\alpha  \rho_0}{2m}\,\left(
\frac{1-e^{-\gamma[-k_j]}e^{-i({\omega}_{d\,-}[-{k}_j]+{\omega})t}}{{\omega}_{d\,-}[-{k}_j]-i\gamma[-k_j]+\omega}-
\frac{1-e^{-\gamma[-k_j]}e^{-i({\omega}_{d\,+}[-{k}_j]+{\omega})t}}{\omega_{d\, +}[-{k}_j]-i\gamma[-k_j]+\omega}\right)\no\\
&\times\frac{(-{k}_j)^2}{\Delta \omega_{d\, \pm}[-{k}_j]}\,\chi_{p}^{j,out\ast}(x,t)
\end{align}
\end{subequations}
with
\begin{align*}
\Delta\omega_{d\, \pm}[k_j]={\omega}_{d\,-}[{k}_j]-\omega_{d\, +}[{k}_j]=-2\sqrt{c_d^2{k}_j^2+{k}_j^4/4m^2}
\end{align*}
for the channels $j$ including $j=ur,ul,vl$ with the wave number $k_j$ of the gapped excitations obtained from a fixed $\omega$.
%
The Green's function serves as  a window function that smears the effect from the gapped excitations to the gapless excitations.

We are now ready  to calculate the correlation function of the gapless excitations including the corrections from the gapped excitations for a particular frequency  $\omega$
\begin{align}
G_{d \, \omega}(x,t;x',t')&=\langle in\vert \{\delta\hat{n}_d(x,t),\delta\hat{n}_d(x',t')\}\vert in\rangle_\omega\no\\
&=\langle in\vert \{\delta\hat{n}_{d,0}(x,t),\delta\hat{n}_{d,0}(x',t')\}\vert in\rangle_\omega+\langle in\vert \{\delta\hat{n}_{d,1}(x,t),\delta\hat{n}_{d,1}(x',t')\}\vert in\rangle_\omega\no\\
&=G_{d,\omega\,0}(x,t;x',t')+G_{d,\omega\,1}(x,t;x',t'),
\label{co_fun}
\end{align}
where the initial  state  is given by the in-vacuum state of the gapless and gapped excitations  $\vert in\rangle=\vert 0_{in}\rangle_\text{gapless}\otimes \vert 0_{in}\rangle_\text{gapped}$.
The first term is the  correlation function of the unperturbed gapless excitations.
This  is quite a straightforward calculation to compute $G_{d,\omega\,0}(x,t;x',t')$ by starting from the mode expansion of the gapless excitations in (\ref{gapless}) where  the mode functions are written in (\ref{gap_sol2}) in terms of the coefficients in (\ref{mode_amplitude}).
The wave numbers denoted by  {$k_d$} together with the $S$-matrix elements  {$S^d$} associated with the Bogoliubov transformation of the creation/annihilation operators as in (\ref{bogo_b}) can be read off from the gapped excitations in (\ref{ku}) and (\ref{kpm}) and the above $S$-matrix elements by replacing $c_p \rightarrow c_d$ and also setting $m_\text{eff}=0$ or $\Omega=0$ limit.
The second term comes from the modification due to the gapped excitations.
In this study, we consider  both gapless and gapped excitations are under the supersonic-subsonic configuration in the monometric  in \eqref{phi_gapl_eq}-\eqref{phi_gap_eq}  choosing $\Omega=4\rho_0 U_{AB}\ll \rho_0 U$.
Also, for small Rabi-coupling $\Omega$  giving small $m_\text{eff} \propto \sqrt{\Omega} $ in (\ref{meff}), the  sound-speed difference  $c_p-c_d \propto \Omega $  in (\ref{cd}) and (\ref{cs}) is small, and controls the smallness of the wave number difference $k_j- k_{d\, j} \propto \Omega \ll 1/\xi$  in various  modes.
We then can  approximate $w_{\pm}^{j}$ in (\ref{dndp4}) of the gapped excitations by
\begin{subequations}
\begin{align}
&w_+^{j}(x,t)\simeq W_+^{j}(t) \chi_{d}^{j,out}(x,t)\left[1+i\mathcal{O}\left(x({k}_j- k_{ d \, j})\right)\right],\label{wp}\\[10pt]
&w_-^{j}(x,t)\simeq W_-^{j}(t) \chi_{d}^{j,out}(x,t)\left[1-i\mathcal{O}\left(x({k}_j- k_{d \, j})\right)\right],\label{wm}
\end{align}
\end{subequations}
 where
\begin{subequations}  \label{W_fun}
\begin{align}
	&W_+^{j}(t)=-\frac{\alpha\rho_0}{2m}\left(\frac{A_{p\,k_j}+B_{p\,k_j}}{A_{d\,k_j}+B_{d\,k_j}}\right)\no\\
	&\quad\qquad\qquad\times\left(
	\frac{1-e^{-\gamma[k_j]t}e^{-i({\omega}_{d\,-}[{k}_j]-{\omega})t}}{{\omega}_{d\,-}[{k}_j]-i\gamma[k_j]-{\omega}}-
	\frac{1-e^{-\gamma[k_j]t}e^{-i(\omega_{d\, +}[{k}_j]-{\omega})t}}{\omega_{d\, +}[{k}_j]-i\gamma[k_j]-{\omega}}\right)\frac{{k}_j^2}{\Delta\omega_{d\, \pm}[{k}_j]}\,,\\
	&W_-^{j}(t)=W_+^{j\ast}(t)\, .
\end{align}
\end{subequations}
As long as  the length scales in the hydrodynamic approximation  of our interest are of order of the correlation length $\xi$ or beyond it, it leads to the considerably small error to be ignored safely.
Therefore, one can rewrite the functions $\chi_{p}^{j,out}(x,t)$ in (\ref{np}) of the gapped excitations defined in (\ref{chi_phi}), (\ref{gap_sol2}) and (\ref{mode_amplitude2}) in terms of $A_{p\,k_j}$ and $B_{p\,k_j}$ \eqref{mode_amplitude2} by the functions $\chi_{d}^{j,out}(x,t)$ of the gapless excitations in terms of $A_{d\,k_j}$ and $B_{d\,k_j}$ in (\ref{mode_amplitude})
instead.

According to different quadrants in the $(x,x')$ plane,  we summarize the correlation functions as follows.
For the upper-right quadrant $x>0,x'>0$ in the interval $\omega_\text{min}<\omega\le\omega_\text{max}$,
\begin{align}
G_{d \, \omega}(x,x';t)\simeq& \left[(1+2\vert {S}^d_{ur,-l}\vert^2)+(1+2\vert {S}_{ur,-l}\vert^2)W_+^{ur}(t)W_-^{ur}(t)\right] \chi_{d}^{ur,out}(x,t)\chi_{d}^{ur,out\ast}(x',t)\no\\
&+\text{c.c.},
\label{mco_fun1}
\end{align}
and in the interval $0<\omega\le\omega_\text{min}$,
\begin{align}
G_{d\, \omega }(x,x';t)\simeq (1+2 {S}^d_{ur,-l}\vert^2)\chi_{d}^{ur,out}(x,t)\chi_{d}^{ur,out\ast}(x',t)+\text{c.c.}.
\label{mco_fun11}
\end{align}
For the upper-left quadrant $x<0,x'>0$ in the interval $\omega_\text{min}<\omega\le\omega_\text{max}$,
\begin{align}
G_{d\, \omega }(x,x';t)\simeq &\left(2 {S}^d_{ur,-l}{S}_{ul,-l}^{d\, \ast}+2S_{ur,-l}S_{ul,-l}^\ast W_+^{ur}(t)W_-^{ul}(t)\right) {\chi}_{d}^{ur,out}(x',t){\chi}_{d}^{ul,out\ast}(x,t)\no\\
&\qquad+\left(2 {S}^d_{ur,-l}{S}_{vl,-l}^{d\, \ast}+2S_{ur,-l}S_{vl,-l}^\ast W_+^{ur}(t)W_-^{vl}(t)\right){\chi}_{d}^{ur,out}(x',t){\chi}_{d}^{vl,out\ast}(x,t)\no\\
&+\text{c.c.},
\label{mco_fun2}
\end{align}
and in the interval $0<\omega\le\omega_\text{min}$,
\begin{align}
G_{d\,\omega} (x,x';t)\simeq &\left(2 {S}^d_{ur,-l}{S}_{ul,-l}^{d\,\ast}\right) {\chi}_{d}^{ur}(x',t){\chi}_{d}^{ul\ast}(x,t)+\left(2{S}^d_{ur,-l}{S}_{vl,-l}^{d\,\ast}\right){\chi}_{d}^{ur,out}(x',t){\chi}_{d}^{vl,out\ast}(x,t)\no\\
&+\text{c.c.}.
\label{mco_fun2}
\end{align}
Finally, for the lower-right quadrant $x<0,x'<0$ in the interval of $\omega_\text{min}<\omega\le\omega_\text{max}$,
\begin{align}
G_{d\, \omega }(x,x';t)\simeq
&\left\{ \left[1+2\left(\vert{S}^d_{ul,vr}\vert^2+\vert {S}^d_{ul,+l}\vert^2\right)\right]+\left[1+2\left(\vert{S}_{ul,vr}\vert^2+\vert{S}_{ul,+l}\vert^2\right)\right]W_+^{ul}(t)W_-^{ul}(t)\right\}\no\\
&\times{\chi}_{\omega \,d}^{ul,out}(x,t){\chi}_{d}^{ul,out\ast}(x',t)+\left[ \left(1+2\vert{S}^d_{vl,-l}\vert^2\right)+\left(1+2\vert{S}_{vl,-l}\vert^2\right)W_+^{vl}(t)W_-^{vl}(t)\right]\no\\
&\times{\chi}_{\omega \,d}^{vl,out}(x,t){\chi}_{d}^{vl,out\ast}(x',t)+\text{c.c.},
\end{align}
and in the interval $0<\omega\le\omega_\text{min}$,
\begin{align}
G_{d\, \omega} (x,x';t)\simeq &\left\{ \left[1+2\left(\vert{S}^d_{ul,vr}\vert^2+\vert{S}^d_{ul,+l}\vert^2\right)\right]+\left(1+2\vert{S}_{ul,+l}\vert^2\right)W_+^{ul}(t)W_-^{ul}(t)\right\}\no\\
&\times \chi_{d}^{ul,out}(x,t)\chi_{d}^{ul,out\ast}(x',t)+\left[ \left(1+2\vert{S}^d_{vl,-l}\vert^2\right)+\left(1+2\vert{S}_{vl,-l}\vert^2\right)W_+^{vl}(t)W_-^{vl}(t)\right]\no\\
&\times\chi_{d}^{vl,out}(x,t)\chi_{d}^{vl,out\ast}(x',t)+\text{c.c.}.
\end{align}

\subsection{Nonseparability of Hawking-partner pairs}

Having gotten the correlation functions, we can extract the effective  mean occupation number $1+2{n}_{\omega}^{ur}$ and $1+2{n}_{\omega}^{ul}$  from the correlation functions in $x>0,x'>0$ and $x<0,x'<0$, and the cross correlation terms $2{c}_{\omega}^{ur,ul}$ from the correlation function in  $x<0,x'>0$ \cite{Busch2013,Robertson2015}.
For a given frequency of the pair mode of the gapless excitations, due to the bilinear coupling between the gapless and gapped excitations, the corrections from the gapped excitations also come from the modes with frequency $\omega$.
Thus, when $\omega_\text{min}<\omega\le\omega_\text{max}$
\begin{subequations}
\begin{align}
&n_{d\, \omega}^{ur}=\vert  {S}^d_{ur,-l}\vert^2+\left(\,\vert  S_{ur,-l}\vert^2+1/2\right)\vert W_+^{ur}\vert^2,\label{n^ur}\\
&n_{d\,\omega}^{ul}=\left(\,\vert  {S}^d_{ul,vr}\vert^2+\vert  {S}^d_{ul,+l}\vert^2\right)+\left(\,\vert  S_{ul,vr}\vert^2+\vert  S_{ul,+l}\vert^2+1/2\right)\vert W_+^{ul}\vert^2,
\label{n^ul}\\
&c_{d\, \omega }^{ur,ul}= {S}^d_{ur,-l} {S}_{ul,-l}^{d\, \ast}+  S_{ur,-l} S_{ul,-l}^\ast W_+^{ur} W_-^{ul}\, .
\label{c}
\end{align}
\end{subequations}
The criterion $\Delta_{d\,\omega}^{ur,ul}$ in (\ref{delta_n_c}) can be computed, using the unitary relations of gapless and gapped excitations leads to
\begin{align}
		\Delta_{d\, \omega}^{ur,ul}\simeq &-\vert {S}^d_{ur,-l}\vert^2 +\vert {S}^d_{ur,-l}\vert^2\left(\frac{1}{2}+\vert {S}_{ul,+l}\vert^2\right)\vert W_+^{ul}\vert^2+\vert {S}^d_{ul,+l}\vert^2\left(\frac{1}{2}+\vert {S}_{ur,-l}\vert^2\right)\vert W_+^{ur}\vert^2\no\\
		&-\left( {S}^d_{ur,-l}{S}_{ul,-l}^{d\,\ast} S_{ur,-l}S_{ul,-l}^\ast W_+^{ur}W_-^{ul}+\text{c.c.}\right)+\mathcal{O}(\vert W_+^j\vert^3)\, .
		\label{delta_large_w}
	\end{align}
With no influence from the gapped excitations, $\Delta_{d\, \omega}^{ur,ul}$ is negative meaning that the Hawking modes $\phi_d^{ur}$ and their partner modes  $\phi_d^{ul}$ are quantum mechanically entangled.
The effects of the gapped excitations contribute the modification of the occupation number density $n_{d\, \omega}^{ur}$ and $n_{d\,\omega}^{ul}$ in (\ref{n^ur}) and (\ref{n^ul}) that increase the value of $\Delta_{d\, \omega}^{ur,ul}$ diminishing the separability.
However,  the cross correlation term $\vert c_{d\, \omega }^{ur,ul} \vert^2$ in (\ref{c}) decreases $\Delta_{d\, \omega}^{ur,ul}$ strengthening the entanglement,  but its effect is  less than that of  $n_{d\, \omega}^{ur}$ and $n_{d\,\omega}^{ul}$ due to the relative smallness of the overlap between $W_+^{ur}$ and $W_-^{ul}$.
The net effects of the gapped excitations raise $\Delta_{d\, \omega}^{ur,ul}$, driving the pair modes toward disentanglement.

However, when $0<\omega\le \omega_\text{min}$, since the $\phi^{ur}$ modes of the gapped excitations are not the propagating modes, the modification only acts on $n_\omega^{ul}$, while the other two terms remain unaffected.
Thus,
\begin{subequations}\label{nnc}
\begin{align}
&n_{d\,\omega}^{ur}=\vert  {S}^d_{ur,-l}\vert^2,\label{n^ur2}\\
&n_{d\,\omega}^{ul}=\left(\,\vert  {S}^d_{ul,vr}\vert^2+\vert  {S}^d_{ul,+l}\vert^2\right)+\left(\,\vert  S_{ul,+l}\vert^2+1/2\right)\vert W_+^{ul}\vert^2,
\label{n^ul2}\\
&c_{d\, \omega }^{ur,ul}= {S}^d_{ur,-l} {S}_{ul,-l}^{d\, \ast}.
\label{c2}
\end{align}
\end{subequations}
\begin{figure}[thb]
	\centering
	\includegraphics[width=0.7\textwidth]{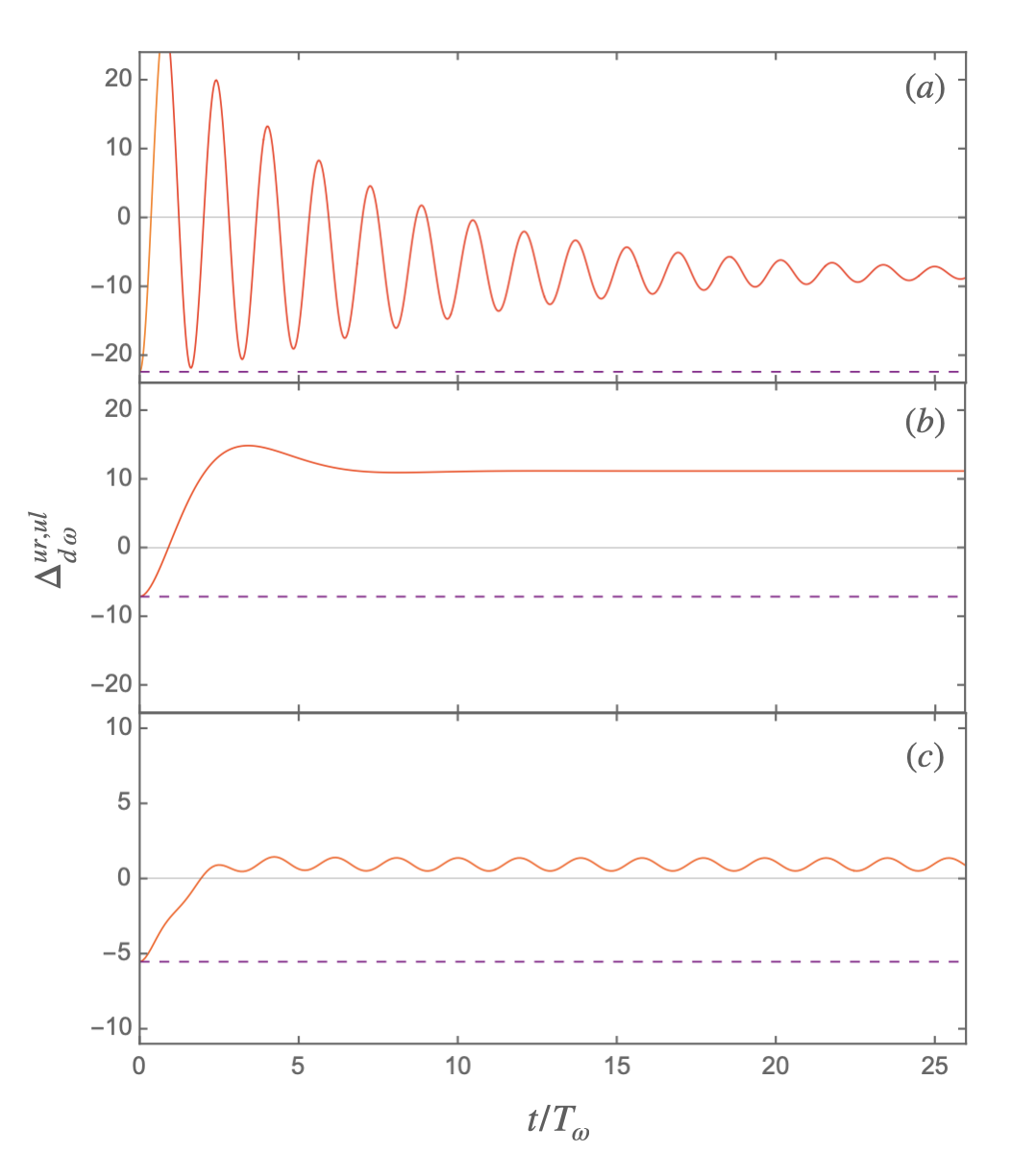}
	\caption{The time evolution of $\Delta_{d,\omega}^{ur,ul}$ as a function of the time  in unit of period $T_\omega=2\pi/\omega$ with the parameters Mach number $m_{p,r}=0.63,\, m_{p,l}=1.63$,  the  coupling constant $\alpha/U_{l}=0.13$, the Rabi frequency $\Omega/\rho_0U_{l}=6.6\times 10^{-4}$.  Considering (a) $\omega=0.041T_H<\omega_{\text{min}}=0.157T_H$,  $\Delta_{d\,\omega}^{ur,ul}$ behaves as an underdamped  oscillation,   (b)  $\omega=0.124T_H<\omega_{\text{min}}=0.157T_H$ ,   $\Delta_{d\,\omega}^{ur,ul}$ behaves as an overdamped  oscillation,  and (c)  $\omega=0.16T_H>\omega_{\text{min}}=0.157T_H$, the behavior of  $\Delta_{d\,\omega}^{ur,ul}$ has mixture of overdamped  oscillation for $ul$ mode and underdamped  oscillation for $ur$ mode.   The dashed lines indicate the values of $\Delta_{d,\omega}^{ur,ul}$ with no corrections from the gapped excitations.
		\label{fig_damping}
	}
\end{figure}
\begin{figure}[h]
	\centering
	\includegraphics[width=0.7\textwidth]{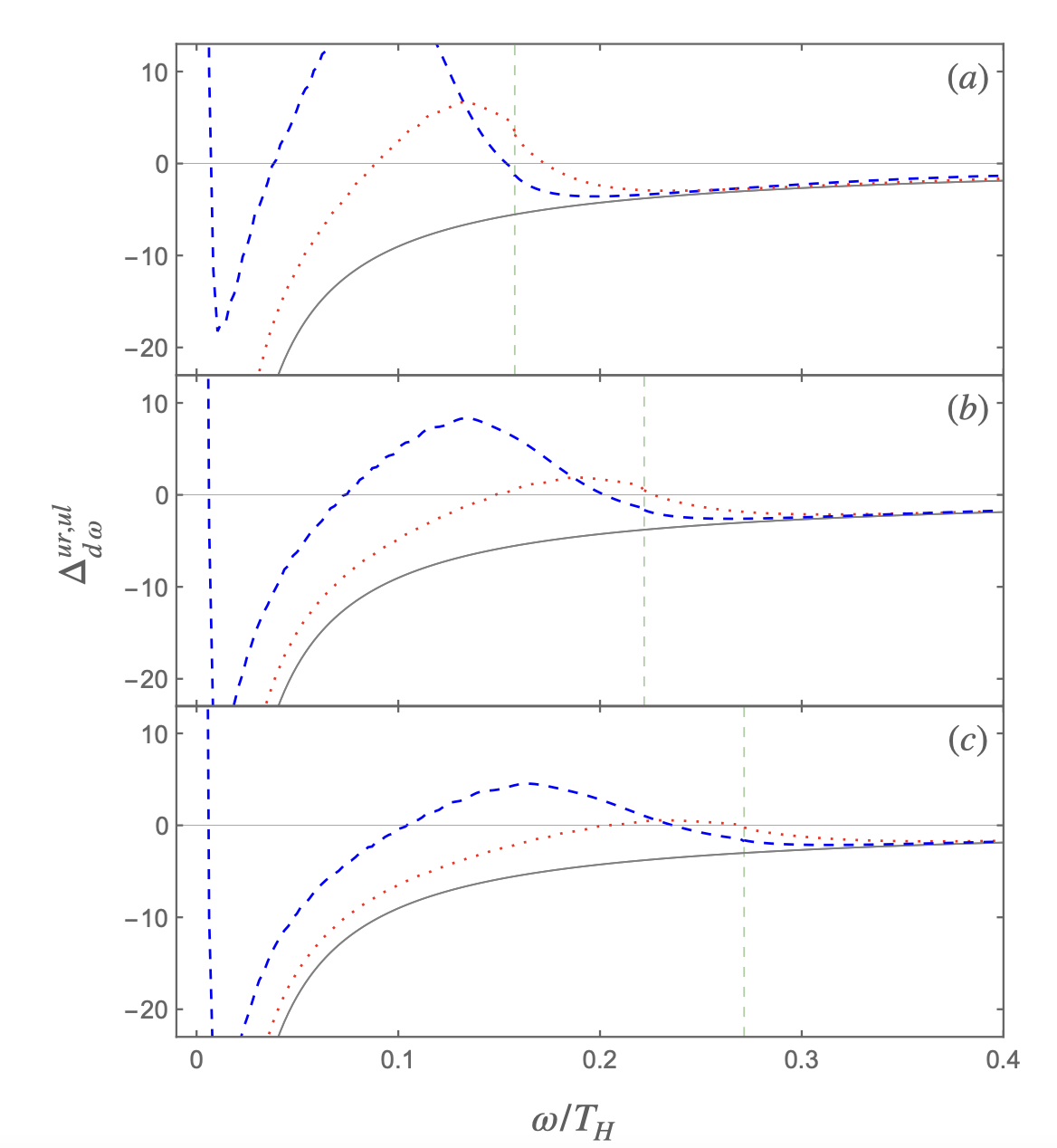}
	\caption{Late-time saturation of  $\Delta_{d\,\omega}^{ur,ul}$ is plotted as function of $\omega$ in unit of effective temperature $T_H$.  By varying the Rabi frequency $\Omega/\rho_0U_{l}$:  (a) $6.6\times10^{-4}$,  (b) $1.33\times10^{-3}$, and  (c) $2\times10^{-3}$, we show $\Delta_{d\, \omega}^{ur,ul}$  with two different values of the mutual coupling constants $\alpha/U_{l}:$ 0.1 (red-dotted line) and 0.17 (blue-dashed line).  Other parameters are chosen as $m_{p,r}=0.63,\,m_{p,l}=1.63$.
	}
	\label{fig_delta_om_alpha}
\end{figure}
%
The criterion $\Delta_{d\, \omega }^{ur,ul}$  given by \eqref{nnc} becomes
\begin{align}
\Delta_{d\, \omega }^{ur,ul}\simeq&-\vert {S}^d_{ur,-l}\vert^2 +\vert {S}^d_{ur,-l}\vert^2\left(\frac{1}{2}+\vert S_{ul,+l}\vert^2\right)\vert W_+^{ul}\vert^2+\mathcal{O}(\vert W_+^{ul}\vert^3)\,
\label{delta_small_w}
\end{align}
also increasing  $\Delta_{d\, \omega}^{ur,ul}$  with the tendency to disentangle the pair of modes.

The time evolution of $\Delta_{d\, \omega }^{ur,ul}$ for various values of $\omega$ is plotted  in Fig.~\ref{fig_damping}. To understand the behavior, one can approximate $\vert W_+^{ur}\vert^2$ and $\vert W_+^{ul}\vert^2$ from (\ref{W_fun})  as
\begin{subequations}
\begin{align}
&\vert W_+^{ur}\vert^2\propto \frac{1+e^{-2\gamma[k_{ur}]t}-2e^{-\gamma[k_{ur}]t}\cos{[(\omega-{\omega}_{d\,+}[k_{ur}])}t]}{\gamma^2[k_{ur}]+(\omega -{\omega}_{d\,+}[k_{ur}])^2}\frac{k_{ur}^4}{ (\Delta {\omega}_{d\,\pm}[k_{ur}])^2},\\[7pt]
&\vert W_+^{ul}\vert^2\propto \frac{1+e^{-2\gamma[k_{ul}]t}-2e^{-\gamma[k_{ul}]t}\cos{[(\omega-{\omega}_{d\,-}[k_{ul}])}t]}{\gamma^2[k_{ul}]+(\omega -{\omega}_{d\,-}[k_{ul}])^2 }\frac{k_{ul}^4}{(\Delta\omega_{d\, \pm}[k_{ul}])^2}\,.
\end{align}
\end{subequations}
For a given frequency $\omega$, the wave numbers of the Hawking modes  and their partners of the gapped excitations $k_{ur}$ and $k_{ul}$ can be given from (\ref{ku}) in the setting of the supersonic-subsonic configuration.
For $\omega < \omega_\text{min}$, the corrections to $\Delta_{d\, \omega }^{ur,ul}$  in (\ref{delta_small_w}) only come from the term $\vert W_+^{ul}\vert^2$.
However, when $\omega > \omega_\text{min}$, the terms of $ W_+^{ur}$ and $ W_+^{ul}$ contributes to the corrections in (\ref{delta_large_w}). Due to the fact that the damping factor $\gamma[k_{ul}]> \gamma[k_{ur}]$ for a given $\omega$, the saturation of $\Delta_{d\, \omega }^{ur,ul}$ will take longer time.
As for the saturated values in Fig. \ref{fig_delta_om_alpha}, it is evident that the large corrections occur in frequency that respectively minimize  $\gamma^2[k_{ur}]+(\omega -{\omega}_{d\,+}[k_{ur}])^2$ and $\gamma^2[k_{ul}]+(\omega -{\omega}_{d\,-}[k_{ul}])^2$ when $\omega \sim \omega_\text{min}$
as anticipated. And for large enough value $\alpha$, the behavior of $1/\omega$ in the thermal spectrum $n_{p \,\omega}^{ur}=\vert S_{ur,-l} \vert^2$ given by (\ref{np_ur})
will significantly deteriorate the quantum entanglement of the pair modes of the gapless excitations as seen in the expression (\ref{delta_small_w}) and Fig. \ref{fig_delta_om_alpha}.

The presence of the environmental field at finite temperature that deteriorate the quantumness of the pair modes
of the gapless excitations in the BEC system has been studied in \cite{Busch2014}.
Here the effects the gapped excitations serving as the environmental degrees of the freedom
in two-component BEC systems even though they are in their vacuum states are to provide the
stochastic noise that not only induces the sound cone fluctuations 
in \cite{Syu2019} but also reduces the quantum entanglement of  gapless excitation pairs.
The similar reduction mechanism on the quantumness of the modes will expectedly be seen
in the BEC/BCS crossover models of the ultra-cold Fermi systems as  the noise manifested from quantum density fluctuations in condensates of ultra-cold Fermi gases of the gapped modes
is found to lead to fluctuations in phonon times-of-flight in the BEC regime \cite{Lee2020}.

\section{Summary and outlook}\label{ABH6}
We investigate the properties of the condensates of cold atoms at zero temperature in the tunable binary Bose-Einstein condensate system  with  the Rabi transition between  atomic hyperfine states where the system can be represented by a coupled two-field model of gapless excitations and gapped excitations, analogous to the Goldstone modes and the gapped Higgs modes in particle physics.
We then set up the configuration of the supersonic and subsonic regimes with the acoustic horizon between them by means of  the spatially dependent coupling constant between two hyperfine states in the elongated two-component Bose-Einstein condensates. The aim is to try to mimic Hawking radiations, in particular due to the gapped excitations with the tunable energy-gap term induced by the Rabi-coupling constant.
In this work, the simplified steplike sound speed profile is adopted to implement the subsonic-supersonic transition  so that such a system be analytically treatable.
Two sets of the wave functions of the gapped excitations can be determined by matching them at the acoustic horizon using the obtained dispersion relations.
The existence of the negative norm states in the supersonic regime leads to the mixing between the  creation operator and the annihilation operator through the Bogoliubov transformation, and thus triggers the particle production, the analog Hawking radiation.
The effective-energy gap term in the dispersion relation of the gapped excitations gives a threshold frequency $\omega_\text{min}$ in the subsonic regime in the analog Hawking modes, below which the propagating modes do not exist.
Thus, the particle spectrum of the corresponding Hawking modes significantly deviates  from the gapless cases near the threshold frequency resulting from the modified gray-body factor, which vanishes as the mode frequency is below the threshold frequency.
The other feature is that the energy gap term introduces  the zero-frequency modes, which leads to  the density-density correlation function with the peculiar pattern of the undulations in the supersonic regime.
Their wavelength can be determined by the effective-energy gap with cone shaped boundaries depending on the March number of the condensate fluid in the supersonic regime.

The created radiations of the gapped modes will influence the quantumness of the pair of the Hawking mode in the subsonic regime and its partner in the supersonic regime of the gapless excitations by turning on the interactions between the gapless and gapped excitation through tuning their own atomic coupling constants.
We consider the limit of the small Rabi-coupling constant giving the small difference in the wave numbers for a given frequency  between the gapless and gapped excitations as compared with the inverse of the healing length $\xi$. In the hydrodynamic approximation, the corrections to the density-density correlation function of the gapless excitations due to the gapped excitations can be written in terms of the wave functions of the gapless excitations, resulting in the effective density-density correlation function.
Then, from there the effective $S$-matrix elements of the gapless excitations with the corrections from the gapped excitations can be extracted.
The measure of the quantum entanglement is according to the PHS  criterion to be computed from the obtained effective $S$-matrix elements.
The negative value of the {PHS} measure of the pair modes indicates the nature of the quantum entanglement.
It shows that the presence of the gapped excitations created from the $in$-vacuum state, although they are quantum entangled, will significantly deteriorate the quantumness of the pair modes of the gapless excitations created also from the $in$-vacuum state when the frequency of the pair modes is around the threshold frequency, $\omega \sim \omega_\text{min}$.
On top of that, when the coupling constant between the gapless and gapped excitations becomes large enough, the huge particle density of the gapped excitations in the small $\omega$ regime
will significantly disentangle  the pair modes of the gapless excitations.

Here the damping term is added from the type of the coupling between the gapless and gapped excitations and can be derived via the influence functional approach following the work of \cite{Syu2019}.
After integrating out the degrees of freedom of the gapped excitations, the corresponding Langevin equations for describing the gapless excitations can be derived with the damping term and the accompanying noise term that manifests quantum fluctuations of the gapped excitations.
One can find the time-dependent PHS measure from which to decode the more precise time scales after which  the measure is settled to some final value.
Additionally,  we may extend our analysis from 1+1  spacetime to 2+1 spacetime, to mimic the  effects from the acoustic spinning black holes.

\begin{acknowledgements}
This work was supported in part by the
Ministry of Science and Technology, Taiwan, R.O.C. under grant number 110-2112-M-259-003.
\end{acknowledgements}

\newpage
\bibliography{ref_ATBEC}

\begin{thebibliography}{51}%
\makeatletter
\providecommand \@ifxundefined [1]{%
 \@ifx{#1\undefined}
}%
\providecommand \@ifnum [1]{%
 \ifnum #1\expandafter \@firstoftwo
 \else \expandafter \@secondoftwo
 \fi
}%
\providecommand \@ifx [1]{%
 \ifx #1\expandafter \@firstoftwo
 \else \expandafter \@secondoftwo
 \fi
}%
\providecommand \natexlab [1]{#1}%
\providecommand \enquote  [1]{``#1''}%
\providecommand \bibnamefont  [1]{#1}%
\providecommand \bibfnamefont [1]{#1}%
\providecommand \citenamefont [1]{#1}%
\providecommand \href@noop [0]{\@secondoftwo}%
\providecommand \href [0]{\begingroup \@sanitize@url \@href}%
\providecommand \@href[1]{\@@startlink{#1}\@@href}%
\providecommand \@@href[1]{\endgroup#1\@@endlink}%
\providecommand \@sanitize@url [0]{\catcode `\\12\catcode `\$12\catcode
  `\&12\catcode `\#12\catcode `\^12\catcode `\_12\catcode `\%12\relax}%
\providecommand \@@startlink[1]{}%
\providecommand \@@endlink[0]{}%
\providecommand \url  [0]{\begingroup\@sanitize@url \@url }%
\providecommand \@url [1]{\endgroup\@href {#1}{\urlprefix }}%
\providecommand \urlprefix  [0]{URL }%
\providecommand \Eprint [0]{\href }%
\providecommand \doibase [0]{http://dx.doi.org/}%
\providecommand \selectlanguage [0]{\@gobble}%
\providecommand \bibinfo  [0]{\@secondoftwo}%
\providecommand \bibfield  [0]{\@secondoftwo}%
\providecommand \translation [1]{[#1]}%
\providecommand \BibitemOpen [0]{}%
\providecommand \bibitemStop [0]{}%
\providecommand \bibitemNoStop [0]{.\EOS\space}%
\providecommand \EOS [0]{\spacefactor3000\relax}%
\providecommand \BibitemShut  [1]{\csname bibitem#1\endcsname}%
\let\auto@bib@innerbib\@empty
\bibitem [{\citenamefont {Unruh}(1981)}]{Unruh1981}%
  \BibitemOpen
  \bibfield  {author} {\bibinfo {author} {\bibfnamefont {W.~G.}\ \bibnamefont
  {Unruh}},\ }\bibfield  {title} {\enquote {\bibinfo {title} {Experimental
  black-hole evaporation?}}\ }\href {\doibase 10.1103/PhysRevLett.46.1351}
  {\bibfield  {journal} {\bibinfo  {journal} {Phys. Rev. Lett.}\ }\textbf
  {\bibinfo {volume} {46}},\ \bibinfo {pages} {1351--1353} (\bibinfo {year}
  {1981})}\BibitemShut {NoStop}%
\bibitem [{\citenamefont {Visser}\ and\ \citenamefont
  {Weinfurtner}(2005)}]{Visser2005}%
  \BibitemOpen
  \bibfield  {author} {\bibinfo {author} {\bibfnamefont {Matt}\ \bibnamefont
  {Visser}}\ and\ \bibinfo {author} {\bibfnamefont {Silke}\ \bibnamefont
  {Weinfurtner}},\ }\bibfield  {title} {\enquote {\bibinfo {title} {Massive
  klein-gordon equation from a bose-einstein-condensation-based analogue
  spacetime},}\ }\href {\doibase 10.1103/PhysRevD.72.044020} {\bibfield
  {journal} {\bibinfo  {journal} {Phys. Rev. D}\ }\textbf {\bibinfo {volume}
  {72}},\ \bibinfo {pages} {044020} (\bibinfo {year} {2005})}\BibitemShut
  {NoStop}%
\bibitem [{\citenamefont {Mu{\~n}oz~de Nova}\ \emph {et~al.}(2019)\citenamefont
  {Mu{\~n}oz~de Nova}, \citenamefont {Golubkov}, \citenamefont {Kolobov},\ and\
  \citenamefont {Steinhauer}}]{Jeff2019}%
  \BibitemOpen
  \bibfield  {author} {\bibinfo {author} {\bibfnamefont {Juan~Ram{\'o}n}\
  \bibnamefont {Mu{\~n}oz~de Nova}}, \bibinfo {author} {\bibfnamefont
  {Katrine}\ \bibnamefont {Golubkov}}, \bibinfo {author} {\bibfnamefont
  {Victor~I.}\ \bibnamefont {Kolobov}}, \ and\ \bibinfo {author} {\bibfnamefont
  {Jeff}\ \bibnamefont {Steinhauer}},\ }\bibfield  {title} {\enquote {\bibinfo
  {title} {Observation of thermal hawking radiation and its temperature in an
  analogue black hole},}\ }\href {\doibase 10.1038/s41586-019-1241-0}
  {\bibfield  {journal} {\bibinfo  {journal} {Nature}\ }\textbf {\bibinfo
  {volume} {569}},\ \bibinfo {pages} {688--691} (\bibinfo {year}
  {2019})}\BibitemShut {NoStop}%
\bibitem [{\citenamefont {Kolobov}\ \emph {et~al.}(2021)\citenamefont
  {Kolobov}, \citenamefont {Golubkov}, \citenamefont {Mu{\~n}oz~de Nova},\ and\
  \citenamefont {Steinhauer}}]{Kolobov:2021wd}%
  \BibitemOpen
  \bibfield  {author} {\bibinfo {author} {\bibfnamefont {Victor~I.}\
  \bibnamefont {Kolobov}}, \bibinfo {author} {\bibfnamefont {Katrine}\
  \bibnamefont {Golubkov}}, \bibinfo {author} {\bibfnamefont {Juan~Ram{\'o}n}\
  \bibnamefont {Mu{\~n}oz~de Nova}}, \ and\ \bibinfo {author} {\bibfnamefont
  {Jeff}\ \bibnamefont {Steinhauer}},\ }\bibfield  {title} {\enquote {\bibinfo
  {title} {Observation of stationary spontaneous hawking radiation and the time
  evolution of an analogue black hole},}\ }\href {\doibase
  10.1038/s41567-020-01076-0} {\bibfield  {journal} {\bibinfo  {journal}
  {Nature Physics}\ }\textbf {\bibinfo {volume} {17}},\ \bibinfo {pages}
  {362--367} (\bibinfo {year} {2021})}\BibitemShut {NoStop}%
\bibitem [{\citenamefont {Diatlyk}(2021)}]{Diatlyk2021}%
  \BibitemOpen
  \bibfield  {author} {\bibinfo {author} {\bibfnamefont {O.}~\bibnamefont
  {Diatlyk}},\ }\bibfield  {title} {\enquote {\bibinfo {title} {Hawking
  radiation of massive fields in 2d},}\ }\href {\doibase
  10.1103/PhysRevD.104.065011} {\bibfield  {journal} {\bibinfo  {journal}
  {Phys. Rev. D}\ }\textbf {\bibinfo {volume} {104}},\ \bibinfo {pages}
  {065011} (\bibinfo {year} {2021})}\BibitemShut {NoStop}%
\bibitem [{\citenamefont {Jannes}\ \emph {et~al.}(2011)\citenamefont {Jannes},
  \citenamefont {Ma\"{\i}ssa}, \citenamefont {Philbin},\ and\ \citenamefont
  {Rousseaux}}]{Jannes2011}%
  \BibitemOpen
  \bibfield  {author} {\bibinfo {author} {\bibfnamefont {G.}~\bibnamefont
  {Jannes}}, \bibinfo {author} {\bibfnamefont {P.}~\bibnamefont {Ma\"{\i}ssa}},
  \bibinfo {author} {\bibfnamefont {T.~G.}\ \bibnamefont {Philbin}}, \ and\
  \bibinfo {author} {\bibfnamefont {G.}~\bibnamefont {Rousseaux}},\ }\bibfield
  {title} {\enquote {\bibinfo {title} {Hawking radiation and the boomerang
  behavior of massive modes near a horizon},}\ }\href {\doibase
  10.1103/PhysRevD.83.104028} {\bibfield  {journal} {\bibinfo  {journal} {Phys.
  Rev. D}\ }\textbf {\bibinfo {volume} {83}},\ \bibinfo {pages} {104028}
  (\bibinfo {year} {2011})}\BibitemShut {NoStop}%
\bibitem [{\citenamefont {Coutant}\ \emph {et~al.}(2012)\citenamefont
  {Coutant}, \citenamefont {Fabbri}, \citenamefont {Parentani}, \citenamefont
  {Balbinot},\ and\ \citenamefont {Anderson}}]{Antonin2012}%
  \BibitemOpen
  \bibfield  {author} {\bibinfo {author} {\bibfnamefont {Antonin}\ \bibnamefont
  {Coutant}}, \bibinfo {author} {\bibfnamefont {Alessandro}\ \bibnamefont
  {Fabbri}}, \bibinfo {author} {\bibfnamefont {Renaud}\ \bibnamefont
  {Parentani}}, \bibinfo {author} {\bibfnamefont {Roberto}\ \bibnamefont
  {Balbinot}}, \ and\ \bibinfo {author} {\bibfnamefont {Paul~R.}\ \bibnamefont
  {Anderson}},\ }\bibfield  {title} {\enquote {\bibinfo {title} {Hawking
  radiation of massive modes and undulations},}\ }\href {\doibase
  10.1103/PhysRevD.86.064022} {\bibfield  {journal} {\bibinfo  {journal} {Phys.
  Rev. D}\ }\textbf {\bibinfo {volume} {86}},\ \bibinfo {pages} {064022}
  (\bibinfo {year} {2012})}\BibitemShut {NoStop}%
\bibitem [{\citenamefont {Dudley}\ \emph {et~al.}(2018)\citenamefont {Dudley},
  \citenamefont {Anderson}, \citenamefont {Balbinot},\ and\ \citenamefont
  {Fabbri}}]{Dudley2018}%
  \BibitemOpen
  \bibfield  {author} {\bibinfo {author} {\bibfnamefont {Richard~A.}\
  \bibnamefont {Dudley}}, \bibinfo {author} {\bibfnamefont {Paul~R.}\
  \bibnamefont {Anderson}}, \bibinfo {author} {\bibfnamefont {Roberto}\
  \bibnamefont {Balbinot}}, \ and\ \bibinfo {author} {\bibfnamefont
  {Alessandro}\ \bibnamefont {Fabbri}},\ }\bibfield  {title} {\enquote
  {\bibinfo {title} {Correlation patterns from massive phonons in $1+1$
  dimensional acoustic black holes: A toy model},}\ }\href {\doibase
  10.1103/PhysRevD.98.124011} {\bibfield  {journal} {\bibinfo  {journal} {Phys.
  Rev. D}\ }\textbf {\bibinfo {volume} {98}},\ \bibinfo {pages} {124011}
  (\bibinfo {year} {2018})}\BibitemShut {NoStop}%
\bibitem [{\citenamefont {Tojo}\ \emph {et~al.}(2010)\citenamefont {Tojo},
  \citenamefont {Taguchi}, \citenamefont {Masuyama}, \citenamefont {Hayashi},
  \citenamefont {Saito},\ and\ \citenamefont {Hirano}}]{Tojo2010}%
  \BibitemOpen
  \bibfield  {author} {\bibinfo {author} {\bibfnamefont {Satoshi}\ \bibnamefont
  {Tojo}}, \bibinfo {author} {\bibfnamefont {Yoshihisa}\ \bibnamefont
  {Taguchi}}, \bibinfo {author} {\bibfnamefont {Yuta}\ \bibnamefont
  {Masuyama}}, \bibinfo {author} {\bibfnamefont {Taro}\ \bibnamefont
  {Hayashi}}, \bibinfo {author} {\bibfnamefont {Hiroki}\ \bibnamefont {Saito}},
  \ and\ \bibinfo {author} {\bibfnamefont {Takuya}\ \bibnamefont {Hirano}},\
  }\bibfield  {title} {\enquote {\bibinfo {title} {Controlling phase separation
  of binary bose-einstein condensates via mixed-spin-channel feshbach
  resonance},}\ }\href {\doibase 10.1103/PhysRevA.82.033609} {\bibfield
  {journal} {\bibinfo  {journal} {Phys. Rev. A}\ }\textbf {\bibinfo {volume}
  {82}},\ \bibinfo {pages} {033609} (\bibinfo {year} {2010})}\BibitemShut
  {NoStop}%
\bibitem [{\citenamefont {Modugno}\ \emph {et~al.}(2002)\citenamefont
  {Modugno}, \citenamefont {Modugno}, \citenamefont {Riboli}, \citenamefont
  {Roati},\ and\ \citenamefont {Inguscio}}]{Modugno2002}%
  \BibitemOpen
  \bibfield  {author} {\bibinfo {author} {\bibfnamefont {G.}~\bibnamefont
  {Modugno}}, \bibinfo {author} {\bibfnamefont {M.}~\bibnamefont {Modugno}},
  \bibinfo {author} {\bibfnamefont {F.}~\bibnamefont {Riboli}}, \bibinfo
  {author} {\bibfnamefont {G.}~\bibnamefont {Roati}}, \ and\ \bibinfo {author}
  {\bibfnamefont {M.}~\bibnamefont {Inguscio}},\ }\bibfield  {title} {\enquote
  {\bibinfo {title} {Two atomic species superfluid},}\ }\href {\doibase
  10.1103/PhysRevLett.89.190404} {\bibfield  {journal} {\bibinfo  {journal}
  {Phys. Rev. Lett.}\ }\textbf {\bibinfo {volume} {89}},\ \bibinfo {pages}
  {190404} (\bibinfo {year} {2002})}\BibitemShut {NoStop}%
\bibitem [{\citenamefont {Thalhammer}\ \emph {et~al.}(2008)\citenamefont
  {Thalhammer}, \citenamefont {Barontini}, \citenamefont {De~Sarlo},
  \citenamefont {Catani}, \citenamefont {Minardi},\ and\ \citenamefont
  {Inguscio}}]{Thalhammer2008}%
  \BibitemOpen
  \bibfield  {author} {\bibinfo {author} {\bibfnamefont {G.}~\bibnamefont
  {Thalhammer}}, \bibinfo {author} {\bibfnamefont {G.}~\bibnamefont
  {Barontini}}, \bibinfo {author} {\bibfnamefont {L.}~\bibnamefont {De~Sarlo}},
  \bibinfo {author} {\bibfnamefont {J.}~\bibnamefont {Catani}}, \bibinfo
  {author} {\bibfnamefont {F.}~\bibnamefont {Minardi}}, \ and\ \bibinfo
  {author} {\bibfnamefont {M.}~\bibnamefont {Inguscio}},\ }\bibfield  {title}
  {\enquote {\bibinfo {title} {Double species bose-einstein condensate with
  tunable interspecies interactions},}\ }\href {\doibase
  10.1103/PhysRevLett.100.210402} {\bibfield  {journal} {\bibinfo  {journal}
  {Phys. Rev. Lett.}\ }\textbf {\bibinfo {volume} {100}},\ \bibinfo {pages}
  {210402} (\bibinfo {year} {2008})}\BibitemShut {NoStop}%
\bibitem [{\citenamefont {Papp}\ \emph {et~al.}(2008)\citenamefont {Papp},
  \citenamefont {Pino},\ and\ \citenamefont {Wieman}}]{Papp2008}%
  \BibitemOpen
  \bibfield  {author} {\bibinfo {author} {\bibfnamefont {S.~B.}\ \bibnamefont
  {Papp}}, \bibinfo {author} {\bibfnamefont {J.~M.}\ \bibnamefont {Pino}}, \
  and\ \bibinfo {author} {\bibfnamefont {C.~E.}\ \bibnamefont {Wieman}},\
  }\bibfield  {title} {\enquote {\bibinfo {title} {Tunable miscibility in a
  dual-species bose-einstein condensate},}\ }\href {\doibase
  10.1103/PhysRevLett.101.040402} {\bibfield  {journal} {\bibinfo  {journal}
  {Phys. Rev. Lett.}\ }\textbf {\bibinfo {volume} {101}},\ \bibinfo {pages}
  {040402} (\bibinfo {year} {2008})}\BibitemShut {NoStop}%
\bibitem [{\citenamefont {McCarron}\ \emph {et~al.}(2011)\citenamefont
  {McCarron}, \citenamefont {Cho}, \citenamefont {Jenkin}, \citenamefont
  {K\"oppinger},\ and\ \citenamefont {Cornish}}]{McCarron2011}%
  \BibitemOpen
  \bibfield  {author} {\bibinfo {author} {\bibfnamefont {D.~J.}\ \bibnamefont
  {McCarron}}, \bibinfo {author} {\bibfnamefont {H.~W.}\ \bibnamefont {Cho}},
  \bibinfo {author} {\bibfnamefont {D.~L.}\ \bibnamefont {Jenkin}}, \bibinfo
  {author} {\bibfnamefont {M.~P.}\ \bibnamefont {K\"oppinger}}, \ and\ \bibinfo
  {author} {\bibfnamefont {S.~L.}\ \bibnamefont {Cornish}},\ }\bibfield
  {title} {\enquote {\bibinfo {title} {Dual-species bose-einstein condensate of
  $^{87}\mathrm{Rb}$ and $^{133}\mathrm{Cs}$},}\ }\href {\doibase
  10.1103/PhysRevA.84.011603} {\bibfield  {journal} {\bibinfo  {journal} {Phys.
  Rev. A}\ }\textbf {\bibinfo {volume} {84}},\ \bibinfo {pages} {011603}
  (\bibinfo {year} {2011})}\BibitemShut {NoStop}%
\bibitem [{\citenamefont {Cipriani}\ and\ \citenamefont
  {Nitta}(2013)}]{Cipriani2013}%
  \BibitemOpen
  \bibfield  {author} {\bibinfo {author} {\bibfnamefont {Mattia}\ \bibnamefont
  {Cipriani}}\ and\ \bibinfo {author} {\bibfnamefont {Muneto}\ \bibnamefont
  {Nitta}},\ }\bibfield  {title} {\enquote {\bibinfo {title} {Crossover between
  integer and fractional vortex lattices in coherently coupled two-component
  bose-einstein condensates},}\ }\href {\doibase
  10.1103/PhysRevLett.111.170401} {\bibfield  {journal} {\bibinfo  {journal}
  {Phys. Rev. Lett.}\ }\textbf {\bibinfo {volume} {111}},\ \bibinfo {pages}
  {170401} (\bibinfo {year} {2013})}\BibitemShut {NoStop}%
\bibitem [{\citenamefont {Fischer}\ and\ \citenamefont
  {Sch\"utzhold}(2004)}]{Fischer2004}%
  \BibitemOpen
  \bibfield  {author} {\bibinfo {author} {\bibfnamefont {Uwe~R.}\ \bibnamefont
  {Fischer}}\ and\ \bibinfo {author} {\bibfnamefont {Ralf}\ \bibnamefont
  {Sch\"utzhold}},\ }\bibfield  {title} {\enquote {\bibinfo {title} {Quantum
  simulation of cosmic inflation in two-component bose-einstein condensates},}\
  }\href {\doibase 10.1103/PhysRevA.70.063615} {\bibfield  {journal} {\bibinfo
  {journal} {Phys. Rev. A}\ }\textbf {\bibinfo {volume} {70}},\ \bibinfo
  {pages} {063615} (\bibinfo {year} {2004})}\BibitemShut {NoStop}%
\bibitem [{\citenamefont {Liberati}\ \emph
  {et~al.}(2006{\natexlab{a}})\citenamefont {Liberati}, \citenamefont
  {Visser},\ and\ \citenamefont {Weinfurtner}}]{Liberati2006}%
  \BibitemOpen
  \bibfield  {author} {\bibinfo {author} {\bibfnamefont {Stefano}\ \bibnamefont
  {Liberati}}, \bibinfo {author} {\bibfnamefont {Matt}\ \bibnamefont {Visser}},
  \ and\ \bibinfo {author} {\bibfnamefont {Silke}\ \bibnamefont
  {Weinfurtner}},\ }\bibfield  {title} {\enquote {\bibinfo {title} {Naturalness
  in an emergent analogue spacetime},}\ }\href {\doibase
  10.1103/PhysRevLett.96.151301} {\bibfield  {journal} {\bibinfo  {journal}
  {Phys. Rev. Lett.}\ }\textbf {\bibinfo {volume} {96}},\ \bibinfo {pages}
  {151301} (\bibinfo {year} {2006}{\natexlab{a}})}\BibitemShut {NoStop}%
\bibitem [{\citenamefont {Liberati}\ \emph
  {et~al.}(2006{\natexlab{b}})\citenamefont {Liberati}, \citenamefont
  {Visser},\ and\ \citenamefont {Weinfurtner}}]{Liberati2006_2}%
  \BibitemOpen
  \bibfield  {author} {\bibinfo {author} {\bibfnamefont {Stefano}\ \bibnamefont
  {Liberati}}, \bibinfo {author} {\bibfnamefont {Matt}\ \bibnamefont {Visser}},
  \ and\ \bibinfo {author} {\bibfnamefont {Silke}\ \bibnamefont
  {Weinfurtner}},\ }\bibfield  {title} {\enquote {\bibinfo {title} {Analogue
  quantum gravity phenomenology from a two-component bose{\textendash}einstein
  condensate},}\ }\href {\doibase 10.1088/0264-9381/23/9/023} {\bibfield
  {journal} {\bibinfo  {journal} {Classical and Quantum Gravity}\ }\textbf
  {\bibinfo {volume} {23}},\ \bibinfo {pages} {3129--3154} (\bibinfo {year}
  {2006}{\natexlab{b}})}\BibitemShut {NoStop}%
\bibitem [{\citenamefont {Weinfurtner}\ \emph {et~al.}(2007)\citenamefont
  {Weinfurtner}, \citenamefont {Liberati},\ and\ \citenamefont
  {Visser}}]{Weinfurtner2007}%
  \BibitemOpen
  \bibfield  {author} {\bibinfo {author} {\bibfnamefont {S.}~\bibnamefont
  {Weinfurtner}}, \bibinfo {author} {\bibfnamefont {S.}~\bibnamefont
  {Liberati}}, \ and\ \bibinfo {author} {\bibfnamefont {M.}~\bibnamefont
  {Visser}},\ }\enquote {\bibinfo {title} {Analogue space-time based on
  2-component bose-einstein condensates},}\ in\ \href {\doibase
  10.1007/3-540-70859-6_6} {\emph {\bibinfo {booktitle} {Quantum Analogues:
  From Phase Transitions to Black Holes and Cosmology}}},\ \bibinfo {editor}
  {edited by\ \bibinfo {editor} {\bibfnamefont {William~G.}\ \bibnamefont
  {Unruh}}\ and\ \bibinfo {editor} {\bibfnamefont {Ralf}\ \bibnamefont
  {Sch{\"u}tzhold}}}\ (\bibinfo  {publisher} {Springer Berlin Heidelberg},\
  \bibinfo {address} {Berlin, Heidelberg},\ \bibinfo {year} {2007})\ pp.\
  \bibinfo {pages} {115--163}\BibitemShut {NoStop}%
\bibitem [{\citenamefont {Syu}\ \emph {et~al.}(2020)\citenamefont {Syu},
  \citenamefont {Lee},\ and\ \citenamefont {Lin}}]{Syu2020}%
  \BibitemOpen
  \bibfield  {author} {\bibinfo {author} {\bibfnamefont {Wei-Can}\ \bibnamefont
  {Syu}}, \bibinfo {author} {\bibfnamefont {Da-Shin}\ \bibnamefont {Lee}}, \
  and\ \bibinfo {author} {\bibfnamefont {Chi-Yong}\ \bibnamefont {Lin}},\
  }\bibfield  {title} {\enquote {\bibinfo {title} {Regular and chaotic behavior
  of collective atomic motion in two-component bose-einstein condensates},}\
  }\href {\doibase 10.1103/PhysRevA.101.063622} {\bibfield  {journal} {\bibinfo
   {journal} {Phys. Rev. A}\ }\textbf {\bibinfo {volume} {101}},\ \bibinfo
  {pages} {063622} (\bibinfo {year} {2020})}\BibitemShut {NoStop}%
\bibitem [{\citenamefont {Hamner}\ \emph {et~al.}(2011)\citenamefont {Hamner},
  \citenamefont {Chang}, \citenamefont {Engels},\ and\ \citenamefont
  {Hoefer}}]{Hamner2011}%
  \BibitemOpen
  \bibfield  {author} {\bibinfo {author} {\bibfnamefont {C.}~\bibnamefont
  {Hamner}}, \bibinfo {author} {\bibfnamefont {J.~J.}\ \bibnamefont {Chang}},
  \bibinfo {author} {\bibfnamefont {P.}~\bibnamefont {Engels}}, \ and\ \bibinfo
  {author} {\bibfnamefont {M.~A.}\ \bibnamefont {Hoefer}},\ }\bibfield  {title}
  {\enquote {\bibinfo {title} {Generation of dark-bright soliton trains in
  superfluid-superfluid counterflow},}\ }\href {\doibase
  10.1103/PhysRevLett.106.065302} {\bibfield  {journal} {\bibinfo  {journal}
  {Phys. Rev. Lett.}\ }\textbf {\bibinfo {volume} {106}},\ \bibinfo {pages}
  {065302} (\bibinfo {year} {2011})}\BibitemShut {NoStop}%
\bibitem [{\citenamefont {Hamner}\ \emph {et~al.}(2013)\citenamefont {Hamner},
  \citenamefont {Zhang}, \citenamefont {Chang}, \citenamefont {Zhang},\ and\
  \citenamefont {Engels}}]{Hamner2013}%
  \BibitemOpen
  \bibfield  {author} {\bibinfo {author} {\bibfnamefont {C.}~\bibnamefont
  {Hamner}}, \bibinfo {author} {\bibfnamefont {Yongping}\ \bibnamefont
  {Zhang}}, \bibinfo {author} {\bibfnamefont {J.~J.}\ \bibnamefont {Chang}},
  \bibinfo {author} {\bibfnamefont {Chuanwei}\ \bibnamefont {Zhang}}, \ and\
  \bibinfo {author} {\bibfnamefont {P.}~\bibnamefont {Engels}},\ }\bibfield
  {title} {\enquote {\bibinfo {title} {Phase winding a two-component
  bose-einstein condensate in an elongated trap: Experimental observation of
  moving magnetic orders and dark-bright solitons},}\ }\href {\doibase
  10.1103/PhysRevLett.111.264101} {\bibfield  {journal} {\bibinfo  {journal}
  {Phys. Rev. Lett.}\ }\textbf {\bibinfo {volume} {111}},\ \bibinfo {pages}
  {264101} (\bibinfo {year} {2013})}\BibitemShut {NoStop}%
\bibitem [{\citenamefont {Clark}\ \emph {et~al.}(2015)\citenamefont {Clark},
  \citenamefont {Ha}, \citenamefont {Xu},\ and\ \citenamefont
  {Chin}}]{Clark2015}%
  \BibitemOpen
  \bibfield  {author} {\bibinfo {author} {\bibfnamefont {Logan~W.}\
  \bibnamefont {Clark}}, \bibinfo {author} {\bibfnamefont {Li-Chung}\
  \bibnamefont {Ha}}, \bibinfo {author} {\bibfnamefont {Chen-Yu}\ \bibnamefont
  {Xu}}, \ and\ \bibinfo {author} {\bibfnamefont {Cheng}\ \bibnamefont
  {Chin}},\ }\bibfield  {title} {\enquote {\bibinfo {title} {Quantum dynamics
  with spatiotemporal control of interactions in a stable bose-einstein
  condensate},}\ }\href {\doibase 10.1103/PhysRevLett.115.155301} {\bibfield
  {journal} {\bibinfo  {journal} {Phys. Rev. Lett.}\ }\textbf {\bibinfo
  {volume} {115}},\ \bibinfo {pages} {155301} (\bibinfo {year}
  {2015})}\BibitemShut {NoStop}%
\bibitem [{\citenamefont {Arunkumar}\ \emph {et~al.}(2019)\citenamefont
  {Arunkumar}, \citenamefont {Jagannathan},\ and\ \citenamefont
  {Thomas}}]{Arunkumar2019}%
  \BibitemOpen
  \bibfield  {author} {\bibinfo {author} {\bibfnamefont {N.}~\bibnamefont
  {Arunkumar}}, \bibinfo {author} {\bibfnamefont {A.}~\bibnamefont
  {Jagannathan}}, \ and\ \bibinfo {author} {\bibfnamefont {J.~E.}\ \bibnamefont
  {Thomas}},\ }\bibfield  {title} {\enquote {\bibinfo {title} {Designer spatial
  control of interactions in ultracold gases},}\ }\href {\doibase
  10.1103/PhysRevLett.122.040405} {\bibfield  {journal} {\bibinfo  {journal}
  {Phys. Rev. Lett.}\ }\textbf {\bibinfo {volume} {122}},\ \bibinfo {pages}
  {040405} (\bibinfo {year} {2019})}\BibitemShut {NoStop}%
\bibitem [{\citenamefont {Di~Carli}\ \emph {et~al.}(2020)\citenamefont
  {Di~Carli}, \citenamefont {Henderson}, \citenamefont {Flannigan},
  \citenamefont {Colquhoun}, \citenamefont {Mitchell}, \citenamefont {Oppo},
  \citenamefont {Daley}, \citenamefont {Kuhr},\ and\ \citenamefont
  {Haller}}]{Carli2020}%
  \BibitemOpen
  \bibfield  {author} {\bibinfo {author} {\bibfnamefont {Andrea}\ \bibnamefont
  {Di~Carli}}, \bibinfo {author} {\bibfnamefont {Grant}\ \bibnamefont
  {Henderson}}, \bibinfo {author} {\bibfnamefont {Stuart}\ \bibnamefont
  {Flannigan}}, \bibinfo {author} {\bibfnamefont {Craig~D.}\ \bibnamefont
  {Colquhoun}}, \bibinfo {author} {\bibfnamefont {Matthew}\ \bibnamefont
  {Mitchell}}, \bibinfo {author} {\bibfnamefont {Gian-Luca}\ \bibnamefont
  {Oppo}}, \bibinfo {author} {\bibfnamefont {Andrew~J.}\ \bibnamefont {Daley}},
  \bibinfo {author} {\bibfnamefont {Stefan}\ \bibnamefont {Kuhr}}, \ and\
  \bibinfo {author} {\bibfnamefont {Elmar}\ \bibnamefont {Haller}},\ }\bibfield
   {title} {\enquote {\bibinfo {title} {Collisionally inhomogeneous
  bose-einstein condensates with a linear interaction gradient},}\ }\href
  {\doibase 10.1103/PhysRevLett.125.183602} {\bibfield  {journal} {\bibinfo
  {journal} {Phys. Rev. Lett.}\ }\textbf {\bibinfo {volume} {125}},\ \bibinfo
  {pages} {183602} (\bibinfo {year} {2020})}\BibitemShut {NoStop}%
\bibitem [{\citenamefont {Horodecki}\ \emph {et~al.}(1996)\citenamefont
  {Horodecki}, \citenamefont {Horodecki},\ and\ \citenamefont
  {Horodecki}}]{Horodecki1996}%
  \BibitemOpen
  \bibfield  {author} {\bibinfo {author} {\bibfnamefont {Micha{\l}}\
  \bibnamefont {Horodecki}}, \bibinfo {author} {\bibfnamefont {Pawe{\l}}\
  \bibnamefont {Horodecki}}, \ and\ \bibinfo {author} {\bibfnamefont {Ryszard}\
  \bibnamefont {Horodecki}},\ }\bibfield  {title} {\enquote {\bibinfo {title}
  {Separability of mixed states: necessary and sufficient conditions},}\ }\href
  {\doibase https://doi.org/10.1016/S0375-9601(96)00706-2} {\bibfield
  {journal} {\bibinfo  {journal} {Physics Letters A}\ }\textbf {\bibinfo
  {volume} {223}},\ \bibinfo {pages} {1--8} (\bibinfo {year}
  {1996})}\BibitemShut {NoStop}%
\bibitem [{\citenamefont {Horodecki}(1997)}]{Horodecki1997}%
  \BibitemOpen
  \bibfield  {author} {\bibinfo {author} {\bibfnamefont {Pawel}\ \bibnamefont
  {Horodecki}},\ }\bibfield  {title} {\enquote {\bibinfo {title} {Separability
  criterion and inseparable mixed states with positive partial
  transposition},}\ }\href {\doibase
  https://doi.org/10.1016/S0375-9601(97)00416-7} {\bibfield  {journal}
  {\bibinfo  {journal} {Physics Letters A}\ }\textbf {\bibinfo {volume}
  {232}},\ \bibinfo {pages} {333--339} (\bibinfo {year} {1997})}\BibitemShut
  {NoStop}%
\bibitem [{\citenamefont {Simon}(2000)}]{Simon2000}%
  \BibitemOpen
  \bibfield  {author} {\bibinfo {author} {\bibfnamefont {R.}~\bibnamefont
  {Simon}},\ }\bibfield  {title} {\enquote {\bibinfo {title} {Peres-horodecki
  separability criterion for continuous variable systems},}\ }\href {\doibase
  10.1103/PhysRevLett.84.2726} {\bibfield  {journal} {\bibinfo  {journal}
  {Phys. Rev. Lett.}\ }\textbf {\bibinfo {volume} {84}},\ \bibinfo {pages}
  {2726--2729} (\bibinfo {year} {2000})}\BibitemShut {NoStop}%
\bibitem [{\citenamefont {Syu}\ \emph {et~al.}(2019)\citenamefont {Syu},
  \citenamefont {Lee},\ and\ \citenamefont {Lin}}]{Syu2019}%
  \BibitemOpen
  \bibfield  {author} {\bibinfo {author} {\bibfnamefont {Wei-Can}\ \bibnamefont
  {Syu}}, \bibinfo {author} {\bibfnamefont {Da-Shin}\ \bibnamefont {Lee}}, \
  and\ \bibinfo {author} {\bibfnamefont {Chi-Yong}\ \bibnamefont {Lin}},\
  }\bibfield  {title} {\enquote {\bibinfo {title} {Analogue stochastic gravity
  phenomena in two-component bose-einstein condensates: Sound cone
  fluctuations},}\ }\href {\doibase 10.1103/PhysRevD.99.104011} {\bibfield
  {journal} {\bibinfo  {journal} {Phys. Rev. D}\ }\textbf {\bibinfo {volume}
  {99}},\ \bibinfo {pages} {104011} (\bibinfo {year} {2019})}\BibitemShut
  {NoStop}%
\bibitem [{\citenamefont {Pethick}\ and\ \citenamefont
  {Smith}(2008)}]{pethick2008}%
  \BibitemOpen
  \bibfield  {author} {\bibinfo {author} {\bibfnamefont {C.~J.}\ \bibnamefont
  {Pethick}}\ and\ \bibinfo {author} {\bibfnamefont {H.}~\bibnamefont
  {Smith}},\ }\href {\doibase 10.1017/CBO9780511802850} {\emph {\bibinfo
  {title} {Bose--Einstein Condensation in Dilute Gases}}},\ \bibinfo {edition}
  {2nd}\ ed.\ (\bibinfo  {publisher} {Cambridge University Press},\ \bibinfo
  {year} {2008})\BibitemShut {NoStop}%
\bibitem [{\citenamefont {Carusotto}\ \emph {et~al.}(2008)\citenamefont
  {Carusotto}, \citenamefont {Fagnocchi}, \citenamefont {Recati}, \citenamefont
  {Balbinot},\ and\ \citenamefont {Fabbri}}]{Carusotto2008}%
  \BibitemOpen
  \bibfield  {author} {\bibinfo {author} {\bibfnamefont {Iacopo}\ \bibnamefont
  {Carusotto}}, \bibinfo {author} {\bibfnamefont {Serena}\ \bibnamefont
  {Fagnocchi}}, \bibinfo {author} {\bibfnamefont {Alessio}\ \bibnamefont
  {Recati}}, \bibinfo {author} {\bibfnamefont {Roberto}\ \bibnamefont
  {Balbinot}}, \ and\ \bibinfo {author} {\bibfnamefont {Alessandro}\
  \bibnamefont {Fabbri}},\ }\bibfield  {title} {\enquote {\bibinfo {title}
  {Numerical observation of hawking radiation from acoustic black holes in
  atomic bose{\textendash}einstein condensates},}\ }\href {\doibase
  10.1088/1367-2630/10/10/103001} {\bibfield  {journal} {\bibinfo  {journal}
  {New Journal of Physics}\ }\textbf {\bibinfo {volume} {10}},\ \bibinfo
  {pages} {103001} (\bibinfo {year} {2008})}\BibitemShut {NoStop}%
\bibitem [{\citenamefont {Recati}\ \emph {et~al.}(2009)\citenamefont {Recati},
  \citenamefont {Pavloff},\ and\ \citenamefont {Carusotto}}]{Recati2009}%
  \BibitemOpen
  \bibfield  {author} {\bibinfo {author} {\bibfnamefont {A.}~\bibnamefont
  {Recati}}, \bibinfo {author} {\bibfnamefont {N.}~\bibnamefont {Pavloff}}, \
  and\ \bibinfo {author} {\bibfnamefont {I.}~\bibnamefont {Carusotto}},\
  }\bibfield  {title} {\enquote {\bibinfo {title} {Bogoliubov theory of
  acoustic hawking radiation in bose-einstein condensates},}\ }\href {\doibase
  10.1103/PhysRevA.80.043603} {\bibfield  {journal} {\bibinfo  {journal} {Phys.
  Rev. A}\ }\textbf {\bibinfo {volume} {80}},\ \bibinfo {pages} {043603}
  (\bibinfo {year} {2009})}\BibitemShut {NoStop}%
\bibitem [{\citenamefont {Cominotti}\ \emph {et~al.}(2022)\citenamefont
  {Cominotti}, \citenamefont {Berti}, \citenamefont {Farolfi}, \citenamefont
  {Zenesini}, \citenamefont {Lamporesi}, \citenamefont {Carusotto},
  \citenamefont {Recati},\ and\ \citenamefont {Ferrari}}]{Cominotti2022}%
  \BibitemOpen
  \bibfield  {author} {\bibinfo {author} {\bibfnamefont {R.}~\bibnamefont
  {Cominotti}}, \bibinfo {author} {\bibfnamefont {A.}~\bibnamefont {Berti}},
  \bibinfo {author} {\bibfnamefont {A.}~\bibnamefont {Farolfi}}, \bibinfo
  {author} {\bibfnamefont {A.}~\bibnamefont {Zenesini}}, \bibinfo {author}
  {\bibfnamefont {G.}~\bibnamefont {Lamporesi}}, \bibinfo {author}
  {\bibfnamefont {I.}~\bibnamefont {Carusotto}}, \bibinfo {author}
  {\bibfnamefont {A.}~\bibnamefont {Recati}}, \ and\ \bibinfo {author}
  {\bibfnamefont {G.}~\bibnamefont {Ferrari}},\ }\bibfield  {title} {\enquote
  {\bibinfo {title} {Observation of massless and massive collective excitations
  with faraday patterns in a two-component superfluid},}\ }\href {\doibase
  10.1103/PhysRevLett.128.210401} {\bibfield  {journal} {\bibinfo  {journal}
  {Phys. Rev. Lett.}\ }\textbf {\bibinfo {volume} {128}},\ \bibinfo {pages}
  {210401} (\bibinfo {year} {2022})}\BibitemShut {NoStop}%
\bibitem [{\citenamefont {Larr\'e}\ \emph {et~al.}(2012)\citenamefont
  {Larr\'e}, \citenamefont {Recati}, \citenamefont {Carusotto},\ and\
  \citenamefont {Pavloff}}]{Larre2012}%
  \BibitemOpen
  \bibfield  {author} {\bibinfo {author} {\bibfnamefont {P.-\'E.}\ \bibnamefont
  {Larr\'e}}, \bibinfo {author} {\bibfnamefont {A.}~\bibnamefont {Recati}},
  \bibinfo {author} {\bibfnamefont {I.}~\bibnamefont {Carusotto}}, \ and\
  \bibinfo {author} {\bibfnamefont {N.}~\bibnamefont {Pavloff}},\ }\bibfield
  {title} {\enquote {\bibinfo {title} {Quantum fluctuations around black hole
  horizons in bose-einstein condensates},}\ }\href {\doibase
  10.1103/PhysRevA.85.013621} {\bibfield  {journal} {\bibinfo  {journal} {Phys.
  Rev. A}\ }\textbf {\bibinfo {volume} {85}},\ \bibinfo {pages} {013621}
  (\bibinfo {year} {2012})}\BibitemShut {NoStop}%
\bibitem [{\citenamefont {Michel}\ \emph {et~al.}(2016)\citenamefont {Michel},
  \citenamefont {Coupechoux},\ and\ \citenamefont {Parentani}}]{Michel2016}%
  \BibitemOpen
  \bibfield  {author} {\bibinfo {author} {\bibfnamefont {Florent}\ \bibnamefont
  {Michel}}, \bibinfo {author} {\bibfnamefont {Jean-Fran\ifmmode
  \mbox{\c{c}}\else~\c{c}\fi{}ois}\ \bibnamefont {Coupechoux}}, \ and\ \bibinfo
  {author} {\bibfnamefont {Renaud}\ \bibnamefont {Parentani}},\ }\bibfield
  {title} {\enquote {\bibinfo {title} {Phonon spectrum and correlations in a
  transonic flow of an atomic bose gas},}\ }\href {\doibase
  10.1103/PhysRevD.94.084027} {\bibfield  {journal} {\bibinfo  {journal} {Phys.
  Rev. D}\ }\textbf {\bibinfo {volume} {94}},\ \bibinfo {pages} {084027}
  (\bibinfo {year} {2016})}\BibitemShut {NoStop}%
\bibitem [{\citenamefont {Mayoral}\ \emph {et~al.}(2011)\citenamefont
  {Mayoral}, \citenamefont {Fabbri},\ and\ \citenamefont
  {Rinaldi}}]{Carlos2011}%
  \BibitemOpen
  \bibfield  {author} {\bibinfo {author} {\bibfnamefont {Carlos}\ \bibnamefont
  {Mayoral}}, \bibinfo {author} {\bibfnamefont {Alessandro}\ \bibnamefont
  {Fabbri}}, \ and\ \bibinfo {author} {\bibfnamefont {Massimiliano}\
  \bibnamefont {Rinaldi}},\ }\bibfield  {title} {\enquote {\bibinfo {title}
  {Steplike discontinuities in bose-einstein condensates and hawking radiation:
  Dispersion effects},}\ }\href {\doibase 10.1103/PhysRevD.83.124047}
  {\bibfield  {journal} {\bibinfo  {journal} {Phys. Rev. D}\ }\textbf {\bibinfo
  {volume} {83}},\ \bibinfo {pages} {124047} (\bibinfo {year}
  {2011})}\BibitemShut {NoStop}%
\bibitem [{\citenamefont {Balbinot}\ \emph {et~al.}(2019)\citenamefont
  {Balbinot}, \citenamefont {Fabbri}, \citenamefont {Dudley},\ and\
  \citenamefont {Anderson}}]{Roberto2019}%
  \BibitemOpen
  \bibfield  {author} {\bibinfo {author} {\bibfnamefont {Roberto}\ \bibnamefont
  {Balbinot}}, \bibinfo {author} {\bibfnamefont {Alessandro}\ \bibnamefont
  {Fabbri}}, \bibinfo {author} {\bibfnamefont {Richard~A.}\ \bibnamefont
  {Dudley}}, \ and\ \bibinfo {author} {\bibfnamefont {Paul~R.}\ \bibnamefont
  {Anderson}},\ }\bibfield  {title} {\enquote {\bibinfo {title} {Particle
  production in the interiors of acoustic black holes},}\ }\href {\doibase
  10.1103/PhysRevD.100.105021} {\bibfield  {journal} {\bibinfo  {journal}
  {Phys. Rev. D}\ }\textbf {\bibinfo {volume} {100}},\ \bibinfo {pages}
  {105021} (\bibinfo {year} {2019})}\BibitemShut {NoStop}%
\bibitem [{\citenamefont {Macher}\ and\ \citenamefont
  {Parentani}(2009)}]{Macher2009}%
  \BibitemOpen
  \bibfield  {author} {\bibinfo {author} {\bibfnamefont {Jean}\ \bibnamefont
  {Macher}}\ and\ \bibinfo {author} {\bibfnamefont {Renaud}\ \bibnamefont
  {Parentani}},\ }\bibfield  {title} {\enquote {\bibinfo {title} {Black-hole
  radiation in bose-einstein condensates},}\ }\href {\doibase
  10.1103/PhysRevA.80.043601} {\bibfield  {journal} {\bibinfo  {journal} {Phys.
  Rev. A}\ }\textbf {\bibinfo {volume} {80}},\ \bibinfo {pages} {043601}
  (\bibinfo {year} {2009})}\BibitemShut {NoStop}%
\bibitem [{\citenamefont {Anderson}\ \emph {et~al.}(2014)\citenamefont
  {Anderson}, \citenamefont {Balbinot}, \citenamefont {Fabbri},\ and\
  \citenamefont {Parentani}}]{Anderson2014}%
  \BibitemOpen
  \bibfield  {author} {\bibinfo {author} {\bibfnamefont {Paul~R.}\ \bibnamefont
  {Anderson}}, \bibinfo {author} {\bibfnamefont {Roberto}\ \bibnamefont
  {Balbinot}}, \bibinfo {author} {\bibfnamefont {Alessandro}\ \bibnamefont
  {Fabbri}}, \ and\ \bibinfo {author} {\bibfnamefont {Renaud}\ \bibnamefont
  {Parentani}},\ }\bibfield  {title} {\enquote {\bibinfo {title} {Gray-body
  factor and infrared divergences in 1d bec acoustic black holes},}\ }\href
  {\doibase 10.1103/PhysRevD.90.104044} {\bibfield  {journal} {\bibinfo
  {journal} {Phys. Rev. D}\ }\textbf {\bibinfo {volume} {90}},\ \bibinfo
  {pages} {104044} (\bibinfo {year} {2014})}\BibitemShut {NoStop}%
\bibitem [{\citenamefont {Fabbri}\ \emph {et~al.}(2016)\citenamefont {Fabbri},
  \citenamefont {Balbinot},\ and\ \citenamefont {Anderson}}]{Fabbri2016}%
  \BibitemOpen
  \bibfield  {author} {\bibinfo {author} {\bibfnamefont {Alessandro}\
  \bibnamefont {Fabbri}}, \bibinfo {author} {\bibfnamefont {Roberto}\
  \bibnamefont {Balbinot}}, \ and\ \bibinfo {author} {\bibfnamefont {Paul~R.}\
  \bibnamefont {Anderson}},\ }\bibfield  {title} {\enquote {\bibinfo {title}
  {Scattering coefficients and gray-body factor for 1d bec acoustic black
  holes: Exact results},}\ }\href {\doibase 10.1103/PhysRevD.93.064046}
  {\bibfield  {journal} {\bibinfo  {journal} {Phys. Rev. D}\ }\textbf {\bibinfo
  {volume} {93}},\ \bibinfo {pages} {064046} (\bibinfo {year}
  {2016})}\BibitemShut {NoStop}%
\bibitem [{\citenamefont {Coutant}\ and\ \citenamefont
  {Weinfurtner}(2018)}]{Antonin2018}%
  \BibitemOpen
  \bibfield  {author} {\bibinfo {author} {\bibfnamefont {Antonin}\ \bibnamefont
  {Coutant}}\ and\ \bibinfo {author} {\bibfnamefont {Silke}\ \bibnamefont
  {Weinfurtner}},\ }\bibfield  {title} {\enquote {\bibinfo {title}
  {Low-frequency analogue hawking radiation: The bogoliubov-de gennes model},}\
  }\href {\doibase 10.1103/PhysRevD.97.025006} {\bibfield  {journal} {\bibinfo
  {journal} {Phys. Rev. D}\ }\textbf {\bibinfo {volume} {97}},\ \bibinfo
  {pages} {025006} (\bibinfo {year} {2018})}\BibitemShut {NoStop}%
\bibitem [{\citenamefont {Belgiorno}\ \emph {et~al.}(2020)\citenamefont
  {Belgiorno}, \citenamefont {Cacciatori}, \citenamefont {Farahat},\ and\
  \citenamefont {Vigan\`o}}]{Belgiorno2020}%
  \BibitemOpen
  \bibfield  {author} {\bibinfo {author} {\bibfnamefont {F.}~\bibnamefont
  {Belgiorno}}, \bibinfo {author} {\bibfnamefont {S.~L.}\ \bibnamefont
  {Cacciatori}}, \bibinfo {author} {\bibfnamefont {A.}~\bibnamefont {Farahat}},
  \ and\ \bibinfo {author} {\bibfnamefont {A.}~\bibnamefont {Vigan\`o}},\
  }\bibfield  {title} {\enquote {\bibinfo {title} {Analog hawking effect: Bec
  and surface waves},}\ }\href {\doibase 10.1103/PhysRevD.102.105004}
  {\bibfield  {journal} {\bibinfo  {journal} {Phys. Rev. D}\ }\textbf {\bibinfo
  {volume} {102}},\ \bibinfo {pages} {105004} (\bibinfo {year}
  {2020})}\BibitemShut {NoStop}%
\bibitem [{\citenamefont {Larr{\'{e}}}\ and\ \citenamefont
  {Pavloff}(2013)}]{Larr2013}%
  \BibitemOpen
  \bibfield  {author} {\bibinfo {author} {\bibfnamefont {P.-{\'{E}}.}\
  \bibnamefont {Larr{\'{e}}}}\ and\ \bibinfo {author} {\bibfnamefont
  {N.}~\bibnamefont {Pavloff}},\ }\bibfield  {title} {\enquote {\bibinfo
  {title} {Hawking radiation in a two-component bose-einstein condensate},}\
  }\href {\doibase 10.1209/0295-5075/103/60001} {\bibfield  {journal} {\bibinfo
   {journal} {{EPL} (Europhysics Letters)}\ }\textbf {\bibinfo {volume}
  {103}},\ \bibinfo {pages} {60001} (\bibinfo {year} {2013})}\BibitemShut
  {NoStop}%
\bibitem [{\citenamefont {Peres}(1996)}]{Peres1996}%
  \BibitemOpen
  \bibfield  {author} {\bibinfo {author} {\bibfnamefont {Asher}\ \bibnamefont
  {Peres}},\ }\bibfield  {title} {\enquote {\bibinfo {title} {Separability
  criterion for density matrices},}\ }\href {\doibase
  10.1103/PhysRevLett.77.1413} {\bibfield  {journal} {\bibinfo  {journal}
  {Phys. Rev. Lett.}\ }\textbf {\bibinfo {volume} {77}},\ \bibinfo {pages}
  {1413--1415} (\bibinfo {year} {1996})}\BibitemShut {NoStop}%
\bibitem [{\citenamefont {Busch}\ and\ \citenamefont
  {Parentani}(2014)}]{Busch2014}%
  \BibitemOpen
  \bibfield  {author} {\bibinfo {author} {\bibfnamefont {Xavier}\ \bibnamefont
  {Busch}}\ and\ \bibinfo {author} {\bibfnamefont {Renaud}\ \bibnamefont
  {Parentani}},\ }\bibfield  {title} {\enquote {\bibinfo {title} {Quantum
  entanglement in analogue hawking radiation: When is the final state
  nonseparable?}}\ }\href {\doibase 10.1103/PhysRevD.89.105024} {\bibfield
  {journal} {\bibinfo  {journal} {Phys. Rev. D}\ }\textbf {\bibinfo {volume}
  {89}},\ \bibinfo {pages} {105024} (\bibinfo {year} {2014})}\BibitemShut
  {NoStop}%
\bibitem [{\citenamefont {Adamek}\ \emph {et~al.}(2013)\citenamefont {Adamek},
  \citenamefont {Busch},\ and\ \citenamefont {Parentani}}]{Adamek2013}%
  \BibitemOpen
  \bibfield  {author} {\bibinfo {author} {\bibfnamefont {Julian}\ \bibnamefont
  {Adamek}}, \bibinfo {author} {\bibfnamefont {Xavier}\ \bibnamefont {Busch}},
  \ and\ \bibinfo {author} {\bibfnamefont {Renaud}\ \bibnamefont {Parentani}},\
  }\bibfield  {title} {\enquote {\bibinfo {title} {Dissipative fields in de
  sitter and black hole spacetimes: Quantum entanglement due to pair production
  and dissipation},}\ }\href {\doibase 10.1103/PhysRevD.87.124039} {\bibfield
  {journal} {\bibinfo  {journal} {Phys. Rev. D}\ }\textbf {\bibinfo {volume}
  {87}},\ \bibinfo {pages} {124039} (\bibinfo {year} {2013})}\BibitemShut
  {NoStop}%
\bibitem [{\citenamefont {Busch}\ and\ \citenamefont
  {Parentani}(2013)}]{Busch2013}%
  \BibitemOpen
  \bibfield  {author} {\bibinfo {author} {\bibfnamefont {Xavier}\ \bibnamefont
  {Busch}}\ and\ \bibinfo {author} {\bibfnamefont {Renaud}\ \bibnamefont
  {Parentani}},\ }\bibfield  {title} {\enquote {\bibinfo {title} {Dynamical
  casimir effect in dissipative media: When is the final state nonseparable?}}\
  }\href {\doibase 10.1103/PhysRevD.88.045023} {\bibfield  {journal} {\bibinfo
  {journal} {Phys. Rev. D}\ }\textbf {\bibinfo {volume} {88}},\ \bibinfo
  {pages} {045023} (\bibinfo {year} {2013})}\BibitemShut {NoStop}%
\bibitem [{\citenamefont {Robertson}\ and\ \citenamefont
  {Parentani}(2015)}]{Robertson2015}%
  \BibitemOpen
  \bibfield  {author} {\bibinfo {author} {\bibfnamefont {Scott}\ \bibnamefont
  {Robertson}}\ and\ \bibinfo {author} {\bibfnamefont {Renaud}\ \bibnamefont
  {Parentani}},\ }\bibfield  {title} {\enquote {\bibinfo {title} {Hawking
  radiation in the presence of high-momentum dissipation},}\ }\href {\doibase
  10.1103/PhysRevD.92.044043} {\bibfield  {journal} {\bibinfo  {journal} {Phys.
  Rev. D}\ }\textbf {\bibinfo {volume} {92}},\ \bibinfo {pages} {044043}
  (\bibinfo {year} {2015})}\BibitemShut {NoStop}%
\bibitem [{\citenamefont {Lang}\ \emph {et~al.}(2020)\citenamefont {Lang},
  \citenamefont {Sch\"utzhold},\ and\ \citenamefont {Unruh}}]{Lang2020}%
  \BibitemOpen
  \bibfield  {author} {\bibinfo {author} {\bibfnamefont {Sascha}\ \bibnamefont
  {Lang}}, \bibinfo {author} {\bibfnamefont {Ralf}\ \bibnamefont
  {Sch\"utzhold}}, \ and\ \bibinfo {author} {\bibfnamefont {William~G.}\
  \bibnamefont {Unruh}},\ }\bibfield  {title} {\enquote {\bibinfo {title}
  {Quantum radiation in dielectric media with dispersion and dissipation},}\
  }\href {\doibase 10.1103/PhysRevD.102.125020} {\bibfield  {journal} {\bibinfo
   {journal} {Phys. Rev. D}\ }\textbf {\bibinfo {volume} {102}},\ \bibinfo
  {pages} {125020} (\bibinfo {year} {2020})}\BibitemShut {NoStop}%
\bibitem [{\citenamefont {Boyanovsky}\ and\ \citenamefont
  {Jasnow}(2017)}]{Boyanovsky2017}%
  \BibitemOpen
  \bibfield  {author} {\bibinfo {author} {\bibfnamefont {Daniel}\ \bibnamefont
  {Boyanovsky}}\ and\ \bibinfo {author} {\bibfnamefont {David}\ \bibnamefont
  {Jasnow}},\ }\bibfield  {title} {\enquote {\bibinfo {title} {Coherence of
  mechanical oscillators mediated by coupling to different baths},}\ }\href
  {\doibase 10.1103/PhysRevA.96.012103} {\bibfield  {journal} {\bibinfo
  {journal} {Phys. Rev. A}\ }\textbf {\bibinfo {volume} {96}},\ \bibinfo
  {pages} {012103} (\bibinfo {year} {2017})}\BibitemShut {NoStop}%
\bibitem [{\citenamefont {Syu}\ \emph {et~al.}(2021)\citenamefont {Syu},
  \citenamefont {Lee},\ and\ \citenamefont {Yeh}}]{Syu2021}%
  \BibitemOpen
  \bibfield  {author} {\bibinfo {author} {\bibfnamefont {Wei-Can}\ \bibnamefont
  {Syu}}, \bibinfo {author} {\bibfnamefont {Da-Shin}\ \bibnamefont {Lee}}, \
  and\ \bibinfo {author} {\bibfnamefont {Chen-Pin}\ \bibnamefont {Yeh}},\
  }\bibfield  {title} {\enquote {\bibinfo {title} {Entanglement of quantum
  oscillators coupled to different heat baths},}\ }\href {\doibase
  10.1088/1361-6455/abde53} {\bibfield  {journal} {\bibinfo  {journal} {Journal
  of Physics B: Atomic, Molecular and Optical Physics}\ }\textbf {\bibinfo
  {volume} {54}},\ \bibinfo {pages} {055501} (\bibinfo {year}
  {2021})}\BibitemShut {NoStop}%
\bibitem [{\citenamefont {Lee}\ \emph {et~al.}(2020)\citenamefont {Lee},
  \citenamefont {Lin},\ and\ \citenamefont {Rivers}}]{Lee2020}%
  \BibitemOpen
  \bibfield  {author} {\bibinfo {author} {\bibfnamefont {Da-Shin}\ \bibnamefont
  {Lee}}, \bibinfo {author} {\bibfnamefont {Chi-Yong}\ \bibnamefont {Lin}}, \
  and\ \bibinfo {author} {\bibfnamefont {Ray~J}\ \bibnamefont {Rivers}},\
  }\bibfield  {title} {\enquote {\bibinfo {title} {Large phonon time-of-flight
  fluctuations in expanding flat condensates of cold fermi gases.}}\ }\href
  {\doibase 10.1088/1361-648X/aba388} {\bibfield  {journal} {\bibinfo
  {journal} {J Phys Condens Matter}\ }\textbf {\bibinfo {volume} {32}},\
  \bibinfo {pages} {435101} (\bibinfo {year} {2020})}\BibitemShut {NoStop}%
\end{thebibliography}%
\end{document}